\documentclass[10pt,final,journal]{IEEEtran}
\ifCLASSINFOpdf
  \usepackage[pdftex]{graphicx}
\else
\fi
\usepackage{amsmath}
\usepackage{amsfonts}

\usepackage{caption}

\ifCLASSOPTIONcompsoc
 \usepackage[caption=false,font=normalsize,labelfont=sf,textfont=sf]{subfig}
\else
 \usepackage[caption=false,font=footnotesize]{subfig}
\fi
\usepackage{dblfloatfix}

\let\MYoriglatexcaption\caption
\renewcommand{\caption}[2][\relax]{\MYoriglatexcaption[#2]{#2}}
\usepackage{hyperref}

\usepackage{authblk}
\usepackage{bm}
\usepackage{dsfont}
\usepackage{tabularx}
\usepackage{algorithm}
\usepackage[noend]{algpseudocode}
\usepackage{multirow}
\usepackage[export]{adjustbox}
\usepackage{xcolor}
\usepackage[T1]{fontenc}

\newcommand\Tstrut{\rule{0pt}{2.6ex}}         
\newcommand\Bstrut{\rule[-0.9ex]{0pt}{0pt}}   

\begin{document}
%
\title{Adaptive multi-channel event segmentation and feature extraction for monitoring health outcomes} 
%
%
%

\author[1]{Xichen She$^*$}
\author[2]{Yaya Zhai$^*$}
\author[3]{Ricardo Henao}
\author[3]{Christopher W. Woods}
\author[4]{Christopher Chiu}
\author[3]{Geoffrey S. Ginsburg}
\author[1]{Peter X.K. Song}
\author[5]{Alfred O. Hero ~\IEEEmembership{Fellow,~IEEE}}
\affil[1]{Department of Biostatistics, University of Michigan, Ann Arbor, MI 48109}
\affil[2]{Department of Computational Medicine and Bioinformatics, University of Michigan, Ann Arbor, MI 48109}
\affil[3]{Center for Applied Genomics and Precision Medicine, Duke University, Durham, NC 27708}
\affil[4]{Department of Infectious Disease, Imperial College London, London W2 1PG}
\affil[5]{Departments of Electrical Engineering and Computer Science, Biomedical Engineering, and Statistics, University of Michigan, Ann Arbor, MI 48109}

\markboth{ }%
{She, Zhai, \MakeLowercase{\textit{et al.}}: 
Adaptive multi-channel event segmentation and feature extraction for monitoring health outcomes  }
%



\maketitle
\footnotetext{$^*$ Principal authors making equal contributions.}

\begin{abstract}
   \textbf{\textit{Objective}}: To develop a multi-channel device event segmentation and feature extraction algorithm that is robust to changes in data distribution.  
   \textbf{\textit{Methods}}: We introduce an adaptive transfer learning algorithm
   to classify and segment events from non-stationary multi-channel temporal data. Using a multivariate hidden Markov model (HMM) and Fisher's linear discriminant analysis (FLDA) the algorithm adaptively adjusts to shifts in distribution over time. The proposed algorithm is  unsupervised and learns to label events without requiring \textit{a priori} information about true event states.  
   The procedure is illustrated on experimental data collected from a cohort in a human viral challenge (HVC) study, where certain subjects have disrupted wake and sleep patterns after exposure to an H1N1 influenza pathogen. 
  \textbf{\textit{Results}}: 
   Simulations establish that the proposed adaptive algorithm significantly outperforms other event classification methods. When applied to early time points in the HVC data, the algorithm extracts sleep/wake features that are predictive of both infection 
   and infection onset time. 
   \textbf{\textit{Conclusion}}:  The proposed transfer learning event segmentation  method is robust to temporal shifts in data distribution and can be used to produce highly discriminative event-labeled features for health monitoring.   
   \textbf{\textit{Significance}}: Our integrated multisensor signal processing and transfer learning method is applicable to many ambulatory monitoring applications. 
   \footnote{This research was partially supported by the Prometheus Program of the Defense Advanced Research Projects Agency (DARPA), grant number N66001-17-2-4014. CC is supported by the NIHR Biomedical Research Centre (BRC) award to Imperial College Healthcare NHS Trust. Infrastructure support was provided by the NIHR Imperial BRC and the NIHR Imperial Clinical Research Facility. The views expressed are those of the authors and not necessarily those of the NHS, the NIHR or the Department of Health and Social Care.
  }
\end{abstract}


%
\IEEEpeerreviewmaketitle


\section{Introduction}
%
%
%
%

\IEEEPARstart{M}{any} physiological time series
involve dynamic transition among event states. For example, the transitions in human circadian cycling alternate between sleep and wake states over a 24-hour period. The sleep state can itself be a dynamic process, switching over different stages of rapid eye movement (REM) and non-REM sleep states \cite{carskadon2005normal}.   As another example, the transitions in menstrual cycles go through follicular and luteal phase \cite{silverthorn2010human},  transitioning to an ovulation state. With the emergence of cheap wearable multi-channel physiological monitoring devices, there has been much interest in automating the detection of physiological event states. The potential impact of automated event detection is far reaching, potentially improving a person's health awareness and aiding the management of disease.

A principal impediment to automating event detection for wearable devices is the intrinsically high variability of the measured physiological signals over time. An especially challenging situation is when the wearer of the device undergoes a strong perturbation, such as exposure to a pathogen that results in infection. Most event detection algorithms that are trained over a healthy baseline time period will have difficulty adapting when the device's signals become strongly perturbed away from baseline. Modern machine learning approaches to training that incorporate transfer learning can mitigate these difficulties. This paper introduces an adaptive transfer learning method for automating event detection for wearable devices and demonstrates its ability to adapt to strong perturbations from a healthy baseline in the context of a human viral challenge study. 

Another challenge to automating event detection in time series data is the lack of available ground truth event labels. The training of the automated algorithm must thus be unsupervised, based only on the observable physiological signals measured by the device. Effective unsupervised learning is only possible when features extracted from these signals have statistical distributions that strongly depend on the event states. Selection of strong discriminating features is therefore a crucial step in designing an unsupervised learning algorithm. The adaptive transfer learning algorithm introduced here generates a set of strongly discriminating event-labeled features that are based on statistical summaries of the device signals at multiple time scales.    

A state-of-the-art unsupervised algorithm for detecting unobserved event state sequences from time series data is  the hidden Markov model (HMM) \cite{zucchini2017hidden}. The HMM is a generative model in which the hidden states are latent random variables that condition the joint probability distribution of the data. 
HMM methods have been widely applied to many health applications \cite{boostani2017comparative,kortelainen2010sleep,yu2012health,trabelsi2013unsupervised,bazot2012bayesian}. 
However, the  HMM is not well adapted to data whose statistical distribution may shift over time. When there are external factors that cause perturbations to the distribution, the training data may be mismatched to the post-training data and the HMM will fail to perform as well as expected.  In the machine learning literature this phenomenon is known as  {\em data covariate shift} \cite{moreno2012unifying}. Many methods have been proposed to address the covariate shift problem, principally for supervised learning  \cite{widmer1996learning, klinkenberg2000detecting, kuncheva2009window, pan2010survey, gama2014survey}. 

The adaptive unsupervised transfer learning method for event state classification introduced in this paper is designated by the acronym HMM-FLDA. It involves learning a multivariate  HMM latent state prediction model that initializes a sequential version of Fisher's linear discriminant analysis (FLDA) to adapt the initialized HMM to perturbed post-baseline data.   Our adaptive transfer learning method is introduced as the key component of a complete data processing pipeline that includes pre-processing  for local feature extraction and abnormal signal detection, adaptive event classification using HMM-FLDA, and post-processing for constructing strong high dimensional event-labeled features for inferring perturbation-dependent events, e.g., health outcome after exposure to a viral pathogen.    

We illustrate the proposed analysis pipeline for a human viral challenge (HVC) experiment in which physiological data from multiple subjects is collected over multiple days from wearable wristband devices.  On the second day of the experiment the subjects were exposed to a H1N1 flu virus that caused some to become infected, as clinically determined by viral shedding which does not occur before the 4th day.  For each subject, the HMM-FLDA is trained on the first two days and nights of wearable data to generate sleep/wake features. Using these features we are able to predict if the subject will get infected and the onset time, defined as the first day that shedding is detected.

The remainder of this paper is organized as follows: we introduce in detail the proposed transfer learning algorithm in Section~\ref{sec:sleep_detect}. Section~\ref{sec:simulation} presents a numerical simulation study emulating the HVC experimental data, but with ground truth event labels. 
Section~\ref{sec:analysis} applies the proposed analysis pipeline to wearable data collected in the experimental HVC study. 
In Section~\ref{sec:discussion} we end the paper with discussion and conclusions. Supporting information on the algorithm, the pipeline, and the HVC data is presented in the supplementary Appendices.

\section{Adaptive unsupervised event monitoring}\label{sec:sleep_detect}
\def\bfX{{\bm{X}}}
\def\bfx{{\bm{x}}}
\def\bfy{{\bm{y}}}
\def\bfA{{\bm{A}}}

\begin{figure*}[!ht]
	\centering
	  \subfloat[]{\includegraphics[width = 0.9\linewidth, valign=t]{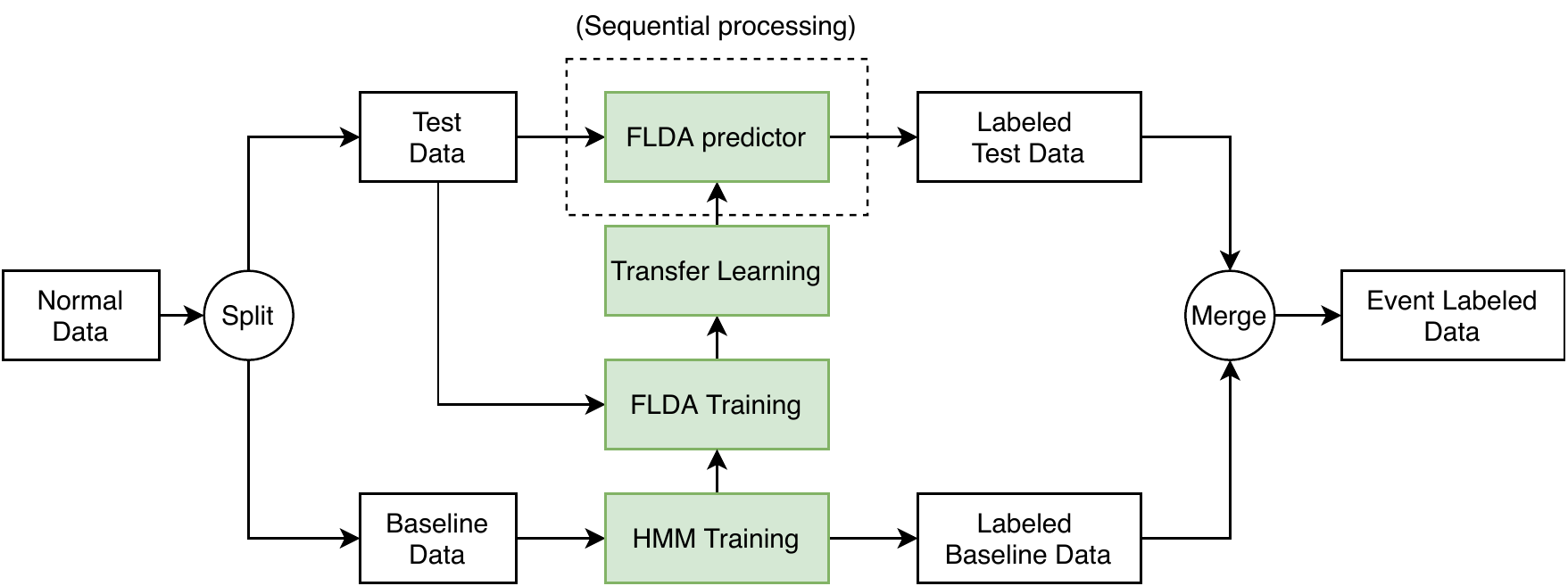}
	  \label{fig:HMMFLDA}}
	\hfil
    	\subfloat[]{\includegraphics[width = 0.9\linewidth, valign=t]{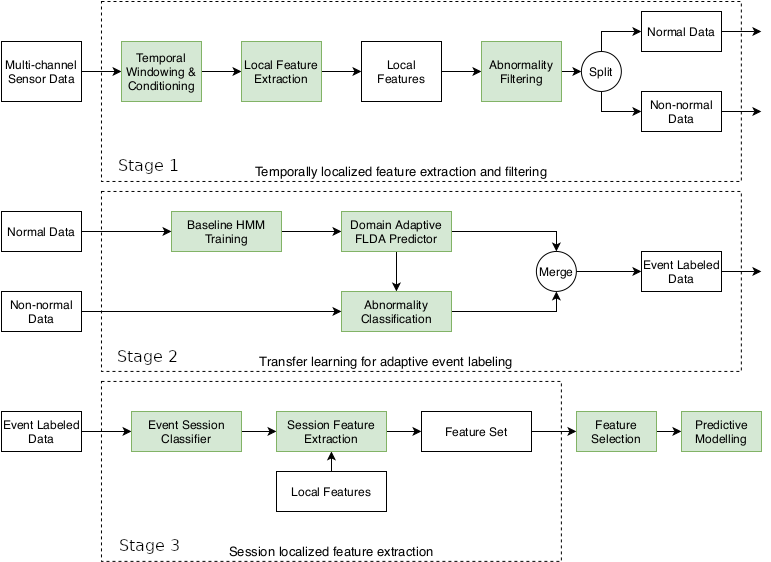}
    	\label{fig:pipe}} 
	\caption{\textbf{(a)} The proposed HMM-FLDA unsupervised transfer learning algorithm based on a multivariate hidden Markov model (HMM) and Fisher's linear discriminant analysis (FLDA) to segment event states; \textbf{(b)} The three stages of proposed data processing pipeline, discussed in Section \ref{sec:experimentdesign}. The HMM-FLDA procedure depicted in \textbf{(a)} appears in the second stage of the pipeline.}
\end{figure*}

Figure \ref{fig:HMMFLDA} summarizes the proposed adaptive event labeling procedure and Algorithm \ref{alg1} provides pseudo-code for its principal steps. Figure \ref{fig:pipe} shows the entire data processing and analysis pipeline, from sensor data capture to event-labeled feature extraction to predictive models using these features. 

\begin{algorithm*}[!ht]
	\caption{HMM-FLDA gradual self-training event state classifier} 
	\begin{algorithmic}[1]
	\State Initialize with: $\{\hat{y}_{t},\bm{x}_t\}_{t=1}^{t_{N_0}}$   \Comment{Use HMM to classify baseline data}
	\While{$t_{N_0}<t < t_N$}    \Comment{Continue over all batches of samples}
		\For{$l$ in $1$ to $L$}  \Comment{Iterate over candidate training sizes $d_l$}
				\Procedure{Fisher's LDA}{train = $(t - d_l,t]$, test = $(t, t + \Delta_t]$}
					\State $\bm{w} = \bm{S}_W^{-1}(\bar{\bm{x}}_{1\cdot} -  \bar{\bm{x}}_{0\cdot})$       \Comment{Update FLDA weights on train data}
					\State $z_{t} = \bm{w}^T \bm{x}_{t}$
					\State $\hat{y}_{t}(d_l) = \mathds{1}\bigl\{(z_{t}-\bar{z}_{0\cdot})^2/\hat\sigma^2_0 - (z_{t}-\bar{z}_{1\cdot})^2/\hat\sigma^2_1 > \log(\gamma \, \hat\sigma_1^2/\hat\sigma_0^2) \bigr\}$ \Comment{Classify test data with updated weights}
					\State $\mbox{SI}(d_l) = \mbox{SI}(\mbox{train}, \mbox{test}; d_l)$   \Comment{Compute separability index (\ref{eqn:si})}
				\EndProcedure
		\EndFor
		\State $d_{(1)} = \mbox{\textbf{which.max}}(\mbox{SI}(d_l))$ \Comment{Select optimal training window length}
		\State $\hat{y}_{t} = \hat{y}_{t}( d_{(1)})$ \Comment{Classified labels of current batch}
		\State $t = t + \Delta_t$  \Comment{Move on to the next  batch}
	\EndWhile
	\end{algorithmic}
	\label{alg1}
\end{algorithm*}


The pre-processed data comes in the form of a $p \times N$ matrix  $\bm{X}=[\bm{x}_{t_1}, \dots, \bm{x}_{t_{N}} ]$ where $\bm{x}_t\in \mathbb R^p$ is a $p$-dimensional real valued feature vector, available at a set of time instants $\{t_i\}_{i=1}^{N}$.
We assume a statistical time series model for $\bm{x}_t$ with hidden states that determine the joint probability distribution $p_{\bm{X}} (\bm{x}_1,\ldots, \bm{x}_{N}).$  Specifically, associated with each time sample $\bm{x}_t$ is a latent (hidden) variable 
$y_t \in \{0, 1, \ldots, K\}$, which labels the  event associated with $\bm{x}_t$. Later in the paper we will specialize to the binary case where $K=2$ and the events correspond to a person's sleep or wake state at a particular time. We denote  $\bfy=[y_{t_1}, \ldots, y_{t_N}]$ the time sequence of a subject's hidden event states. 
In the time series setting, $\bm{y}$ determines the conditional probability distribution $p_{\bm{X}|\bm{y}}$ and the marginal probability distribution $p_{\bm{X}}$. We construct an event classifier function based on an HMM with a parametric (Gaussian) conditional distribution of $\bm{x_t}$ given $y_t$ that may gradually shift over time, e.g., due to a perturbation after exposure to a pathogen. The parameters of the classifier function are estimated from the data in a two phase process involving: (1) an initial training phase where the HMM  model is fitted to a batch of (healthy) baseline data, followed by (2) an adaptation phase where a naive Bayes Gaussian model is adapted to future batches of possibly perturbed data using Fisher Linear Discriminant Analysis (FLDA) in a sequential transfer learning framework. 


\subsection{Baseline training: HMM initial segmentation}
For $N_0<N$ define the {\em baseline} (BL) segment of the data $\bfX$ as the first $N_0$ time samples $\bfX_{\rm{BL}} =[\bm{x}_{t_1}, \ldots, \bm{x}_{t_{N_0}}]$ and the {\em post-baseline} (PBL) segments of the data as the remaining part $\bfX_{\rm{PBL}}= [\bm{x}_{t_{N_0+1}}, \ldots, \bm{x}_{t_N}]$. The BL segment is used as a training set for learning the parameters of a multivariate HMM \cite{visser2010depmixs4}. The HMM is a Markov model: modeling the observations as an $N$-length segment of a Markov random process  $\{(\bm{x}_{t},y_{t})\}_{t=-\infty}^{\infty}$, where transitions $y_{t-1} \rightarrow y_{t}$  between event states have probabilities specified by a  ${K\times K}$ state transition matrix $\bfA$. The Markov property implies that the joint distribution $p_{\bfX,\bfy}$ factorizes as:  $p_{\bfX,\bfy} =p_{\bfx_{t_1},y_{t_1}}\prod_{i=2}^{N_0} p_{\bfx_{t_i},y_{t_i}|\bfx_{t_{i-1}},y_{t_{i-1}}}$, implying an analogous factorization of the marginal $p_{\bfX}$ and
the conditionals $p_{\bfX|\bfy}$. In the special case of a Gaussian  HMM model each factor $p_{\bfx_t|y_t}$ is a multivariate Gaussian conditional distribution, denoted as      
\begin{equation}
\bm{x}_t \mid (y_t = k) \sim N(\bm{\mu}_k, \mathbf \Sigma_k), \;\;\; k = 1,\dots,K,  
\label{eq:mGauss}
\end{equation}
where $ \bm{\mu}_k$ is a $p$-dimensional mean parameter and $\bm{\Sigma}_k$ is a $p\times p$ covariance matrix parameter that must be learned in addition to the state transition matrix $\bfA$. 

There are several methods available for learning multivariate HMM's \cite{gao2011composite, zucchini2017hidden} 
which could be adapted to our setting. Most methods are iterative and many use a variant of the expectation-maximization (EM) algorithm to find maximum likelihood estimates of the unknown parameters. Important practical considerations are whether the HMM algorithm provides a final estimate of labeling accuracy, important for self diagnostics,  if it converges sufficiently rapidly to a global maximum,  and if it has fast computation time per iteration, which tends to increase as $K$ and $p$ increase. In some applications, it may be desirable to estimate the number of event states $K$, for which automated model selection versions of HMM are available. For a comprehensive review of HMM implementations the reader can refer to  \cite[Ch. 8]{zucchini2017hidden}. In our pipeline we use a reasonably fast multivariate HMM algorithm implemented by EM iterations that approximates the maximum likelihood estimator of the HMM parameters. In the Expectation (E) step, the marginal likelihood is computed by a variant of the forward-backward algorithm \cite{lystig2002exact}, which calculates the gradient of the log-likelihood (score) function in a single pass.

\subsection{Post-baseline adaptation: sequential transfer learning}

If the feature distribution varies over time, the static baseline-trained HMM will have difficulty classifying and segmenting events in the post-baseline data $\bfX_{\rm{PBL}}$. To address this difficulty we introduce a transfer learning strategy that is initialized with the HMM on the initial-training data and sequentially updates the event classifier over successive batches of test data, continually adapting to changes in distribution.  In transfer learning the batches are called target domains and the objective is to design a classifier that continually adapts to them, a property called domain adaptation \cite{karlen2011adaptive}, \cite{moreno2012unifying}. To achieve this objective, we use an unsupervised {\em gradual self-training} procedure. In supervised transfer learning, ``self-training" means that domain adaptation is done with an unlabeled test set and ``gradual" means that the adaptation is done sequentially over time, updating over successive batch pairs.  As compared to direct self-training procedures, which try to adapt to all test batches at once,  gradual self-training procedures are better suited to online applications and are more robust to smoothly varying shifts in distribution.  Recent theory establishes that gradual procedures are provably better when the class distributions on successive pairs of batches are close in Wasserstein distance  \cite{kumar2020understanding}. This closeness condition is satisfied when the distributions are quasi-stationary, an assumption common in time series analysis \cite{adak1998time} and online supervised learning \cite{shalev2012online}.   


Our unsupervised gradual self-training procedure consists of three components: 1) an unsupervised version of Fisher's LDA that uses class labels predicted by the HMM to project the data $\bfX_{\rm{BL}}$ to a one dimensional space that maximally separates these classes.  2) application  of this projection to the next batch of samples in $\bfX_{\rm{PBL}}$ followed by event classification  using a naive Bayes classifier; 3) using the class labels to update the projection by reapplying Fisher's LDA to the batch filtered by a tapered sliding-window. This process is sequentially repeated for each successive batch, resulting in a continuous adaptation of the event classifier. 

Fisher's LDA is most commonly applied in the context of supervised learning for dimension reduction and classification \cite{fisher1936use,friedman2001elements} when a labeled sample is available for training. Here we adapt LDA to the unsupervised context of latent event classification. After applying HMM to the baseline sample $\bfx_{t_1},\ldots, \bfx_{t_{N_0}}$ we obtain estimated class labels $\hat{y}_{t_1}, \ldots, \hat{y}_{t_{N_0}}$. We then train Fisher's LDA classifier on the putatively labeled sample $\{(\bfx_{t_1},\hat{y}_{t_1}), \ldots, (\bfx_{t_{N_0}},\hat{y}_{t_{N_0}})\}$. 
Fisher's LDA uses dimensionality reduction to learn a classifier. By applying a weight $\bm{w} \in \mathbb R^p$ to $\bfx_t$ FLDA generates a projection score $z_t=\bm{w}^T\bfx_t$. The weight vector is optimized so that these scores attain the largest possible separation of the classes. Fisher's LDA accomplishes classification of a novel sample $\bfx_{t'}$ by using the trained weights to compute the projected score $\bm{w}^T\bfx_{t'}$ that is used in an optimal naive Bayes LDA classifier to predict its unknown label $y_{t'}$.  

Specifically, for the case of $K=2$ classes $y_t\in\{0,1\}$, in the dimensionality reduction stage of FLDA
the class separability of the projection scores $z_t = \bm{w}^T\bm{x}_t$ is measured by the following ratio of \textit{between-class} variation and \textit{within-class} variation:
\begin{equation}\label{eqn:J1}
  J(\bm{w}) = \frac{(\bar{z}_{1\cdot} - \bar{z})^2 + (\bar{z}_{0\cdot} - \bar{z})^2}{\sum_{t\in G_1} (z_{t} - \bar{z}_{1\cdot})^2 + \sum_{t\in G_0} (z_{t} - \bar{z}_{0\cdot})^2} \; ,
\end{equation}
where $G_k=\{t \leq N_0: y_t=k\}$
denotes the data indices of $\bm{x}_t$ coming from event class $k$,  $\bar{z}_{k\cdot} =  |G_k|^{-1} \sum_{t \in G_k} z_t $ for $k \in\{ 0,  1\}$ are the class-specific sample mean projected score and $\bar{z}$ is the overall sample mean projected score. The optimal $\bm{w}$ that maximizes (\ref{eqn:J1}) is \cite{friedman2001elements}
\begin{equation}\label{eqn:optw}
  \bm{w} = \bm{S}_W^{-1}(\bar{\bm{x}}_{1\cdot} - \bar{\bm{x}}_{0\cdot}),
\end{equation}
where $\bar{\bm{x}}_{k\cdot} = |G_k|^{-1}\sum_{t \in G_k} \bm{x}_t$ is the class-specific sample mean and $\bm{S}_W = \sum_{k = 0,1}\sum_{t\in G_k}(\bm{x}_t - \bar{\bm{x}}_{k\cdot})(\bm{x}_t - \bar{\bm{x}}_{k\cdot})^T$ is the \textit{within-class} scatter matrix of the $\bm{x}_t$'s.

Under the multivariate normal model (\ref{eq:mGauss}) for $\bm{x}_t$, the $z_t$'s are conditionally independent Gaussian distributed random variables with conditional mean $\mu_{k}$ and conditional variance $\sigma^2_{k}$ given $y_t=k$. Assuming independent conjugate priors on $\mu_k$ and $\sigma^2_k$, conditioned on the training data $\bfX_{\rm{BL}}$, the maximum a posteriori (MAP) estimator of the class label $y_{t'}$ of a novel test sample $\bfx_{t'}$  based on its projected score $z_{t'}$ is given by the naive Bayesian classifier
\cite{friedman2001elements}:
\begin{equation}\label{eqn:decision_rule}
y_{t'} =\mathds{1}\left\{\frac{(z_{t'}-\bar{z}_{0\cdot})^2}{\hat\sigma^2_0} - \frac{(z_{t'}-\bar{z}_{1\cdot})^2}{\hat\sigma^2_1} > \log\left(\gamma \,\frac{\hat\sigma_1^2}{\hat\sigma_0^2}\right) \right\},
\end{equation}
where $\mathds{1}(\mathcal A)$ is the binary indicator function equaling 1 when the logical preposition $\mathcal A$ is true.  
Here, as above, $\bar{z}_{k}$ is the class-specific sample mean and 
$\hat{\sigma}^2_k$ the within class sample variance of $\{z_t: t>N_0\}$ 
$ \hat\sigma^2_k = (|G_k|-1)^{-1}\sum_{t\in G_k}(z_t - \bar{z}_{k\cdot})^2$.
The threshold parameter $\gamma$ in (\ref{eqn:decision_rule}) is the prior odds ratio $P(y_{t'}=0)/(1-P(y_{t'}=0))$ and can be adjusted to account for class imbalance and to trade-off the two types of classification errors. 
The naive Bayes classifier classifies the label of the test sample based on its relative distance to each of the class centroids $\bar{z}_k$ weighted by the class dispersions $\hat\sigma^2_k$.

We next describe the proposed unsupervised transfer learning procedure using gradual self-training to sequentially update the HMM-FLDA algorithm as, outlined in Algorithm 1 and depicted in Fig. \ref{fig:HMMFLDA}. 
The post-baseline data is successively divided into disjoint test batches of length  $\Delta_t$ secs., called the test window length, a tuning parameter whose selection is application dependent. 
The current batch is used as a test sample to update the self-trained HMM-FLDA classifier determined from the previous batch.  Let $t_n^{tst}=t_{n-1}^{tst}+\Delta t$ be the start time of the $n$-th test window and  denote the $n$-th test batch as $\mathcal T_{tst}=\{\bm{x}_{t_i}\}_{t_i\in [t_n^{tst}, t_{n}^{tst}+\Delta_t)}$.  The update proceeds as follows. For a specified size $d$, define the $n$-th training set $\mathcal T_{trn}=\{\bm{x}_{t_i}\}_{t_i\in [t_{n}^{tst}-d,t_{n}^{tst})}$. Note that the feature instances in $\mathcal T_{\mathrm trn}$ have already been assigned class labels $\{\hat{y}_{t_i}\}_{t_i\in [t_{n}^{tst}-d,t_{n}^{tst})}$ in the previous update of HMM-FLDA. Hence, the first step in updating the classifier is to compute the updated FLDA weight vector $\bm{w}$, 
as defined in (\ref{eqn:optw}), using  $\mathcal T_{\mathrm trn}$ and its previously assigned class labels.   The second step is to use this updated weight vector to assign predicted labels $\{\hat{y}_{t_i}\}_{t_i\in [t_{n}^{tst},t_{n}^{tst}+\Delta t)}$  to instances in $\mathcal T_{\mathrm tst}$. 

The size $d$ of $\mathcal T_{\mathrm trn}$ is adapted from batch to batch by optimizing a separability index (SI) defined over $\mathcal T_{\mathrm trn} \cup \mathcal T_{\mathrm tst}$.  
Among the many possible SI measures that could be used, we adopt Thornton's SI  \cite{thornton1998separability}, a competitive measure for assessing class separation originally introduced for supervised classification problems \cite{anthony2007comparison,cerdeira2012combinatorial,greene2001feature}.   Also called the geometric separability index (GSI), Thornton's measure has seen wide application in health, robotics, geology and other fields \cite{mishra2010local,coelho2014assessing,schadd2014feature,tuci2010active,rossi2015spatio}.
We use the following simple unsupervised modification of the GSI 
\cite{thornton1998separability}, computed on the merged training and test samples for the batch: 
\begin{equation}
  \mbox{SI}(d) = \frac{\sum_{t \in \mathcal{T}_{trn}(d) \cup \mathcal{T}_{tst}} \{\hat{y}_t + \tilde{y}_t + 1\}\mbox{ \textbf{MOD} }2}{|\mathcal{T}_{trn}(d)| + |\mathcal{T}_{tst}|},
  \label{eqn:si}
\end{equation}
where 
$\tilde{y}_t$ is the predicted class of the nearest neighbor of $\bm{x}_t$ in the set of merged training and test samples. As compared to Thornton's original SI, defined for supervised classification problems where the true labels $y_t$ are known, the unsupervised modification  (\ref{eqn:si}) uses the predicted labels $\hat{y}_t$ in place of $y_t$. 
Here the nearest neighbors are determined by the ``projection distance'' defined, for samples at time $t$ and $t'$, as
$  d(t,t') = |\hat{\bm{w}}^T(\bm{x}_t - \bm{x}_{t'})|.$
Rank ordering  the SI indices $\mbox{SI}(d_{(1)})>\ldots> \mbox{SI}(d_{(L)})$ yields $d_{(1)}$ as the optimal training window length for the batch.

\section{Numerical simulation study} \label{sec:simulation}

We performed a simulation of HMM-FLDA that emulates the experimental study presented in Section \ref{sec:analysis}, but with ground truth event states. Two dimensional data  $\bm{x}_t=[x^{(2)}_{t},x^{(2)}_{t}]$ is simulated from conditional distributions given the latent event state $y_t$, which randomly switches between wake ($y_t=0$) and sleep ($y_t=1$) with a mean cadence of approximately 24 hours. To emulate the perturbation effect of viral infection, each of these conditional distributions are fixed during the baseline training period but may undergo slowly varying time shifts in the post-baseline period. Two different post-baseline scenarios are simulated, called the stable case (no shift) and the unstable case (slow shift). 


For the stable case, we generate the sequence of states as a realization of a Markov process with  state transition probabilities that are empirically matched to those extracted from the HVC data described in the next section. We fix the number of state transitions $T$ and the initial state is set to $y_0=0$. The duration $\Delta\tau_1$, i.e., the time to the first transition, is drawn from $TN(\mu_0, \sigma_0)$, a truncated normal (TN) distribution with location and scale parameters $\mu_0$ and $\sigma_0$. The duration $\Delta\tau_2$ of the second event is independently drawn from a TN distribution with mean $\mu_1=24-\mu_0$ and variance $\sigma_1^2=\sigma_0^2$. This process is repeated until the $T$-th transition variable $\Delta\tau_{T}$ has been drawn. The discretized transition times $\tau_m=\sum_{i=1}^m\Delta \tau_i$, $m=1, \ldots, T$, specify the state sequence  $\{y_t, t=1, \ldots, N\}$. 
A consecutively occurring pair of (wake, sleep) periods 
is called a session. Over a given session $m$, the data channels $x^{(i)}_{t}$, $i=1,2$, were drawn independently from a truncated normal distribution $p(x^{(1)}_{t}|y_t=k)$ of the form $TN(u^{(1)}_{k}, \sigma^{(1)}_{k})$ and a log-normal distribution $p(x^{(2)}_{t}|y_t=k)$ of the form $LN(u^{(2)}_{k}, \sigma^{(2)}_{k})$, whose translation and scale parameters $u$ and $\sigma$  depend on the state $k=0,1$. These two distributions emulate  the median heart rate feature and the accelerometer standard deviation feature used in Section \ref{sec:fulltimecoure_segmentation} for the full time course HVC data.  

For the unstable case, the parameters of the distributions of $y_t$ and $\bm{x}_t$ were changed from session to session to reflect perturbations during the post-baseline period. 
Specifically, the truncated normal parameters $(\mu_{mk}, \sigma_{mk})$ of the event durations $\Delta \tau_m$ were made to be session dependent by matching these parameters to the empirical distributions of the HMM-FLDA estimated duration for the $m$-th session, where the empirical distribution was constructed over the sub-population of symptomatic subjects in the HVC study. 
Furthermore, we introduced a time-varying post-baseline conditional mean $u^{(i)}_{mk} =E[x^{(i)}_t|y_t=k]$ of the $i$-th channel over the $m$-th session for the $k$-th event state. 
In particular, the post-baseline conditional mean was modeled as a quadratically varying function of $m$: 
    $u^{(i)}_{mk} = u^{(i)}_{1k}-b_{k}^{(i)}\left(m-m_o\right)^2/
    \left(1-m_o\right)^2 + b_{k}^{(i)},$ 
for $m = 1, \dots M, \; i = 1, 2, \; k \in\{ 0, 1\}$. Here $m_o$ is the index of the session where the mean $u^{(i)}_{mk}$ reaches a positive or negative apex, and $b_{k}^{(i)} = u^{(i)}_{m_oj} - u^{(i)}_{1k}$ is the difference between the apex value $u^{(i)}_{m_ok}$ and the initial value $u^{(i)}_{mk}$. The pair $(m_o, b_{k}^{(i)})$ controls how the mean values of $(x^{(1)}_t, x^{(2)}_t)$ change over sessions. In the simulation, $m_o$ was randomly drawn from $\{5,\, 6,\, 7\}$ with equal probability, and we considered 2 sets of values for the $b_{k}^{(i)}$'s. 
To achieve a close facsimile to the experimental data analyzed in Section \ref{sec:analysis}, we simulated $M=11$ sessions, and the time between samples was set to $\delta_t = 1/6$ hr, i.e., the 10 minute aggregated sample period used in that section. Furthermore, all of the translation and scale parameters 
used in the simulation were matched to summary statistics obtained from the HMM-FLDA analysis described in Section \ref{sec:fulltimecoure_segmentation}.  Several realizations of the simulated data are shown in Fig. \ref{fig:simu_realizations} in the Appendix.

\begin{table*}[ht]
  \centering
  \caption{Out of sample performance of the proposed adaptive transfer learning algorithm as compared with standard HMM operating on the original data (HMM) and operating on LOESS detrended data (dHMM).}
  \resizebox{0.8\textwidth}{!}{%
  \begin{tabular}{l l l  c c c c}
    \hline\hline
    Setting & Methods & Accuracy  & F1 & Cosine dis. & Onset diff. & Duration diff. \Tstrut\Bstrut\\[+1pt]
    \hline \Tstrut
    stable & HMM  & \textbf{0.9981} & \textbf{0.9987} & \textbf{0.9968} & \textbf{0.0165} & \textbf{0.0307} 
    \\[+0pt]
    \Tstrut
           & dHMM  & 0.9981 & 0.9986 & 0.9967 & 0.0253 & 0.0373
           
     \\[+0pt]
    \Tstrut
           & Proposed  & 0.9980 & 0.9986 & 0.9964 & 0.0168 & 0.0313 
           \\[+1pt]
    \hline\Tstrut\Bstrut
    unstable++ & HMM  &  0.8959 & 0.9209 & 0.8559 & 1.6704 & 2.9169 
    \\[+0pt]
               \Tstrut
               & dHMM  &  0.9127 & 0.9231 & \textbf{0.8957} & 2.4709 & 2.2653  
    \\[+0pt]
               \Tstrut
               & Proposed  & \textbf{0.9356} & \textbf{0.9544} & 0.8904 & \textbf{1.0679} & \textbf{1.5756}  
    \\[+0pt]
    \hline\Tstrut\Bstrut
    unstable+- & HMM  & 0.9369 & 0.9510 & 0.9187 & \textbf{0.2763} & 1.9750 
    \\[+0pt]
               \Tstrut
               & dHMM  & 0.9271 & 0.9370 & 0.9114 & 1.0170 & 2.0131 
    \\[+0pt]
               \Tstrut
               & Proposed  &  \textbf{0.9758} & \textbf{0.9833} & \textbf{0.9564} & 0.3712 & \textbf{0.3817}  
    \\[+0pt]
    \hline\hline \Tstrut\Bstrut
  \end{tabular}%
  }
  \label{table:simulation_resultOoS}
\end{table*}

Table \ref{table:simulation_resultOoS} compares the empirical performance (a total of 1000 independent trials on 100 different realizations of the model) achieved by the proposed HMM-FLDA procedure to  other procedures (HMM and dHMM) for the stable and unstable cases described above.  Each of the stable and unstable cases is characterized by the value of the four coefficients $(b_{0}^{(1)}, b_{1}^{(1)}, b_{0}^{(2)}, b_{1}^{(2)})$ that define the trend in the post-baseline mean $u^{(i)}_{mk}$. For a stable subject there is no mean trend for either event class and all of these coefficients are zero. Two unstable cases are considered: a case called {\em unstable++}, where the trends for both event classes are concave with a randomly located peak, and a case called {\em unstable+-}, where for each event class the trends go in opposite directions, i.e., one trend is concave while the other is convex.  The table shows the value of 5 performance criteria (averaged over 1000 independent trials) when the method is trained on a single realization of the two signals and tested on an independent out-of-sample realization drawn from the same distribution. The performance criteria are: 
\begin{itemize}
    \item \texttt{Accuracy}: the number of samples with correctly classified event states divided by the total number of samples,
    \item \texttt{F1}: the harmonic mean of the precision and recall classification criteria,  
    \item \texttt{Cosine dis.}: the cosine distance between the vector of classified event states and the ground truth states,
    \item \texttt{Onset diff.} the absolute difference between estimated onset time of a sleep session and the nearest onset time of a true sleep period, averaged over sessions,
    \item \texttt{Duration diff.} the absolute difference between estimated duration of a sleep session and the duration of a true sleep period that overlaps with the predicted one, averaged over sessions. (If the predicted period doesn't overlap with any true period, the difference is taken as the length of the predicted period)
\end{itemize}

Table \ref{table:simulation_resultOoS} compares the mean performance of the proposed method to a standard off-the-shelf two-state HMM state estimator and a variant of HMM, called dHMM, that uses locally estimated scatterplot smoothing (LOESS) \cite{cleveland1979robust} to detrend each of the two data channels as a preprocessing step before training the HMM. The HMM and LOESS were implemented with R packages \texttt{depmixS4} \cite{visser2010depmixs4} and \texttt{fANCOVA}. 
The results shown in Table \ref{table:simulation_resultOoS} establish the benefit of our proposed HMM-FLDA adaptive procedure. 
In terms of event classification performance (Accuracy, F1, and Cosine distance), except for cosine distance in the unstable++ case, where the performances are statistically equivalent,  the proposed method significantly outperforms the competing HMM and dHMM procedures for both unstable++ and unstable+- models (p-value $\ll$0.01 according to one-sided paired t-test). 
It is also worth noting that, in the unstable++ case, LOESS detrending of dHMM  improves on HMM across all criteria except for onset diff., which is the r.m.s. estimation error for sleep onset time. The Supplementary Appendix \ref{sec:simulationstudy_appendix} provides additional details on this simulation.

\begin{table*}[ht]
  \centering
  \caption{Parameter settings for the HVC data processing pipeline}
  \resizebox{0.8\textwidth}{!}{%
  \begin{tabular}{l l l l}
    \hline\hline
    Symbol & Value & Description  & Loc. in pipeline \Tstrut\Bstrut\\[+3pt]
    \hline \Tstrut
    $\delta_t$ & 10 minutes & epoch length & local feature extraction (pre-processing)
     \\[+3pt]
     $t_1$ & Hour 0 & starting time of BL segment & Baseline HMM training 
     \\[+3pt]
     $t_{N_0}$ & Hour 36 & end time of BL segment & Baseline HMM training 
     \\[+3pt]
     $\Delta_t$ & 3 hours & test window length& Domain adaptive FLDA training 
     \\[+3pt]
     $d_l$ & $\{12, 13, \dots, 59, 60\}$ hours &  training window lengths & Domain adaptive FLDA training 
     \\[+3pt]
     $\gamma$ & 1 & prior odds ratio & Domain adaptive FLDA training 
     \\[+3pt]
    \hline\hline
  \end{tabular}%
  }
  \label{table:parameter_settings}
\end{table*}

\section{Application to HVC experimental data}
\label{sec:analysis}

We apply the proposed adaptive transfer learning method to sleep/wake event classification for an experimental dataset undergoing a perturbation after baseline. This dataset of wearable data was collected as part of a human viral challenge (HVC) study where data from a cohort of participants was collected before and after exposure to a viral pathogen. More details on the HVC study are provided in Appendix \ref{sec:HVCstudy_AppA}.  We will show that the proposed FLDA-HMM algorithm, trained individually on each participant without clinical outcome data, is capable of segmenting sleep events and  extracting features that can be used to accurately predict clinical outcomes, specifically whether a subject is infected or not, and the onset time of infection. 


Sleep has repeatedly been found to be associated with immune, cardiovascular, and neuro-cognitive function \cite{miller2015role, irwin2015sleep}, among other functions. Many studies have revealed that changes in sleep pattern can be an important modulator of human response to diseases. For example, in a human viral challenge (HVC) study \cite{drake2000effects}, researchers found that nasal inoculation with rhinovirus type 23 significantly reduced total sleep time among symptomatic individuals during the initial active phase of the illness. In another study \cite{cohen2009sleep}, shorter sleep duration in the weeks preceding an exposure to a rhinovirus was found associated with lower resistance to illness. 
Since physiological signals such as instantaneous heart rate, physical activity, and skin temperature differ substantially during wake and sleep periods, automated sleep/wake labeling is possible. Therefore, the development of effective sleep monitoring methods has been of increasing interest.

While polysomnography (PSG) is the gold standard for sleep monitoring in sleep-related studies \cite{kaplan2017gold, dunn2018wearables}, it is often cumbersome outside of a lab setting as it uses electroencephalogram (EEG), electrocardiogram (ECG), and electromyogram (EMG), and requires a registered technician to perform sleep scoring
\cite{randk}. Recently much effort has been made to remedy the inconvenience of manual scoring and, alternatively, to provide automated analysis of the PSG signals \cite{flexerand2002automatic, malhotra2013performance, kang2018state}.  
Thus there has been growing interest in low-cost alternatives to PSG using small package wearable devices. 
Signals captured by wearable sensors have included movement induced accelerometer signals,  circulating blood volume pulse (BVP), heart rate (HR), electrodermal activity (EDA), temperature, respiration effort (RSP), ambient light and sound. These devices can be wrist-worn, ankle-worn, arm-worn, lapel-worn, chest-worn or embedded in mobile phones \cite{karlen2008improving},\cite{saeb2017scalable}, \cite{dunn2018wearables}. 
The Empatica E4 device (Empatica Inc. USA), used in the HVC study discussed below, captures 4 such signals. 

 \begin{figure}[ht]
 \centering
 \includegraphics[width = 0.85\linewidth, valign=t]{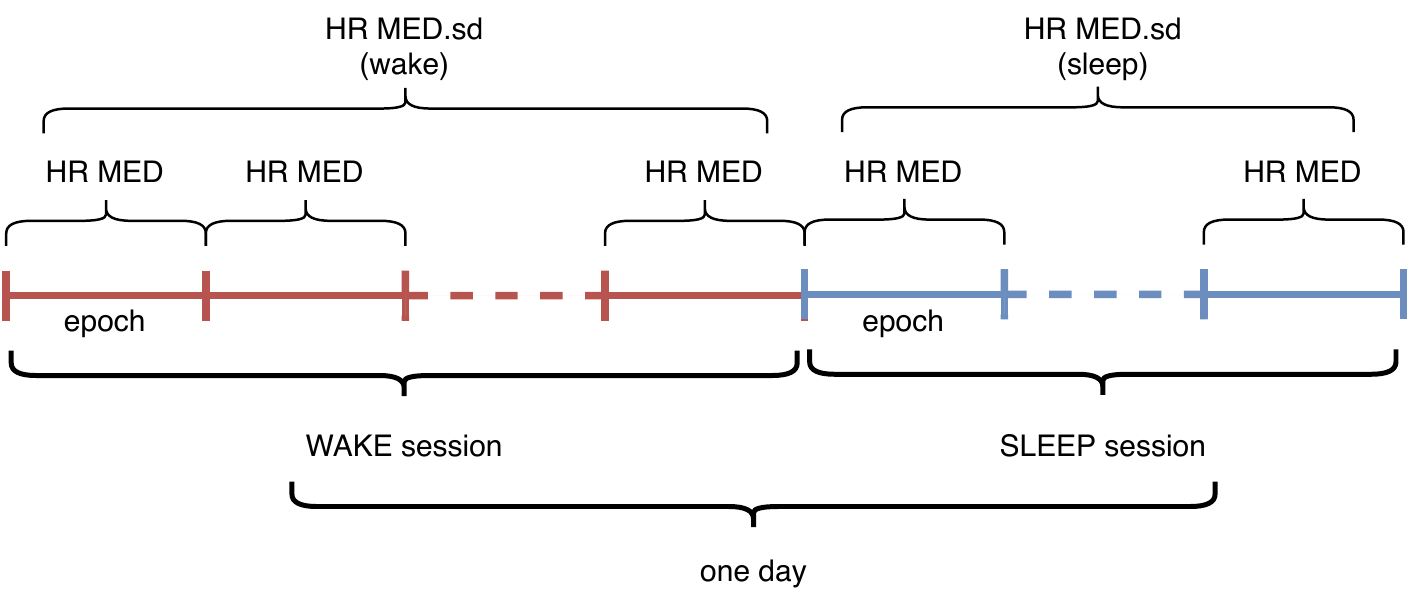}
   \caption{Labeling of events and event sessions from wearable data, illustrated for one of the features (HR MED = heart rate median) and one of the meta-features (HR MED.sd = standard deviation of HR MED) for Empatica E4 device in the Human Viral Challenge (HVC) study.  In the first stage of the pipeline in Fig. \ref{fig:pipe}, epochs of $10$  min duration ($\delta_t$ in Table  \ref{table:parameter_settings}) are used as windows over which temporally localized statistical features (mean, median, standard deviation) of the E4 device signals are extracted. These features are used in the second stage of the pipeline that extracts event labels, wake and sleep in the case of the HVC, and organizes them into contiguous labeled time segments, called sessions. In the third stage of the pipeline event-labeled meta-features (HR MED.sd (wake) and HR MED.sd (sleep)) are extracted as statistical summaries over the wake and sleep sessions.      }
 \label{fig:feature_extraction}
\end{figure}

\subsection{HVC data and processing pipeline}
\label{sec:experimentdesign}

Data was collected from E4 wristband devices worn by 25 participants enrolled in a longitudinal human viral challenge (HVC) study
over the time period: 0 hours to 270 hours (11 days). The study participants were exposed to a pathogen (Inf H1N2) on the second day of the study at 36 hours (noon on day 2).  Over subsequent days some participants developed symptomatic infection, confirmed by viral shedding from PCR assay on nasal lavage once per day. Such tests are often used to detect or confirm viral infection and to identify asymptomatic spreaders \cite{peiris2003clinical,huang2011temporal,zaas2009gene,he2020temporal}. In the HVC study, shedding in infected subjects was first detected either on day 4 at 81 hours, designating early onset shedders, or on day 5 at 106 hours, designating late onset shedders.  

Four channels of physiological signals, including 3-axis accelerometry (ACC sampled at 32 sa/sec), heart rate (HR at 1 sa/sec), skin temperature (TEMP at 4 sa/sec) and electrodermal activity (EDA at 4 sa/sec), were measured by the E4 devices. See Fig. \ref{fig:scatter1}
in Appendix \ref{sec:HVCdata} for representative signals from an infected and a non-infected subject.


We implemented the three-stage data processing pipeline in Fig. \ref{fig:pipe} for sleep segmentation and high dimensional feature extraction from the HVC data. 
The first stage (top branch in Fig. \ref{fig:pipe}) consists of a pre-processing module, performing data conditioning, fine grain temporally-localized feature extraction and abnormality filtering. The second stage (middle branch)  implements the HMM-FLDA module, performing adaptive sleep segmentation on non-abnormal data. 
The third stage (bottom branch) is a post-processing module, aggregating the data into sleep and wake sessions and performing coarse grain session-localized feature extraction.

Table \ref{table:parameter_settings} summarizes the parameter settings we used to implement the pipeline for the HVC data. 
Figure \ref{fig:feature_extraction} depicts how the pipeline constructs sleep/wake sessions and session-localized features from the fine grained features.

\begin{table*}[ht]
  \centering
  \caption{Features (\textbf{196} in total) extracted from sleep  and wake sessions for each day from subjects in the HVC study.}
  \label{table:featuresHVC}
  \resizebox{.95\textwidth}{!}{%
  \begin{tabular}{l l l}
    \hline\hline
    Name  & Number  & Description \Tstrut\Bstrut\\[+3pt]
    \hline
    Duration  & 2   & total sleep, night sleep \Tstrut\\[+3pt]
    Onset/offset  & 2   & night sleep only \\[+3pt]
    HR summary  & 9$\times$2   & 3 (mean, median, s.d.) $\times$ 3 (mean, median, sd within session) $\times$ 2 (sleep, wake)\\[+3pt]
    HR linear coef.   & 6$\times$2 & 3 (mean, median, sd) $\times$ 2 (coef.0, coef.1) $\times$ 2 (sleep, wake)\\[+3pt]
    HR quadratic coef.   & 9$\times$2 & 3 (mean, median, sd) $\times$ 3 (coef.0, coef.1, coef.2) $\times$ 2 (sleep, wake)\\[+3pt]
    TEMP  & 24$\times$2   & same as HR\\[+3pt]
    ACC  & 24$\times$2   & same as HR\\[+3pt]
    EDA  & 24$\times$2   & same as HR\\[+3pt]
    \hline\hline
  \end{tabular}%
  }
\end{table*}

\subsubsection{First stage: local features and abnormality filtering}

Like other wearable sensors, the Empatica E4 captures data at high frequencies and, because of the fact that they are worn by subjects in non-laboratory situations, the raw data collected are often voluminous and noisy. To mitigate the impact of occasional poor readings and reduce computational burden, the pre-processing stage performs data conditioning, local feature extraction and abnormality filtering. First the missingness of a subject's available data is evaluated, resulting in rejection of any subject with more than 40\% missing time points. For each of the remaining subjects, their data is segmented into non-overlapping time intervals, which we call {\it epochs}, of equal length $\delta_t$ secs. The epoch length was set to  $\delta_t=10$ mins to achieve a tradeoff between oversmoothing the sleep/wake transitions (excessive length) and noise sensitivity (insufficient length). See Appendix \ref{sec:variableselectionHVC} for  additional discussion.
Shorter epoch lengths may be more appropriate for classifying other types of event states, e.g., for capturing quality of sleep or detecting transitions between different sleep stages. 
For each of the 4 E4 signals, the module in Fig. \ref{fig:pipe} labeled "Local feature extraction" computes statistical features over each epoch corresponding to the signal mean (MEAN), signal median (MED) and signal standard deviation (SD). If a particular epoch has fewer than 90\% available samples, e.g., due to the E4 not being worn, the epoch is discarded.  

These local features are then processed by the "Abnormality filtering" module to identify time points at which outliers occur, labeled as non-normal data in Fig. \ref{fig:pipe}. Similarly to methods used in network intrusion detection \cite{bhuyan2013network}, the module uses marginal k-means clustering and quantile filtering to label time points as outliers (See Appendix \ref{subsec:abn}). Subjects having less than 60\% normal data cannot be reliably segmented and are omitted from subsequent analysis.

The last pre-processing step consists of selecting a subset of the 12 local features to train the HMM-FLDA sleep segmentation procedure. For this we select putative wake and sleep time intervals in the early evening around 21:00 and in the twilight hours around 4:00. These time periods were chosen since it is expected that most people would be awake at 9pm and sleeping at 4am. The sleep/wake discrimination capability of each feature over these two intervals was measured by a sleep/wake separability index (SWSI) computed over all non-abnormal subjects and over all available days. The SWSI is defined similar to the geometric separability index (\ref{eqn:si}): 
\begin{equation}
 \mbox{SWSI}(k) = \frac{\sum_{t \in \mathcal{T}_{wake} \cup \mathcal{T}_{sleep}} \{\hat{y}_t + \tilde{y}_t(k) + 1\}\mbox{ \textbf{MOD} }2}{|\mathcal{T}_{wake}| + |\mathcal{T}_{sleep}|},
  \label{eqn:swsi}
\end{equation}
where $k=1,\ldots, p$ indexes the features,   $\mathcal{T}_{wake}$ and  $\mathcal{T}_{sleep}$ are respectively the time intervals around 21:00 and 4:00 over all available days of data, $\hat{y}_t$ is the putative label assigned to the $k$-th feature $x_t(k)$, defined as $\hat{y}_t=1$ if $t\in \mathcal{T}_{sleep}$ and $\hat{y}_t=0$ otherwise, and $\tilde{y}_t(k)$ is the label of the nearest neighbor of  $x_t(k)$ among $\left\{x_\tau(k): \tau \in \mathcal{T}_{wake} \cup \mathcal{T}_{sleep}\right\}$. 
The criterion for including the $k$-th feature among those that are used to train the HMM-FLDA is that SWSI(k) be greater than $ 0.7$ for at least $75$\% of the subjects.  

\subsubsection{Second stage: adaptive sleep detection}
\label{sec:adaptivesleepdetection} 
 
Adaptive sleep segmentation was implemented using the HMM-FLDA procedure described in Section \ref{sec:sleep_detect}. 
The initial training window (baseline) for training the HMM-FLDA (Algorithm 1) was defined as the period from 0 to 36 hours, the time of viral inoculation. 
The test window length was chosen to be $\Delta_t = 3$ hours, which is based on our expectations about the time scale of immune response induced changes in the E4 signal distributions subsequent to pathogen exposure. The most rapid time scales of immune response are on the order of a few hours, e.g., inflammatory monocyte recruitment \cite{henderson2003rapid}. Such a choice of $\Delta t$ will minimize the likelihood that the HMM-FLDA algorithm loses track of covariate shifts from batch to batch. 
%
The range of training window lengths $d_1, \ldots, d_L$ used by the optimization loop in Algorithm 1 was restricted to $d_1=12$ to $d_L=60$ hours to guarantee a good balance of wake and sleep samples over the window. 

After the HMM-FLDA procedure labels the normal data, an "Abnormality classifier"  module identifies any abnormal samples (identified in Stage 1) that is physiologically meaningful and reinserts it into the data stream, labeling it with the label of the sample immediately preceding it (See Fig. 	\ref{fig:cat} in Appendix \ref{sec:abnormalityclassification}). Finally, a modified median filter with a 90-min smoothing window is applied to remove short bursty sleep periods of less than 60 mins. Such short sleep periods are likely to lack deep sleep stages 3 and 4, which     
typically start 30 mins. after sleep onset and can last 20 to 40 mins. 
\cite{carskadon2005normal}.

\subsubsection{Third stage: session-level feature set}
In the third stage of the pipeline, the sleep/wake sessions produced by the second stage are aligned to particular calendar days. A full day session is composed of wake sessions and sleep sessions. Any sleep session overlapping with a given day whose onset occurs before 5am (5:00) is associated with the previous day session.  
The binary sleep/wake label is then appended to the local features generated in the the first stage of the pipeline, doubling the total number of local features. Then a set of session-level features are extracted by computing several statistical summaries over daily wake and sleep sessions. This results in $196$ session-level features that include the mean, median, and standard deviation (sd) of the event labeled local features over a sleep or wake session. All features related to standard deviation are log-transformed. The session-level features also include the coefficients 
of linear and quadratic fits to the time course of these features over a sleep or wake session. In addition  session timing features, such as, sleep duration, sleep onset, and sleep offset are included in the session-level feature set. 
See Table \ref{table:featuresHVC}.

\begin{table*}[ht!]
  \centering
  \caption{Top 4 E4 wake/sleep features for predicting clinical outcome within 24 hours (later than day 3) of inoculation day (day 2) based on device data collected over the range 0-60 hours (day 1 and 2).}
  \resizebox{.95\textwidth}{!}{%
  \begin{tabular}{l c c c l c c c c}
    \hline\hline
    \multicolumn{3}{c}{\textit{logistic regression model}} &   & \multicolumn{5}{c}{\textit{continuation-ratio regression model}}\Tstrut\\[+3pt]
    \cline{1-3}\cline{5-9}
    \multirow{2}{*}{Feature}   & \multirow{2}{*}{Coef.}   & \multirow{2}{*}{AUC}   & \multirow{2}{*}{ }   & \multirow{2}{*}{Feature}   & \multirow{2}{*}{Coef.}   & \multicolumn{3}{c}{AUC}\Tstrut\\[+3pt]
    \cline{7-9}
    & & & & & & Early & Late & No onset \Tstrut\\[+3pt]
    \cline{1-3}\cline{5-9}
    HR MED.sd (sleep) & -3.921 & 0.758 & & HR MED.sd (sleep) & -5.073 & 0.882 &0.718 & 0.864\Tstrut\\[+3pt]
    ACC SD.linear.coef1 (wake) & -14.468 & 0.737 & & HR MEAN.sd (sleep) & -4.466	&0.706	&0.628	&0.773\\[+3pt]
    HR MED.quad.coef2 (sleep)	& 4.318 &0.707 & & Total duration & -0.599	&0.676	&0.551	&0.750\\[+3pt]
    Offset	&-1.452	&0.697 & & Night duration	&-0.616	&0.647	& 0.500	&0.705\\[+3pt]
    \hline\hline
  \end{tabular}%
  }
    \label{table:top_featuresHVC}
\end{table*}



\subsection{Online sleep segmentation}
\label{sec:predictive_modeling}

We emulate an online implementation of the pipeline in Fig. \ref{fig:pipe} by reapplying it successively to all available data at the end of each day. Daily updating of the segmentation corresponds to the real-world scenario where data is uploaded from the E4 device once per day, corresponding to the data acquisition rate in the HVC study. Continuous (real-time) updating will only become practical when devices have sufficient power for continuous data transfer or onboard processing. Below we illustrate the online implementation for a case of pre-infection segmentation before shedding occurs, and a full time course segmentation at the end of the study.  For additional details see Appendix \ref{sec:clinicaloutcome_APP}. 

\subsubsection{Illustration: pre-infection segmentation}
Here only the first 60 hours (up to day 3 at 12pm) of the data are available to the pipeline. The first 60 hours include the inoculation time (36 hours) and the first two nights of sleep. Note that no infection (shedding) is detected before day 4 at 81 hours.  
The preprocessing stage of the pipeline removed 5 subjects with excessive missing or abnormal data: 2 of these subjects had more than 40\% of their time points missing and 3 subjects had more than 40\% abnormal time points.   For the remaining 20 subjects local feature selection was accomplished by applying SWSI (\ref{eqn:swsi}) to all 12 variables to contrast early evening periods (19:00-22:00) to twilight periods (2:00-5:00).  
Only three features were found to have SWSI above $0.7$ for at least 75\% of the subjects:  the mean and median heart rate (HR MEAN and HR MEDIAN) and the standard deviation of the magnitude accelerometer (ACC SD). As the mean and median heart rate are highly correlated and the SWSI of HR MEAN has lower 25\% quantile than does the HR MEDIAN, only the HR MEDIAN and ACC SD were selected. Three hour long sleep and wake periods were necessary in order to obtain a sufficient number of representative time samples since only two days and nights are available.


\subsubsection{Illustration: full time course segmentation}
\label{sec:fulltimecoure_segmentation}
The full time course (0-270 hours) of a subject's data was made available to the pipeline. No subjects had more than 40\% missing time points over this period. Stage 1 abnormality filtering resulted in elimination of 5 subjects as excessively abnormal, who were removed. 
To compute SWSI for variable selection on each of the remaining 20 subjects, we used shorter putative wake periods (21:00-22:00) and sleep periods (4:00-5:00) since there are many more available days and nights than in the 60 hour pre-infection time period.  Applying the same SWSI variable selection criterion
three features were selected to train the HMM-FLDA: HR MED, HR SD, and ACC SD.   



\subsubsection{Clinical outcome prediction}
We used the feature set generated by the pre-infection segmentation to perform early detection of infection. Two clinical outcomes were considered: 1) a binary infection state  (whether or not the subject will shed virus anytime after 60 hours) designating the subject as infected or non-infected; and 2) the ternary infection state corresponding to onset time of viral shedding. The three onset times are defined as: early onset shedding (first detected by PCR at 81 hours), late onset shedding (first detected by PCR at 106 hours), and no onset shedding (subject never sheds virus).
Among the 20 subjects whose 0-60 hour data passed the abnormality filtering test, 9 of them were infected and 11 were non-infected. Among those infected, 3 were early shedders and 6 were late shedders.

Two predictor models were used for each of these outcomes, which were trained only on the sleep/wake features collected on inoculation day (wake/sleep sessions for day 2). For prediction of whether a subject will shed or not we applied logistic regression (LR) \cite{friedman2001elements} and for prediction of onset time we used continuation ratio (CR) regression \cite{agresti2010analysis}, 
which can be interpreted as a discrete version of the Cox regression model \cite{hemker2001measurement}.  

 \begin{figure}[ht!]
 \centering
\includegraphics[width = 0.9\linewidth, valign=t]{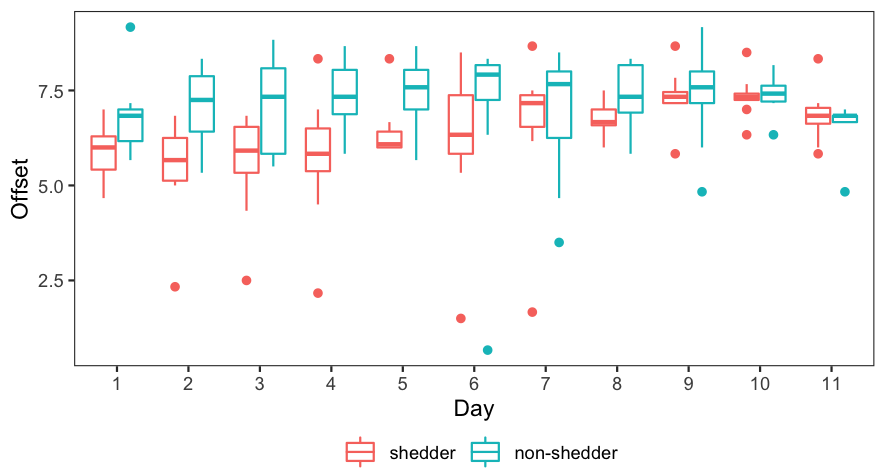}
\\
\includegraphics[width = 0.9\linewidth, valign=t]{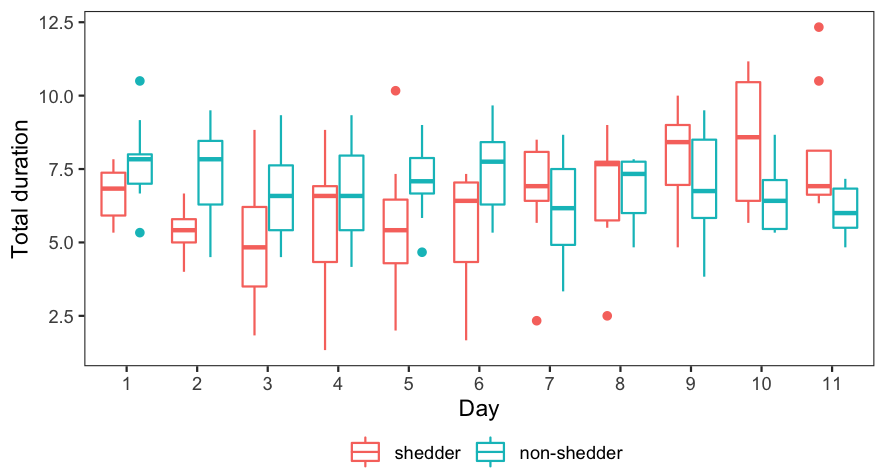}
   \caption{ 
   Offset (wake-up time) and total sleep duration (night and day sleeping) features extracted by the proposed HMM-FLDA pipeline applied to the experimental human viral challenge study (HVC) when all time points (0-270 hours) are available.     Viral inoculation  took place on day 2 and all infected people (shedders) started shedding virus on day 4 or day 5.}
 \label{fig:box_duration}
\end{figure}

For both the logistic and the continuation-ratio models the classification capability of individual features is evaluated by leave-one-out cross-validation. The top 4  discriminative features are reported in Table~\ref{table:top_featuresHVC}, together with their corresponding regression coefficients and a detection performance criterion: the area under the ROC curve  (AUC). AUC is a measure of the accuracy of a binary classifier and a value of AUC near 1 means very high accuracy.  For CR regression, since there exist three classes, three different AUCs are calculated using the ``one-versus-the-rest'' strategy  \cite{bishop2006pattern}. The CR performance is rank ordered in decreasing order of the minimum of the AUC's for Early, Late and No onset. See Appendix \ref{sec:clinicaloutcome_APP} for boxplots of these top 4 features.

The top feature for both LR and CR models is a sleep feature: the standard deviation of the median heart rate (HR MED.sd (sleep)).  Interestingly, most of the 4 top ranked features come from sleep sessions, with the exception of the linear coefficient to a linear fit to the time course of the standard deviation of ACC, denoted ACC SD.linear.coef1 (wake).  
Features related to sleep heart rate variation (HR MED.sd, HR MEAN.sd) and sleep duration (total duration, offset, night duration) are most discriminative.

The results of running full timecourse sleep/wake segmentation reveals striking temporal differences between shedders and non-shedder features.  Figure~\ref{fig:box_duration} indicates qualitatively different total sleep duration and sleep offset behaviors in these two groups over the full 11 day duration of the study. 
The non-shedder group sleeps on the average 2.041 hours longer than does the shedder group (p-value < 0.005 using two-sided t-test) over this time period. This is consistent with studies of the effects of sleep on the course of respiratory infection \cite{drake2000effects}. The mean sleep duration deficit among the shedders gradually decreases over time, and the trend reverses after day 6. The offset feature illustrates that non-shedders tend to wake-up later than the shedders until day 9 of the study.

\section{Discussion and Conclusions}\label{sec:discussion}
We have developed an unsupervised adaptive algorithm for classification of latent event states from multivariate physiological data collected from a wearable device. The algorithm adapts to perturbations of the initial training distributions using a sequential transfer learning model to mitigate covariate shift. The proposed algorithm operates without requiring \textit{a priori} information about true sleep/wake states and is capable of automatically detecting anomalies and abnormal data records. Numerical simulations established significant advantages of our model relative to hidden Markov approaches to hidden event classification.

The results presented in this paper are not without limitations. 
The HVC experiment has the limitation of small sample size and the classes are imbalanced. The negative effect of class imbalance can be compensated, to an extent, using methods such as the synthetic minority over-sampling technique (SMOTE) \cite{chawla2002smote}.  While no substitute for increasing actual sample size, we have demonstrated that SMOTE can improve clinical outcome prediction accuracy as measured by AUC (Table \ref{table:SMOTE} in Appendix E).  
Another limitation is that the HVC experiment lacks ground truth information about the true sleep/wake states of the subjects. The simulation study we presented emulating a similar HVC experiment with ground truth is an in-silico validation but a controlled experiment, e.g., performed in a sleep lab, would provide better confirmation. It would also be worthwhile to test the algorithm in a larger scale experiment that collects self-reported sleep diaries in addition to clinical data. 

There are also limitations of the proposed HMM-FLDA adaptive event segmentation algorithm. First, the algorithm may fail if there is an abrupt and excessively large shift in the event class distributions from time to time.  More generally, loss of track due to abrupt changes is a limitation of the gradual self-training approach \cite{kumar2020understanding},\cite{chen2011co} commonly adopted in transfer learning.  
In extreme cases this limitation may be insurmountable as there are fundamental theoretical limits that limit tracking ability of any adaptive algorithm \cite{helmbold1994tracking}.  
If the abrupt shift in distributions persists over time, a possible remedy would be to episodically re-initialize the HMM-FLDA algorithm during the adaptation phase.  
Secondly, the classification accuracy of FLDA may be poor if the event classes are not linearly separable. At the possible cost of reduced simplicity of implementation, use of a non-linear classifier in place of FLDA would overcome this limitation, e.g., using kernelized FLDA or a Support Vector Machine (SVM) \cite{friedman2001elements}.


We conclude by pointing out that the proposed framework for adaptive multi-channel event segmentation and feature extraction easily generalizes beyond the setting of the binary sleep/wake segmentation illustrated in this paper. Monitoring different stages of sleep or different wake activity types would be a natural non-binary extension. With the continuing advances in the capabilities of wearable devices for digital health, many new applications will be enabled by continuous multi-event tracking. These could include accurate behavioral and health assessment tools that will advance personalized health care.

The R code and HVC data used for this paper has been made  publicly available at GitLab (\url{gitlab.eecs.umich.edu/yayazhai/shezhai_bme2020}).  

\section{Disclosures}

\noindent{\bf Human Subjects}:
The data analysis reported in Section IV of this paper followed a protocol that was submitted to the Internal Review Boards of Duke University and the University of Michigan. On 4/25/2017 the Internal Review Board of Duke University Health Sciences (DUHS IRB) determined that the protocol meets the definition of research not involving human subjects, as described in 45CFR46.102(f), 21 CFR56.102(e) and 21CFR812.3(p), and that the protocol satisfies the Privacy Rule, as described in 45CFR164.514.  On 8/4/2017 the Internal Review Board of the University of Michigan (UM IRB) determined that the research performed under this protocol was not regulated.  

\vspace{0.1in}
\noindent{\bf Conflict-of-interest}:
The authors have no conflicts of interest to disclose.

\ifCLASSOPTIONcaptionsoff
  \newpage
\fi



\bibliographystyle{IEEEtranS}
\bibliography{IEEEabrv,myref}

\clearpage
\onecolumn

\appendices
\setcounter{page}{1}


%
\title{Adaptive multi-channel event segmentation and feature extraction for monitoring health outcomes: \\ Appendices}
\maketitle

\section{Human Viral Challenge (HVC) Study}
\label{sec:HVCstudy_AppA}

A human viral challenge (HVC) study was conducted in 2018 as a collaborative effort between Duke University and University College London under a grant from the Defense Advanced Research Projects Agency (DARPA) under the PROMETHEUS program. Thirty nine healthy volunteers between the ages of 18 and 55 were 
enrolled as participants in the study, which took place in the United Kingdom. 

The HVC was divided into outpatient and confinement phases. During the confinement phase participants stayed overnight for a period of 8-11 days in total 
from the morning before the day the viral challenge was administered to the end of confinement. 
During the outpatient phase subjects were evaluated for health conditions by tests including  ear, nose and throat (ENT), respiratory and cardiac assessment. 
In addition to other data types not relevant to this paper, wearable device data (Empatica E4) and clinical infection status data (viral shedding) were collected from the participants. All data was anonymized prior to transfer to Duke and Michigan for the analysis described in this paper.  


{\noindent \em Participant exclusion criteria}:
Chronic respiratory disease (asthma, COPD, rhinitis, sinusitis) in adulthood.
Inhaled bronchodilator or steroid use within the last 12 months.
Use of any medication or other product (prescription or over-the-counter) for symptoms of rhinitis or nasal congestion within the last 3 months.
Acute upper respiratory infection (URI or sinusitis) in the past 6 weeks.
Smoking in the past 6 months or $>$5 pack-year lifetime history.
Subjects with allergic symptoms present at baseline. 
Clinically relevant abnormality on chest X-ray. 
Any ECG abnormality.
Those in close domestic contact (i.e. sharing a household with, caring for, or daily face to face contact)
with children under 3 years, the elderly (>65 years), immunosuppressed persons, or those with
chronic respiratory disease.
Subjects with known or suspected immune deficiency.
Receipt of systemic glucocorticoids (in a dose $\geq$ 5 mg prednisone daily or equivalent) within one month,
or any other cytotoxic or immunosuppressive drug within 6 months prior to challenge.
Known IgA deficiency, immotile cilia syndrome, or Kartagener’s syndrome.
History of frequent nose bleeds.
Any significant medical condition or prescribed drug deemed by the study doctor to make the participant
unsuitable for the study.
Pregnant or breastfeeding women.
Positive urine drug screen.
Detectable baseline antibody titres against influenza challenge strains.
History of hypersensitivity to eggs, egg proteins, gentamicin, gelatin or arginine, or with life-threatening
reactions to previous influenza vaccinations.

{\em Confinement phase study}:
The eight days of the confinement phase consisted of a 36 hour healthy reference time period (baseline), inoculation at 36 hours (exposure), and a post-baseline time period. The E4 data of the participants was collected over the entire period and viral shedding was measured once per day in the morning.  
On the second day (day 2) at approximately noon  (36 hours from start of study) each participant was challenged by a GMP influenza A/California/04/2009-like (H1N1) 
virus strain. 
The  inoculation was administered 
by inserting intra-nasal drops on a single occasion with diluted inoculum with an average dose of $10^6$ TCID50 in 1mL PBS
divided equally between the two nostrils. This resulted in an average attack rate of 44\%. 
Following inoculation, advice regarding hand hygiene was given and subjects
were provided with alcohol hand gel and face-masks if they moved between the inoculation room
and the quarantine ward.
    
\noindent{\em Viral shedding assay}: Over the confinment phase of the study, viral shedding was measured once per day through a nasal lavage. 
The collected fluid was aliquoted into sterile microfuge tubes and centrifuged for analysis of cells,
and lavage fluid was later analysed by singleplex PCR to quantify the degree of viral shedding of the inoculated strain. Multiplex PCR was
performed on the pre-inoculation lavage and post-inoculation lavage collected during the study to
exclude the presence of other respiratory viruses.

\noindent{\em Wearable device and protocol}:    
Over the confinement phase of the study, subjects agreed to comply with the following wearable device protocol: 1) they wear the Empatica E4 device properly, i.e., comfortably tight on the wrist of their dominant hand; 2) they take care to maintain the device and protect it from shocks, water immersion, and other damage; 3) they wear the Empatica continuously without interruption except for periods that they were showering, recharging, or uploading data. 
The E4 has several sensors that measure physiological parameters including blood volume pulse, skin conductance, temperature, and movement. The E4 was recharged once per day during which time each participant's data were uploaded to a cloud server for processing using Empatica proprietary software. 

The result of this processing was reported at sub-second temporal resolution as the following variables:  heart rate (1sa/sec), skin temperature (4sa/sec), electrodermal activity (4sa/sec), and 3 axis accelerometer (32sa/sec). These were mapped to a vector of four variables at each time point: HR, TEMP, EDA and ACC, respectively, where ACC was computed as the Euclidean norm (magnitude) of the 3 dimensional acceleration vector.  Subjects were trained on best practices for wearing and maintaining the E4 devices over the course of the confinement phase of the study. 

Of the 39 participants enrolled in the HVC, only 25 had sufficient quality E4 data to be included in the analysis presented in Section \ref{sec:analysis}.
A histogram of the demographic data of these 25 participants is shown in the figure below. 
The other participants had wearable data that suffered from factors such as device failure, excessive missingness or data corruption, making their data unusable for our analysis. Non-compliance with wearable device protocols, data upload errors, and device malfunction were the cause of most of these problems.   

\begin{figure*}[!ht]
  \centering
  \includegraphics[width = 0.8\linewidth,valign=t]{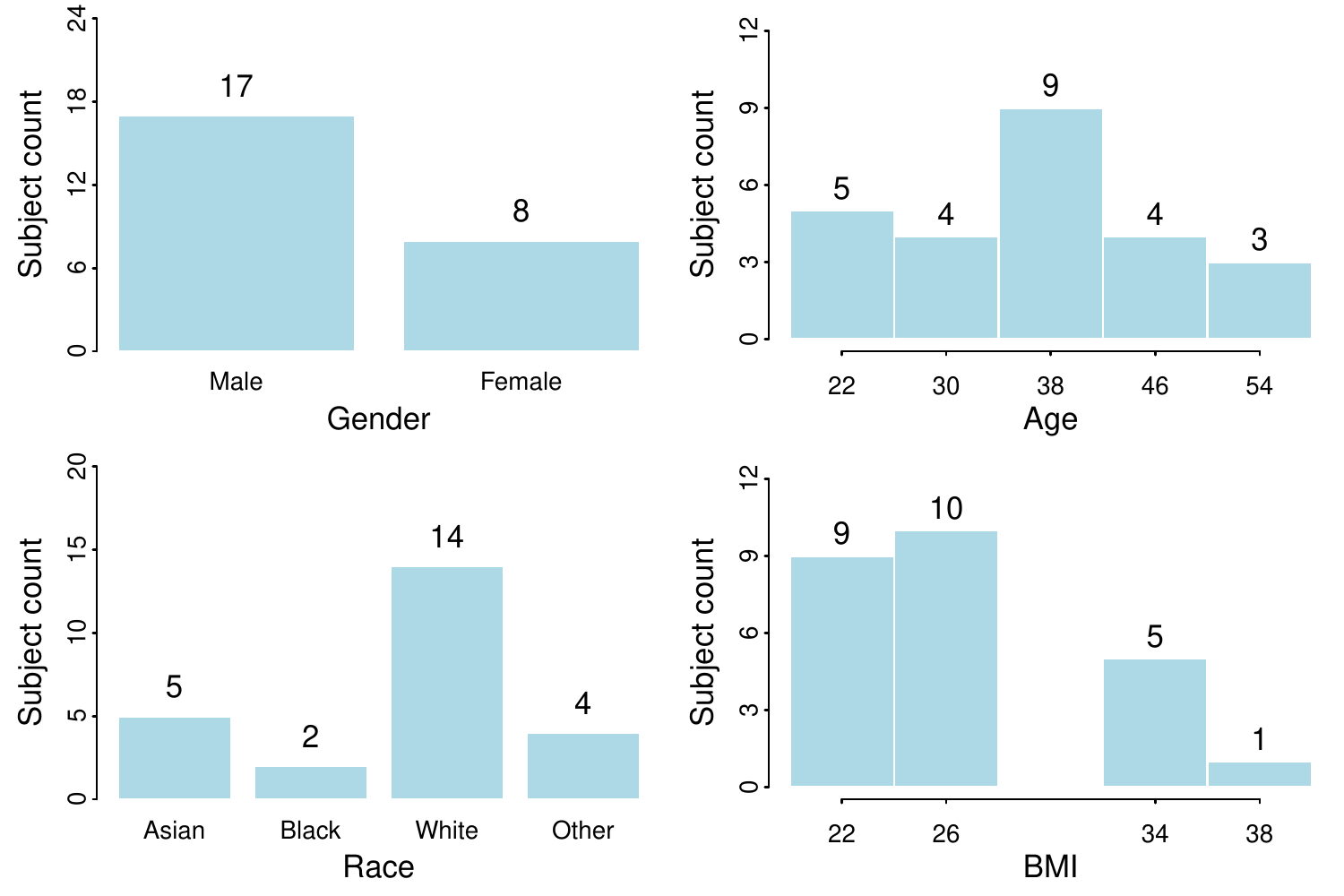}
  \caption*{
    Demographics of 25 participants in HVC study.
  } 
  \label{fig:demographics}
\end{figure*}



\section{HVC data}
\label{sec:HVCdata}
\begin{figure*}[!ht]
	\centering
	\subfloat[]{\includegraphics[width=.7\linewidth]{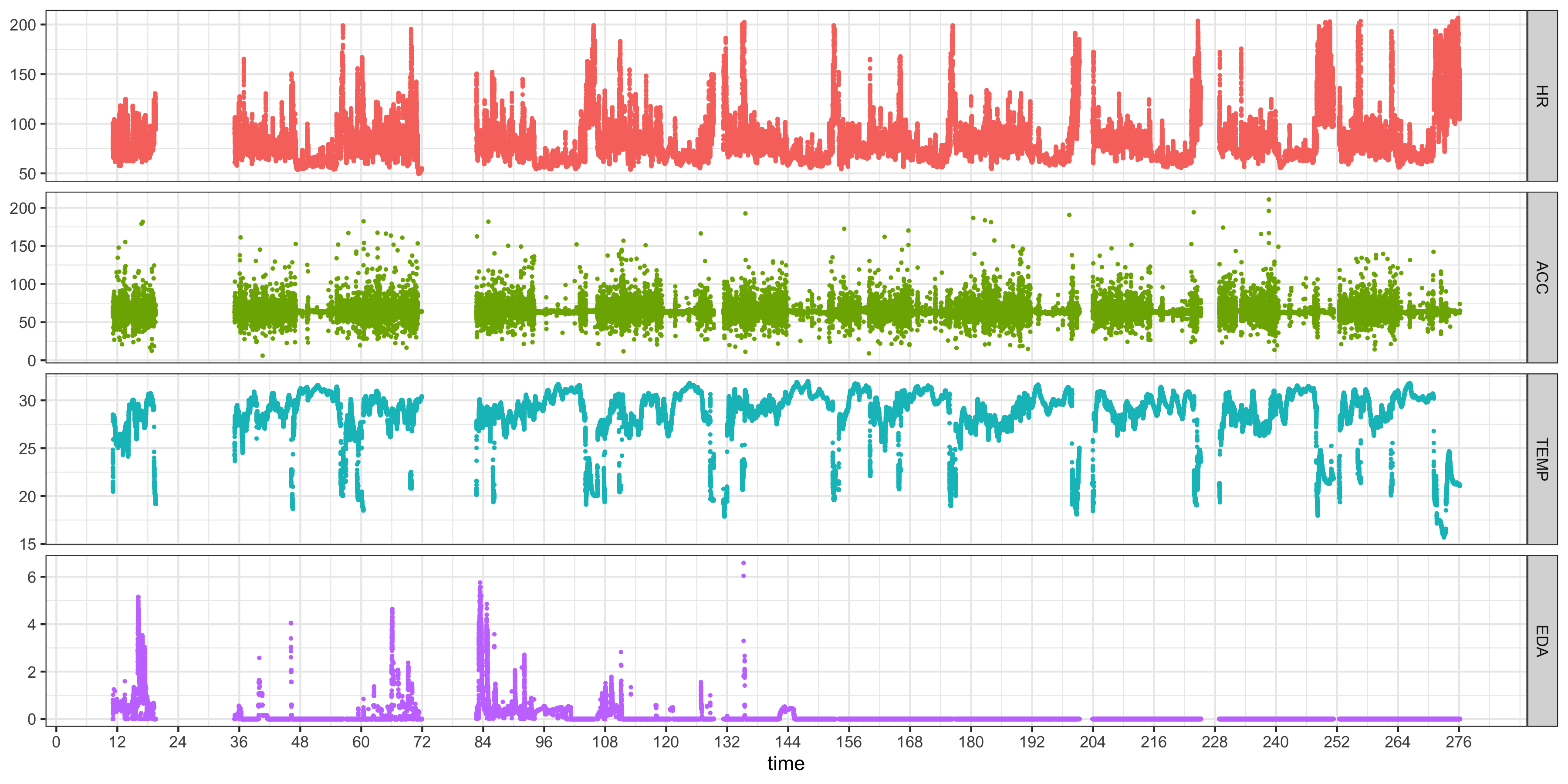}
	\label{fig:scatter1}}
	\hfil
	\\
	\subfloat[]{\includegraphics[width=.7\linewidth]{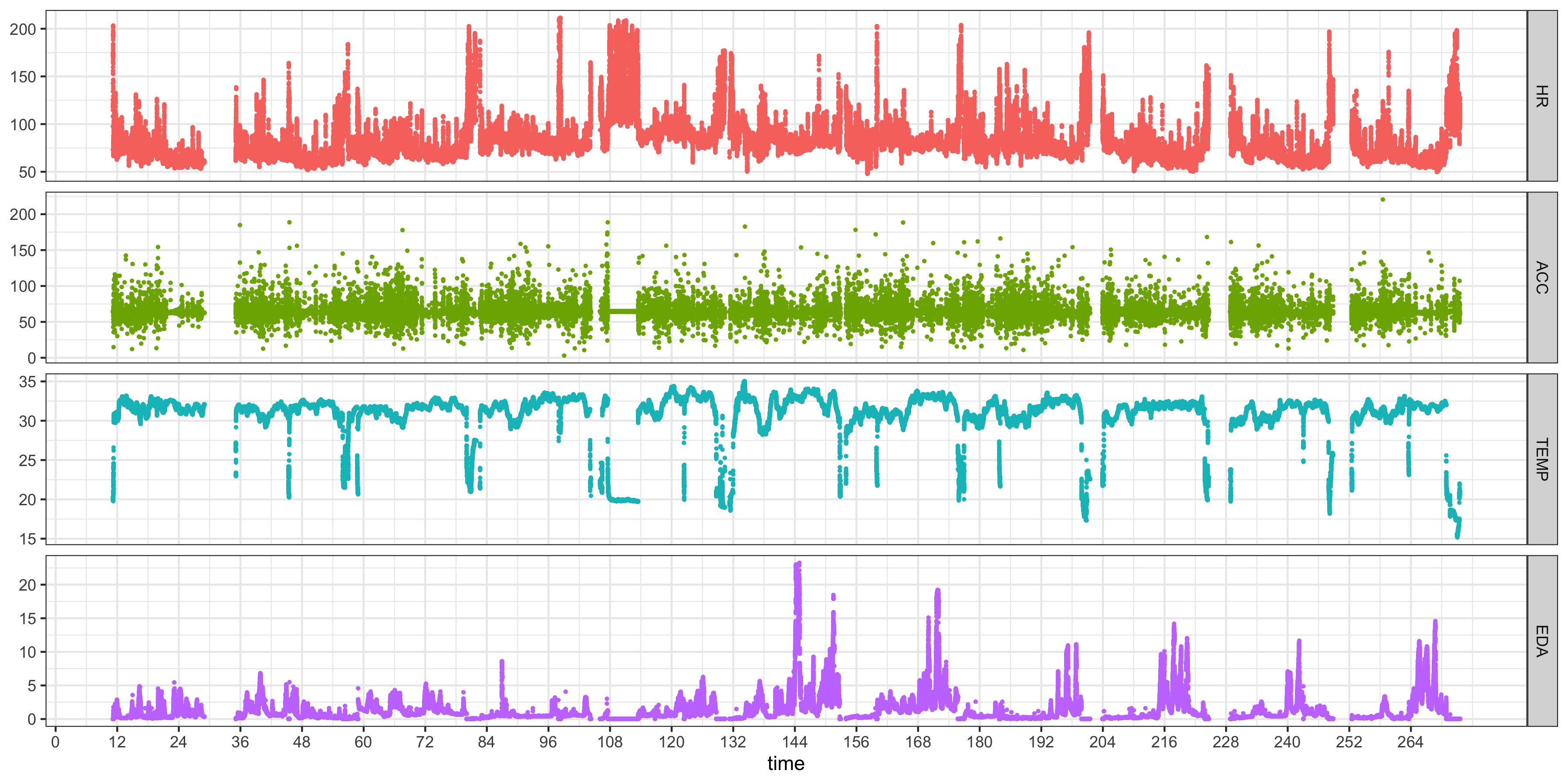}
	\label{fig:scatter2}}
	\caption{Four channel device data (Empatica E4) from human viral challenge (HVC) study for two subjects (Subjects 1 and 2). The four channels are: HR (red), ACC (green), TEMP (blue) and EDA (purple). Time 0hrs corresponds to 12am local time on the first day (day 1) of the study. Viral inoculations were administered to all subjects on the morning of the second day (day 2), i.e., between 32hrs and 36hrs. \textbf{(a)} data from a Subject 1 who had no detected shedding (Non-infected class); 
	\textbf{(b)} data from Subject 2 for whom shedding was detected (Infected class).} 
	\label{fig:timecourseHVC}
\end{figure*}

The raw time course E4 data of two representative subjects in the HVC study are shown in Fig.~\ref{fig:timecourseHVC}. Subject 1 (Fig. \ref{fig:scatter1}) did not become infected after inoculation on the second day of the study (day 2), i.e., this subject had no detectable level of viral shedding at any time over the 8-11 days of the study, while Subject 2 (Fig. \ref{fig:scatter2}) became infected.  The four E4 signals shown in the figure are heart rate (HR), accelerometer (ACC), temperature (TEMP), and electrodermal activity (EDA). The data clearly shows diurnal differences in signal behavior, corresponding to the cycling of sleep and wake states of these subjects.  A quadratic trend in heart rate and temperature is clearly visible in Subject 2, the trend peaking at around 144 hours.   Robustness to this trend is desirable and motivated the proposed HMM-FLDA sequential adaptive sleep/wake segmentation algorithm.      


\section{Simulation study}
\label{sec:simulationstudy_appendix}
In this section we include additional figures supporting the simulation study reported in Table \ref{table:simulation_resultOoS} and described in Section \ref{sec:simulation} of the main text. We also report on in-sample performance comparisons between the proposed HMM-FLDA and the competing HMM and dHMM algorithms.
 
\begin{figure*}[ht]
  \centering
   \subfloat[]{\includegraphics[width = 0.8\linewidth, valign=t]{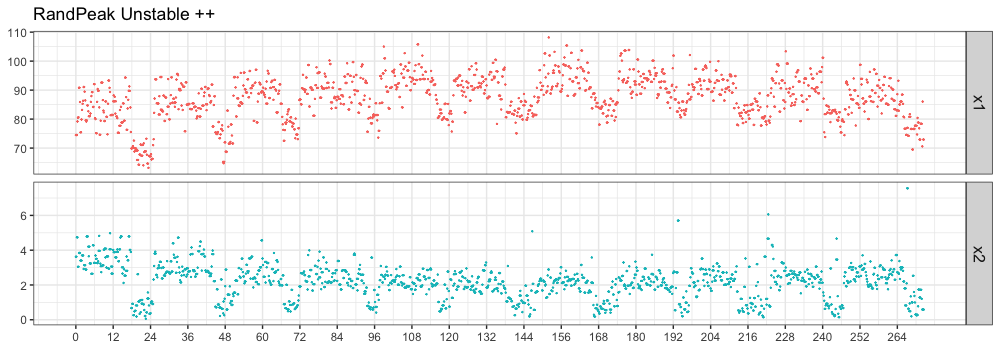}
 \label{fig:unstable1}}
 \\
    \subfloat[]{\includegraphics[width = 0.8\linewidth, valign=t]{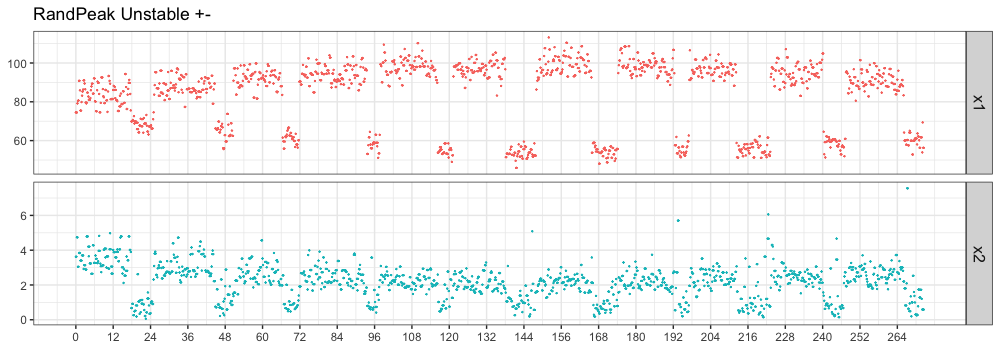}
 \label{fig:unstable2}}
 \\
    \subfloat[]{\includegraphics[width = 0.8\linewidth, valign=t]{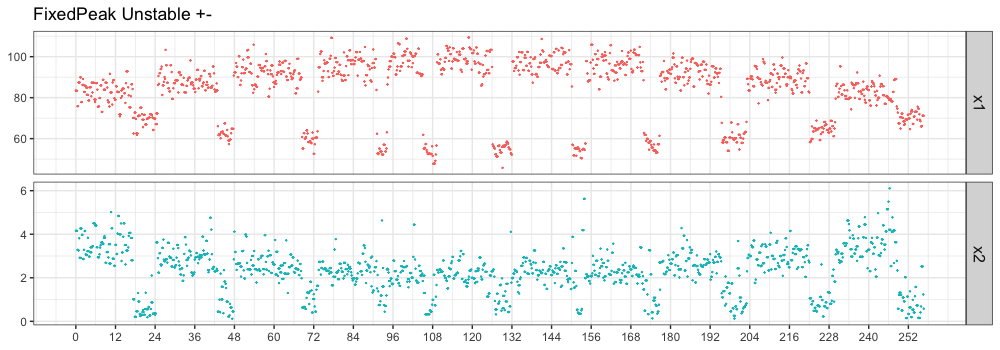}
 \label{fig:unstable3}}
 \caption{ Realizations of simulated signals for: (\textbf{a}) Unstable++ case for which the sleep and wake session both have concave trends after 36 hours in channel $X1$;  (\textbf{b}) Unstable+- case where the sleep and wake sessions have convex and concave trends, respectively, in channel $X1$; (\textbf{c}) another realization of Unstable+- case with sleep and wake sessions whose trends hit their apogee at a different time.}
  \label{fig:simu_realizations}
\end{figure*}

As mentioned in Sec. III these simulations were intended to emulate the experimental HVC data used in Sec IV.  In Fig. \ref{fig:simu_realizations} we show realizations of two simulated device channels, $X1$ and $X2$, emulating the HR MED and ACC SD features in the HVC study, under the unstable++, and unstable+- models for mean trends occurring after 36 hours. See Figs. \ref{fig:scatter1} and \ref{fig:scatter2} for comparison to real E4 data.

Table  \ref{table:simulation_resultOoSapp} provides a more comprehensive report of the results of our simulations of the proposed HMM-FLDA event segmentation than does Table \ref{table:simulation_resultOoS} in the main text.  Three segmentation algorithms are compared: the proposed HMM-FLDA, the HMM and a detrended HMM (dHMM) using LOESS pre-filter to detrend prior to applying HMM. The simulation parameters for the random  wake/sleep event transitions, event conditioned data distributions, and the covariate shift are the same as were used for Table \ref{table:simulation_resultOoS}.  

The top part of the table shows in-sample performance and the bottom part shows out-of-sample performance showing both mean and standard deviation (SD) of the performance measured over 100 simulation trials. Only the means of the OoS part of this table were shown in Table \ref{table:simulation_resultOoS}. The performance of the best performing method in terms of the MEAN performance is bolded.  

For the in-sample results (top part of table), event segmentation performance on the training set is shown. For the out-of-sample results (bottom part of table), an hold-out dataset was created from an independent simulation using the same model as the training set. The performance of each segmentation algorithm, trained only on the training set, was evaluated on the hold-out set.  
Note that the proposed HMM-FLDA would never be implemented in this way as such an implementation would turn off the adaptation mechanism on future data.  
The bottom part (OoS) of Table  \ref{table:simulation_resultOoSapp} provides evidence for the accuracy and robustness of the proposed approach.  


The bolded entries in Table \ref{table:simulation_resultOoSapp} denote the best mean performance per model (stable, unstable++, unstable+-) and performance criterion (Accuracy, F1, Cosine dist, Onset diff, and Duration diff). 
The asterisk on an MEAN entry in the OoS part of Table \ref{table:simulation_resultOoSapp} indicates the best performing method for each of the five criteria where best was determined using a one-sided paired t-test of significance applied to the set of 1000 simulations thresholded at a 0.01 level of significance.
The OoS part of Table \ref{table:simulation_resultOoSapp} has an additional row, P-VALUE, that is the p-value of the one-sided paired t-test that the proposed HMM-FLDA has better performance than both the HMM and dHMM. Only two P-VALUE entries are not significant, the case unstable++ for cosine distance and the case  unstable+- for onset diff. For the former case, dHMM is better than HMM but dHMM cannot be characterized as better than the proposed HMM-FLDA (0.01 level of significance). For the latter, HMM is better than the proposed for onset diff estimation. Hence, except for the onset diff criterion in the unstable+- case, the proposed method is equivalent or better than the other methods at a 0.01 level of significance. 



\begin{table*}[ht]
  \centering
  \caption{Performance of the proposed adaptive transfer learning algorithm as compared with standard HMM operating on the original data (HMM) and operating on LOESS detrended data (dHMM).}
  \resizebox{0.8\textwidth}{!}{%
  \begin{tabular}{l l l c c c c c}
    \hline\hline
    Setting & Methods &  & Accuracy  & F1 & Cosine dis. & Onset diff. & Duration diff. \Tstrut\Bstrut\\[+3pt]
    \hline \Tstrut
    (0, 0, 0, 0) & HMM & MEAN & \textbf{0.9982} & \textbf{0.9987} & \textbf{0.9968} & \textbf{0.0236} & 0.0427 \\[+3pt]
    stable     &     & SD & \textbf{0.0012} & \textbf{0.0009} & \textbf{0.0021} & \textbf{0.0199} & 0.0287 \\[+3pt]
          \Tstrut
          & dHMM & MEAN & 0.9981 & 0.9986 & 0.9966 & 0.0271 & 0.0450 \\[+3pt]
          &    & SD & 0.0012 & 0.0009 & 0.0022 & 0.0209 & 0.0288 \\[+3pt]
          \Tstrut
          & Proposed & MEAN & 0.9981 & 0.9987 & 0.9967 & 0.0259 & \textbf{0.0420} \\[+3pt]
          &    & SD & 0.0012 & 0.0009 & 0.0021 & 0.0218 & \textbf{0.0278} \\[+3pt]
          \hline\Tstrut\Bstrut
    (15, 10, 0.5, -0.5) & HMM & MEAN & 0.8744 & 0.9052 & 0.8234 & 2.4786 & 3.6780 \\[+3pt]
    unstable++           &     & SD & 0.0642 & 0.0520 & 0.0754 & 1.5541 & 2.2121 \\[+3pt]
                                \Tstrut
                                & dHMM & MEAN & \textbf{0.9558} & \textbf{0.9680} & \textbf{0.9306} & 0.9912 & \textbf{1.1574} \\[+3pt]
                                &      & SD & \textbf{0.0502} & \textbf{0.0375} & \textbf{0.0713} & 1.6795 & \textbf{1.4117} \\[+3pt]
                                \Tstrut
                                & Proposed & MEAN & 0.9487 & 0.9636 & 0.9146 & \textbf{0.9813} & 1.4605 \\[+3pt]
                                &         & SD   & 0.0498 & 0.0363 & 0.0781 & \textbf{0.9889} & 1.3812 \\[+3pt]
          \hline\Tstrut\Bstrut
  (-15, 15, 0.5, -0.5) & HMM & MEAN & 0.9371 & 0.9513 & 0.9185 & 0.4329 & 2.0992 \\[+3pt]
    unstable+-           &     & SD & 0.0836 & 0.0660 & 0.1043 & 0.7836 & 2.7722 \\[+3pt]
                                \Tstrut
                                & dHMM & MEAN & 0.9483 & 0.9615 & 0.9253 & 0.4909 & 1.4544 \\[+3pt]
                                &      & SD & 0.0666 & 0.0505 & 0.0928 & 0.9295 & 1.8854 \\[+3pt]
                                \Tstrut
                                & Proposed & MEAN & \textbf{0.9923} & \textbf{0.9946} & \textbf{0.9868} & \textbf{0.1631} & \textbf{0.1699} \\[+3pt]
                                &         & SD & \textbf{0.0221} & \textbf{0.0157} & \textbf{0.0365} & \textbf{0.4648} & \textbf{0.4651} \\[+3pt]
        \hline\hline\Tstrut\Bstrut
    stable & HMM (OoS) & MEAN & \textbf{0.9981} & \textbf{0.9987} & \textbf{0.9968} & \textbf{0.0165} & \textbf{0.0307} \\[+3pt]
          &    & SD & \textbf{0.0005} & \textbf{0.0004} & \textbf{0.0009} & \textbf{0.0051} & \textbf{0.0079} \\[+3pt]
    \Tstrut
          & dHMM (OoS) & MEAN & 0.9981 & 0.9986 & 0.9967 & 0.0253 & 0.0373 \\[+3pt]
          &    & SD & 0.0015 & 0.0011 & 0.0027 & 0.0163 & 0.0211 \\[+3pt]
    \Tstrut
          & Proposed (OoS) & MEAN & 0.9980 & 0.9986 & 0.9964 & 0.0168 & 0.0313 \\[+3pt]
          &    & SD & 0.0015 & 0.0011 & 0.0027 & 0.0163 & 0.0211 \\[+3pt]
          &    & P-VALUE & 1.0000 & 1.0000 & 1.0000 & 0.7850 & 0.8735 \\[+3pt]
    \hline\Tstrut\Bstrut
    unstable++ & HMM (OoS) & MEAN & 0.8959 & 0.9209 & 0.8559 & 1.6704 & 2.9169 \\[+3pt]
          &    & SD & 0.0679 & 0.0547 & 0.0792 & 1.0972 & 2.2120 \\[+3pt]
              \Tstrut
              & dHMM (OoS) & MEAN & 0.9127 & 0.9231 & \textbf{0.8957}* & 2.4709 & 2.2653  \\[+3pt]
              &               & SD & 0.1284 & 0.1377 & \textbf{0.1089} & 2.6125 & 3.1967 \\[+3pt]
              \Tstrut
              & Proposed (OoS) & MEAN & \textbf{0.9356}* & \textbf{0.9544}* & 0.8904* & \textbf{1.0679}* & \textbf{1.5756}*  \\[+3pt]
              &               & SD & \textbf{0.0353} & \textbf{0.0263} & 0.0533 & \textbf{0.7614} & \textbf{1.0664} \\[+3pt]
              &               & P-VALUE & \textbf{9.94E-07} & \textbf{1.07E-10} & 0.9521 & \textbf{4.04E-28} & \textbf{3.73E-08} \\[+3pt]
    \hline\Tstrut\Bstrut
    unstable+- & HMM (OoS) & MEAN & 0.9369 & 0.9510 & 0.9187 & \textbf{0.2763}* & 1.9750 \\[+3pt]
              &    & SD & 0.0810 & 0.0635 & 0.1019 & \textbf{0.4079} & 2.5778 \\[+3pt]
              \Tstrut
              & dHMM (OoS) & MEAN & 0.9271 & 0.9370 & 0.9114 & 1.0170 & 2.0131  \\[+3pt]
              &               & SD & 0.1151 & 0.1298 & 0.1112 & 2.1413 & 3.2464 \\[+3pt]
              \Tstrut
              & Proposed (OoS) & MEAN & \textbf{0.9758}* & \textbf{0.9833}* & \textbf{0.9564}* & 0.3712 & \textbf{0.3817}*  \\[+3pt]
              &               & SD & \textbf{0.0326} & \textbf{0.0226} & \textbf{0.0591} & 0.6225 & \textbf{0.5472} \\[+3pt]
              &               & P-VALUE & \textbf{1.63E-31} & \textbf{9.69E-37} & \textbf{5.80E-16} & 0.9999 & \textbf{1.34E-61} \\[+3pt]
    \hline\hline \Tstrut\Bstrut
  \end{tabular}%
  }
  \label{table:simulation_resultOoSapp}
\end{table*}

\begin{figure*}[ht!]
  \centering
  \subfloat[]{\includegraphics[width = 0.75\linewidth, valign=t]{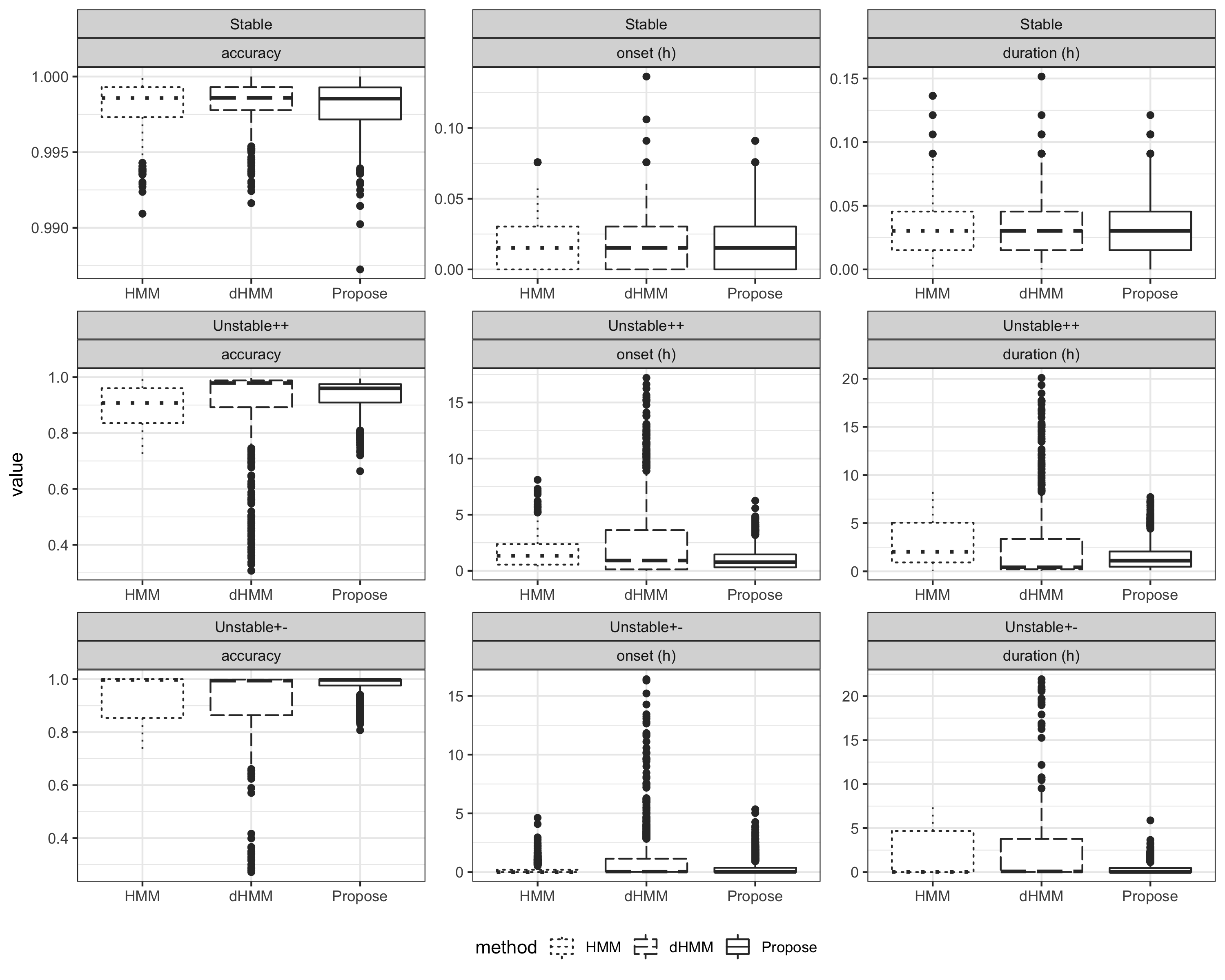}
  \label{fig:table1_boxplot}}
  \\
  \subfloat[]{\includegraphics[width = 0.75\linewidth, valign=t]{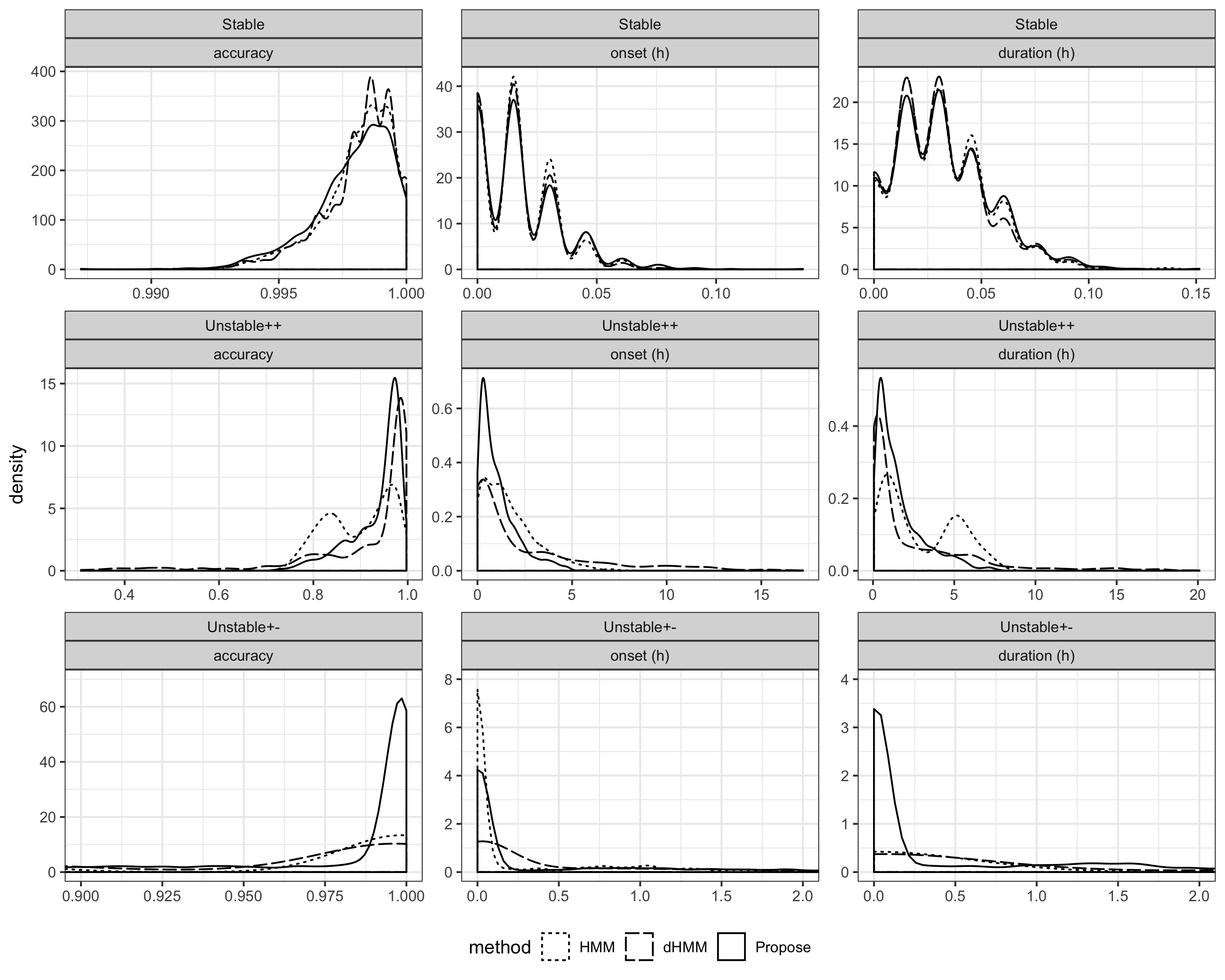}
  \label{fig:table1_density}}
  \caption{(\textbf{a}) Boxplots for label accuracy, onset estimation error, duration estimation error for simulations shown in Table \ref{table:simulation_resultOoS} for out-of-sample (OoS) simulation performance (excluding two outliers by dHMM method in the stable case).  (\textbf{b})  corresponding interpolated densities associated with (\textbf{a}) (for unstable+- case, the density plots are shown for a zoomed in region to better compare density concentration).} 
  \label{fig:table1boxplots}
\end{figure*}

In Fig. \ref{fig:table1boxplots} is shown boxplots and density plots of the out-of-sample (OoS) simulation of all three scenarios, whose mean is shown in Table \ref{table:simulation_resultOoS}. These plots show the distribution of errors committed by the various methods compared in terms of event labeling accuracy, onset estimation error, and duration estimation error when the trained methods are applied to independent sample trajectories drawn from the same distribution. As compared to the others the proposed method has a distribution that is more highly concentrated and has fewer outliers. 

\clearpage

Table \ref{table:simulation_runtime} shows runtime comparisons. The proposed method runs slower than HMM and dHMM. However, 1 minute run time of HMM-FLDA is still small as compared to the 8-11 days time period of HVC data. Furthermore, the method is not optimized in terms of runtime and can likely be accelerated.

\begin{table}[ht!]
  \centering
  \caption{Average run time (seconds) per replication of the proposed adaptive transfer learning algorithm as compared with standard HMM operating on the original data (HMM) and operating on LOESS detrended data (dHMM) on a 2.3 GHz Dual-Core Intel Core i5 processor with 16 GB RAM.}
  \resizebox{0.4\textwidth}{!}{%
  \begin{tabular}{c c c c}
    \hline\hline \Tstrut
        & stable & unstable++ & unstable+-\\[+3pt]
    \hline \Tstrut
    HMM & 0.4347 & 0.5612 & 0.4913 \\[+3pt]
    dHMM & 0.7396 & 0.8351 & 0.8347 \\[+3pt]
    Proposed & 53.5126 & 54.3654 & 55.2601\\[+3pt]
    \hline\hline \Tstrut\Bstrut
  \end{tabular}%
  }
  \label{table:simulation_runtime}
\end{table}


\section{Analysis pipeline implementation for HVC study}
\label{sec:diagnostics}
\label{sec:variableselectionHVC}

Here we provide additional details on our implementation of the pipeline of Fig. \ref{fig:pipe}, in the context of the Human Viral Challenge Study Empatica E4 data. We illustrate the implementation for  both the pre-infection time-line (0-60 hours) and the full time-line (0-270 hours).  First we describe the features used for abnormality detection.

Wearable devices are subject to outliers, anomalies and other abnormal sensor readings. Some types of outliers are physiological and are important to include in the final event-segmented data stream. Other  types of outliers are technical and can be due to device malfunction or improper wearing of the device. These technical outliers must be removed early in the analysis pipeline so as to not compromise downstream event detection and labeling performance. We use a two stage procedure for isolating such anomalies in the pre-processing stage (Stage 1) of the pipeline in Fig. \ref{fig:pipe} and classifying them as physiological vs technical outliers for possible re-insertion in the transfer learning stage (Stage 2).  The first stage of the procedure is called {\em abnormality filtering} and the second stage is called {\em abnormality classification}, which will be discussed below. Both procedures apply standard outlier detection methods to a set of predefined features.  However, since the purposes of these procedures are different, the outlier detection thresholds are different.

\begin{figure*}[!ht]
	\centering
\includegraphics[width=.6\linewidth]{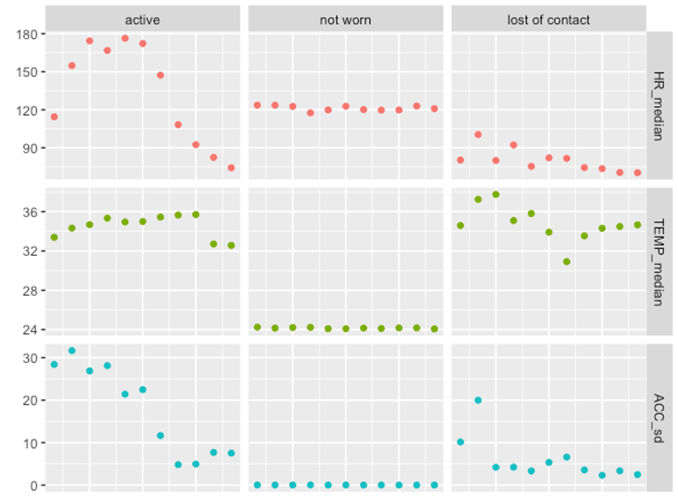}
	\\
	Time
	\caption{Experimental Empatica E4 data collected from one of the co-authors of this paper under three different conditions: active, not worn and loss of contact. Shown are measured values of the variables used in our selected abnormality feature set: HR median, TEMP median and ACC sd.  Note the very different values of these three parameters between the active, not worn and loss of contact classes. }
		\label{fig:E4anomaly_experiment}
\end{figure*}

\noindent{\bf Features for abnormality filtering (Stage 1) and classification (Stage 2)}:
The best features to use for detection of abnormal samples will be experiment dependent and device dependent. Here we explain how the abnormality filtering features were selected for the HVC experiment where participants wore the Empatica E4 device, as discussed in Section \ref{sec:analysis}. Based on experiments on an Empatica E4 performed in our laboratory we determined that there are three principal causes for abnormal measurements, each manifesting abnormality in different combinations of channels. See Fig. \ref{fig:E4anomaly_experiment}. For abnormality filtering and classification physical intuition motivated us to select three temporally localized features, HR MED, TEMP MED and ACC SD, as they are especially affected by the types of abnormalities described below.  
\begin{itemize}
  \item Device not worn (NW). {\em Effect}: Skin temperature sensor (TEMP) reads ambient temperature and activity sensor (ACC) records little or no physical movement. {\em Primary features affected}: median of TEMP and standard deviation of ACC magnitudes are abnormal. 
  \item  Device loss of contact (LOC).  {\em Effect}: Intermittent skin contact causes spurious signal dropout.  {\em Primary features affected}: median of TEMP and HR and standard deviation of ACC magnitudes are abnormal. 
  \item Subject engages in intense activity (Active). {\em Effect}: heart rate (HR) and activity sensor (ACC) readings increase significantly over burst of physical activity. {\em Features primarily affected}:  median of HR and standard deviation of ACC magnitudes are abnormal.  
\end{itemize}
In the HVC study we used HR MED and TEMP MED features for abnormality filtering while we used HR MED, TEMP MED and ACC MED for abnormality classification. The combination of HR MED and ACC MED is especially important for classifying abnormalities that are due to physiological causes,e.g., when a subject is engaged in intense exercise which is legitimate wake session activity that should be reinserted in the final segmented data stream in Stage 2 of the pipeline.  

\begin{figure*}[ht]
  \centering
  \subfloat[$\mbox{SI}_1$ = 0.470, $\mbox{SI}_2$ = 0.605]{\includegraphics[width = 0.32\linewidth, valign=t]{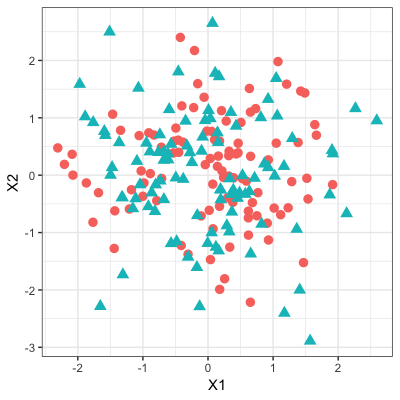}
  \label{fig:si1}}
  \subfloat[$\mbox{SI}_1$ = 0.750, $\mbox{SI}_2$ = 0.785]{\includegraphics[width = 0.32\linewidth, valign=t]{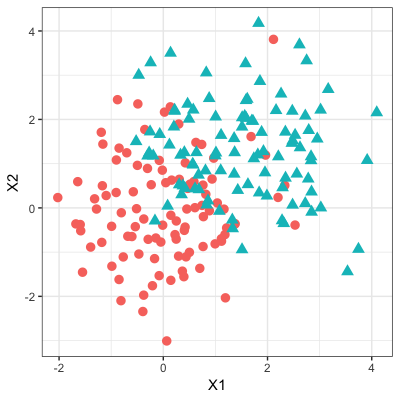}
 \label{fig:si2}}
  \subfloat[$\mbox{SI}_1$ = 1.000, $\mbox{SI}_2$ = 0.990]{\includegraphics[width = 0.32\linewidth, valign=t]{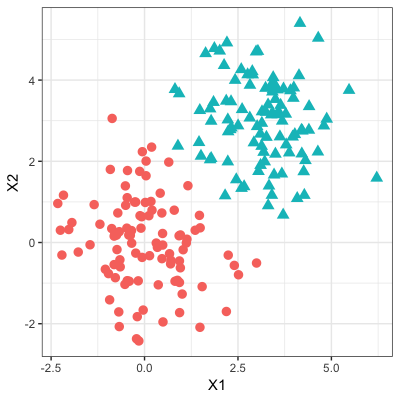}
  \label{fig:si3}}
 \caption{Separability indices under three simulated scenarios: \textbf{(a)} non-separable, \textbf{(b)} weakly separable, \textbf{(c)} strongly separable, where $\mbox{SI}_1$ is based on projection distance and $\mbox{SI}_2$ is based on Euclidean distance.}
 \label{fog:si}
\end{figure*}

Figure~\ref{fog:si} illustrates three simulated cases with different levels of separability. Two variables $(X_1, X_2)$ are generated from $\text{BVN}(\mu_1, \mu_2, 1, 1, 0)$. For samples in Class 0 (indicated by red dots), $\mu_1 = \mu_2 = 0$, while for samples in Class 1 (indicated by green triangles), we considered three settings: (1) $\mu_1 = \mu_2 = 0$; (2) $\mu_1 = \mu_2 = 1.5$; (3) $\mu_1 = \mu_2 = 3$, corresponding to non-, weakly and strongly separable scenarios. SI values based on both projection distance and Euclidean distance are reported for each case. We observe that SI is indeed able to effectively characterize separability, and that the projection distance is preferred since it gives value closer to 0.5 in the non-separable case and value closer to 1 in the strongly separable case.

\subsection{Online feature extraction pipeline: applied to 0-60 hours of data}
\label{sec:pipeline0_60}

Here we describe  details of the proposed pipeline, illustrating with the pre-infection (0-60 hours) data. The result of applying the pipeline to the full data (0-270 hours) is described afterward. 

\subsubsection{Stage 1: pre-processing}
\label{subsec:abn}
\label{subsec:abnormalityfiltering}

Pre-processing in stage 1 of the pipeline accomplishes three tasks: temporal windowing and conditioning; local feature extraction, and abnormality filtering. Here we illustrate each of these tasks when the pipeline is applied to the pre-infection data, i.e.,  only the first 60 hours (2 days and two nights) are available. Results of applying the pipeline to  the full data from 0 to 270 hours are presented for comparison in Section \ref{sec:pipeline0_270}.

\noindent{\em Stage 1: temporal windowing and conditioning}

The tuning parameters and their settings for Stage 1 of the pipeline are
\begin{itemize}
    \item Temporal window (epoch) length: 10 mins
    \item Subject availability threshold: 60\%
    \item Subject abnormality threshold: 40\%
\end{itemize}

\begin{figure*}[!ht]
	\centering
	\includegraphics[width=.5\linewidth]{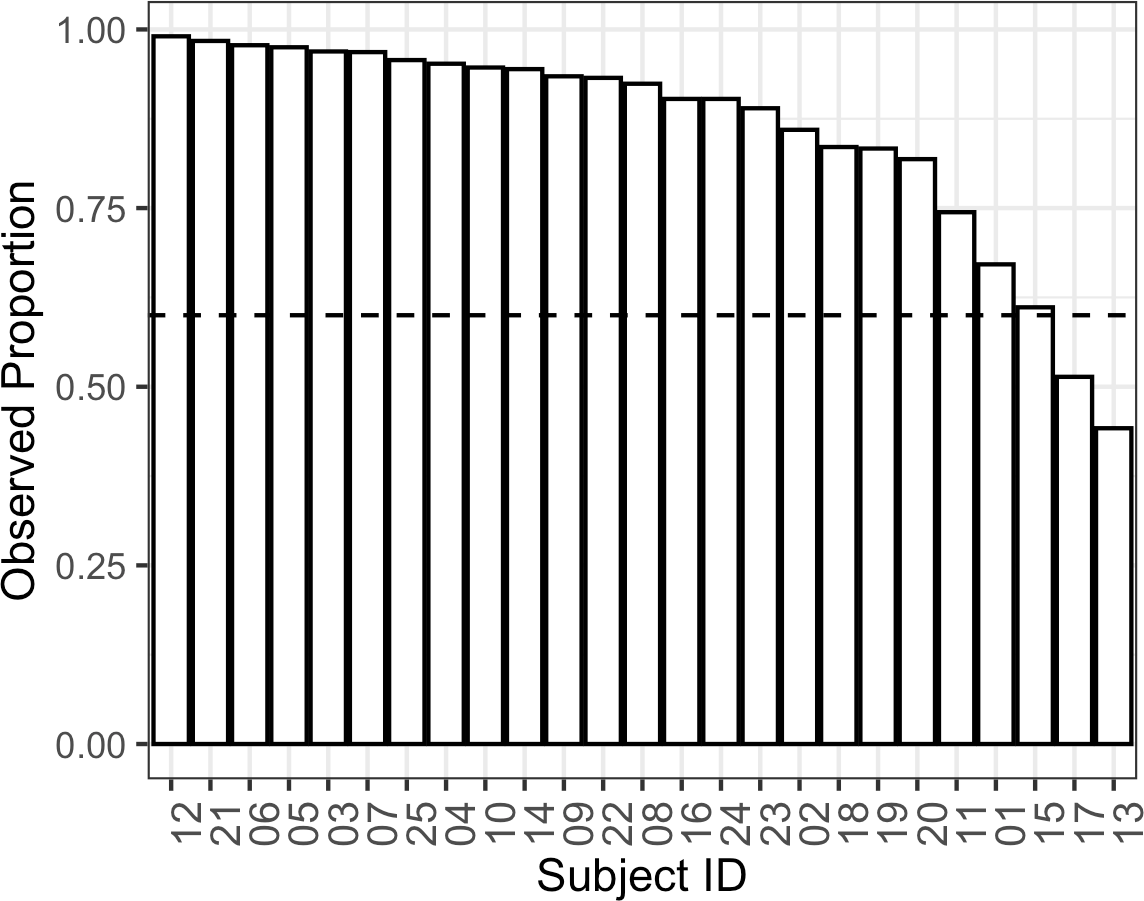}
	\caption{Subject availability for the sleep/wake segmentation when the first 60 hours of data is available for subjects in the HVC study.  Two subjects (13 and 17) have excessive ($>40\%$) missingness. The observed proportion is defined as the ratio of the number of time samples in the subject's data record and the total number of sampling times that should be available over the $60$ hour period. }
	\label{fig:subjectavailability_pre_inf}
\end{figure*}

The subject availability threshold of 60\% is applied to filter out subjects that have E4 data missingness of greater than 40\%. Subject availability is quantified the proportion of time points not missing among the full set of sampling times (number of seconds) over the 0-60 hour period.  Subjects with less than 60\% data availability are not further processed. Figure \ref{fig:subjectavailability_pre_inf} shows that 2 subjects in the HVC cohort (Subjects with identified 13 and 17) have insufficient data availability over this time period.

\noindent{\em Stage 1: abnormality filtering}

\begin{figure*}[!ht]
	\centering
	\includegraphics[width=.5\linewidth]{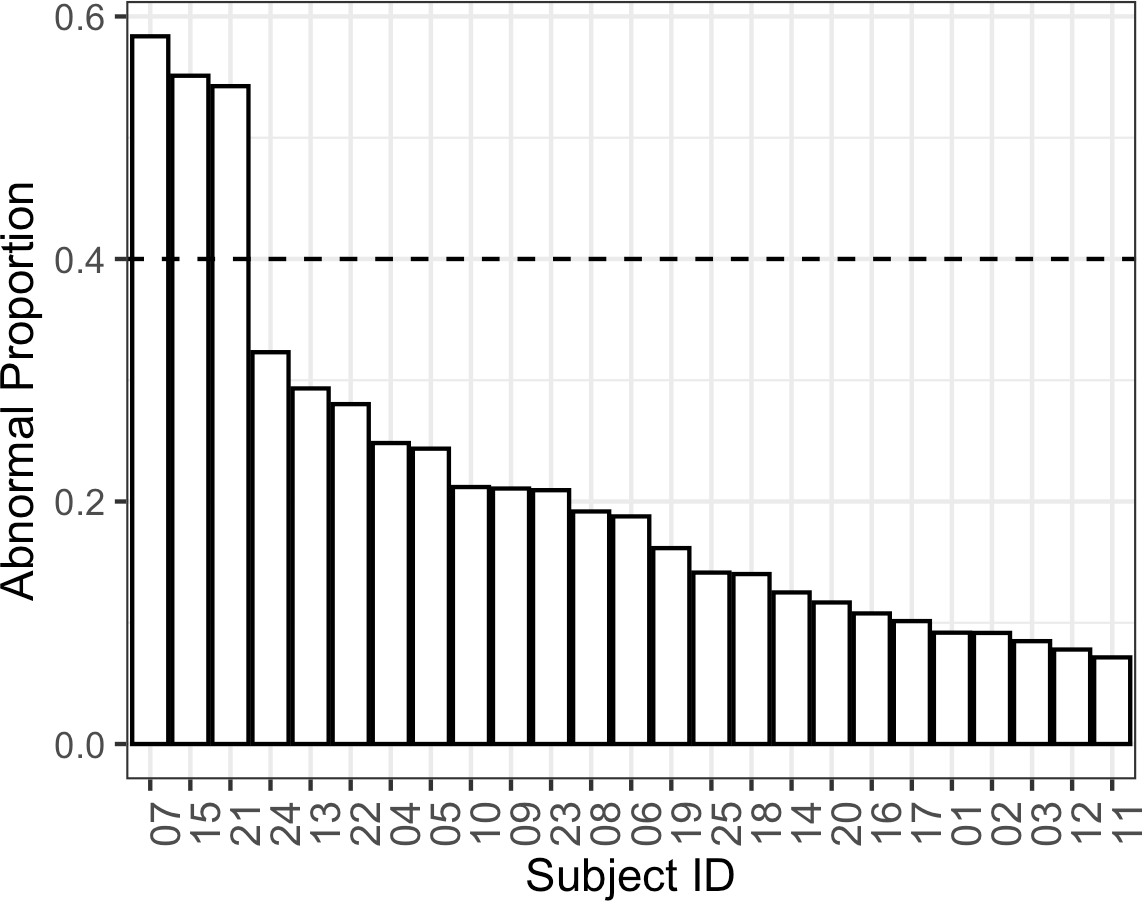}
	\caption{Abnormal proportion for the sleep/wake segmentation over 0-60 hours for subjects in the HVC study. Three subjects have an excessive number of abnormal time samples (greater than 40\%). }
	\label{fig:subjectabnormal_pre_inf}
\end{figure*}

The abnormality filtering module operates as follows. After temporally localized summary statistic features are extracted from the device for each subject, these features are tested for abnormality using a clustering-based anomaly detection procedure. The procedure labels the abnormal samples as {\em non-normal data} and these are temporarily removed from the data stream. 

The clustering-based anomaly detection procedure is as follows.  Define the vector-valued local features for the $i$-th subject at the $t$-th time instant
$\tilde{\bm{X}}_{i,t} =(\tilde{X}_{1i,t}, \dots, \tilde{X}_{Mi,t})$, where here the feature dimension is $M=12$. As explained above, only two of these features was used for abnormality filtering: HR MED and TEMP MED. For each subject $i$,  a combination of k-means clustering and quantile thresholding is used to determined a normal region $\mathcal{C}_i$ in this two dimensional feature space. 
Subject $i$'s feature instance at time $t$ is declared non-normal if it is not in $ \mathcal{C}_i$. 
While we also investigated abnormality using k-means clustering in the full 2 dimensional feature space of HR MED and TEMP MED (See Fig. \ref{fig:kmeans}), the abnormality filtering method we adopted in the pipeline of Fig. \ref{fig:pipe} performs k-means clustering separately on each each of the feature dimensions.  

For the HVC data, where the events of interest are sleep and wake, the k-means algorithm is set to extract 3 clusters corresponding to sleep and wake (normal) and non-normal classes.  Let $\{S_{im1}, S_{im2}, S_{im3}\}$ be the resulting clusters with centers (centroids) $\{\mu_{im1}, \mu_{im2}, \mu_{im3}\}$. 
When the normal and non-normal cluster classes are well separated it is easy to construct a normal region $\mathcal C_i$, e.g., all points in the feature space having a majority of k-nearest neighbors in $\{\tilde{\bm(X)}_{i,t}\}_t$ outside of the non-normal cluster.     
More often, however, the separation between the normal and non-normal cless is not sufficient 
and a different method is needed. We determine $\mathcal C_i$ as the set complement of the hyper-rectangle of minimal volume whose empirical coverage probability is 95\%. The rectangular edge lengths and position are thus determined by the marginal sample quantiles along each feature dimension. 

Specifically, we define a sequence of subject-dependent cutoff values $\{c_{i1}, \dots, c_{iM}\}$ for abnormality.    
The rectangular normal region is designated as the Cartesian product $S_i^{nor}=S_{i1}^{nor}\times \ldots \times S_{iM}^{nor}$, where   
\[
  S_{im}^{nor} = 
  \begin{cases}
    S_{im1} \cup S_{im2} & \text{if } |\mu_{im2} - \mu_{im1}| < |\mu_{im3} - \mu_{im2}|, \\
    S_{im1} & \text{o.w.}. \\
  \end{cases}
\]
Cutoff values defining the normal set are defined as 
\[
  c_{im} = 
  \begin{cases}
    Q_{0.025}(\{\tilde{X}_{mi,t} \in S_{im}^{nor}\}) & \text{if } \mu_{i, nor} > \mu_{i, abn},\\
    Q_{0.975}(\{\tilde{X}_{mi,t} \in S_{im}^{nor}\}) & \text{o.w.}, \\
  \end{cases}
\]
where $Q_q(\cdot)$ is the $q-$th quantile of the samples, and $\mu_{i, nor}$, $\mu_{i, abn}$ are the centroids of the normal and non-normal set, respectively. 
For the experimental HVC data we used the 95\% outlier rule to define the lower and upper quantiles as 2.5\% and 97.5\%, but less stringent threshold values may be adequate for other datasets. 

For the subjects in HVC study with the first 60 hours of data available, Figure~\ref{fig:subjectabnormal_pre_inf} shows a bar-plot of the resulting non-normal data proportion for each subject. 
Three subjects (7, 15 and 21) had excessive abnormal data and were eliminated from further processing.

Thus, when taken with subjects who had excessive missingness, a total of five subjects were excluded from further analyses.
Some of these subjects may come back into the analysis when a greater amount of data is available (see subsection C for the case where 0-270 hours are available).

\begin{figure*}[!ht]
	\centering
	\includegraphics[width=.6\linewidth]{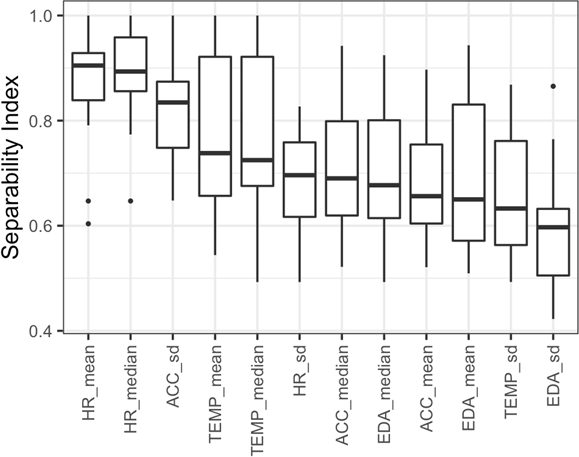}
	\caption{ Box-plot of sleep/wake separability index (\ref{eqn:swsi}) for pre-infection (0-60 hour) data, showing the spread of the 12 local features extracted from wearable data in the first stage of the processing pipeline. 
	Excluding HR MEAN, which is highly correlated with HR MED and has a smaller 25\% quantile, two local features (HR MED, ACC SD) are selected.   
	}
	\label{fig:marginal_gsi_pre}
\end{figure*}

\noindent{\bf Feature selection for training adaptive segmentation algorithm}:
As explained in the main text, we extracted a subset of the 12 local features using the sleep/wake separability index (SWSI) (\ref{eqn:swsi}).  The SWSI is computed over all subjects, to contrast each feature over the putative sleep periods  2:00-5:00 and the wake periods 19:00-22:00 in the 0-60 hour time interval of the pre-infection data window.  
Figure \ref{fig:marginal_gsi_pre} shows the resulting SWSI for each of the 12 features, in  descending order of median. Applying the selection rule that a selected feature must be greater than 0.7 for at least 75\% of the subjects, three features are  above threshold:  HR MEAN, HR MED and HR SD and, as HR MEAN and HR MED are highly correlated, we chose eliminated HR MEAN since it has a lower 75\% quantile than does HR MED.

\subsubsection{Stage 2: adaptive  segmentation}
\label{sec:separability_index}
\label{sec:abnormalityclassification}

The HMM-FLDA procedure is implemented for a given subject on all the time points that pass through the abnormality filter. The details of the HMM-FLDA are given in Sec \ref{sec:adaptivesleepdetection}.

\vskip0.1in
\noindent{\em Stage 2: Separability index}
\vskip0.1in

By definition (\ref{eqn:si}), SI is the proportion of samples that share the same label with their nearest neighbors, and hence SI $\in [0,1]$. Intuitively, when samples from two classes form two tight, well-separated clusters with little overlap, the nearest neighbor of one sample from, say, Class 0, will most likely belong to Class 0 as well, which will result in a large SI value close to $1$. In contrast, when samples from two classes follow exactly the same distribution, i.e., completely non-separable, then the nearest neighbor of one sample will have equal probability of being Class 0 or Class 1, and thus, the SI of these samples is close to $0.5$. A large SI value implies strong separability of classes, and is usually an indication of reliable prediction. SI also depends on the measure of distance used to determine the nearest neighbors. The projection distance captures the difference among samples on the optimal direction $\bm{w}$ that is most relevant to distinguishing between the two classes, and thus is better than Euclidean distance in regard to characterizing separability, as demonstrated by Figure~\ref{fog:si}.


Once the HMM-FLDA event classification procedure terminates the initial sleep/wake segmentation will have missing time points that have been removed by the abnormality filter. Some of these abnormal time points, e.g., those due to rare  physiological events like an exercise session, are reinserted into the data stream. The method to do this is based on an abnormality classification procedure discussed in the next subsection.

\vskip0.1in
\noindent{\em Stage 2: Abnormality classification and reinsertion}
\vskip0.1in

The purpose of the abnormality classification module is to identify, re-insert and assign an event label to physiologically meaningful non-normal data, i.e., abnormalities that are not due to technical issues associated with device failure or improper wearing of the device. Any such non-normal samples re-inserted into the data stream inherit the event class label of the session into which the sample's time stamp falls.      

\begin{figure*}[!ht]
	\centering
	\includegraphics[width=.37\linewidth]{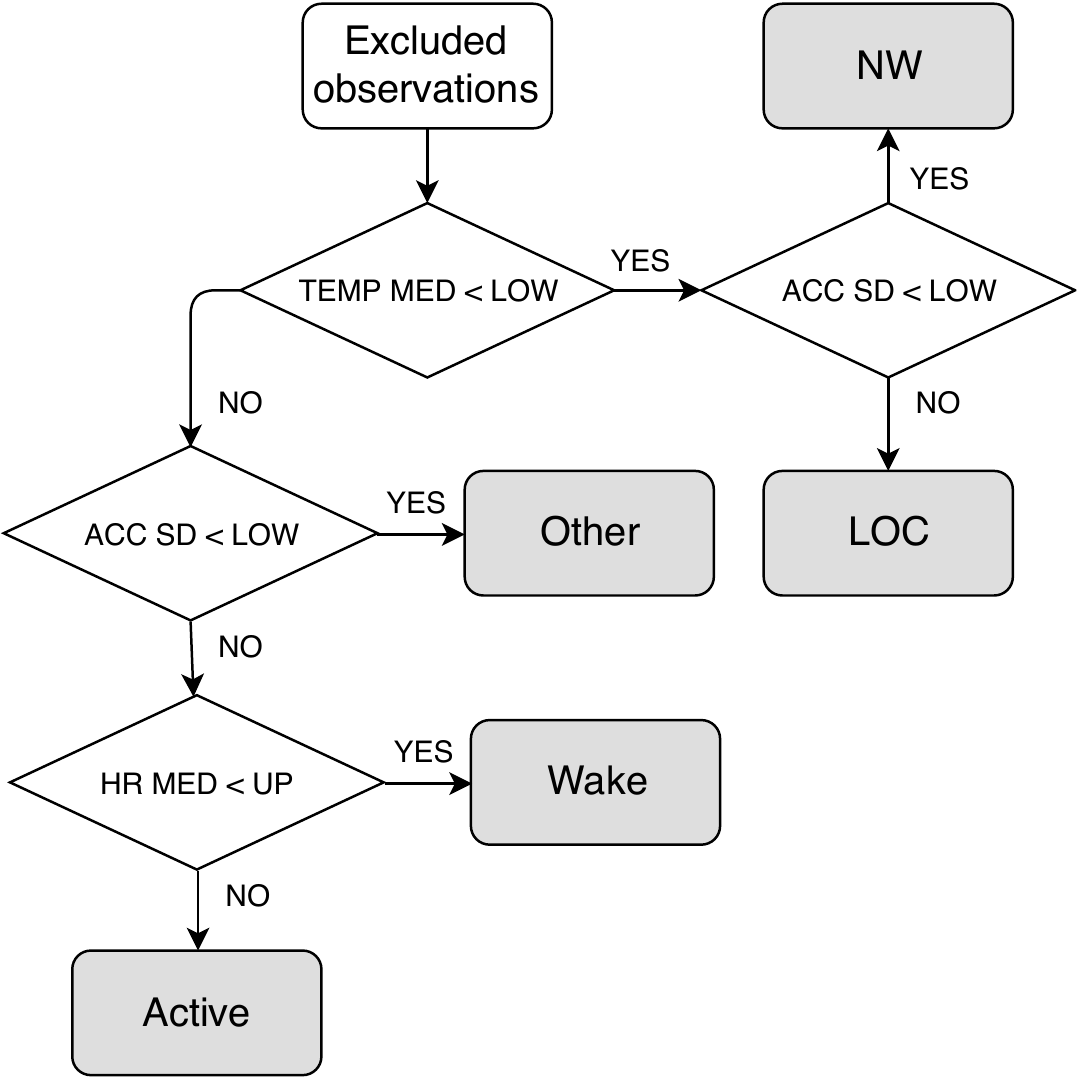}
	\caption{
	Decision tree for abnormality classification module for the HVC data. Non-normal samples (excluded observations) identified by the abnormality filter in first stage of the pipeline of Fig. \ref{fig:pipe} are excluded from the training set used by the second stage transfer learning algorithm. The abnormality classifier re-evaluates these samples for possible re-insertion and labeling of physiologically meaningful abnormalities.  The classifier classifies the samples into final categories of: device not worn (NW), loss of contact (LOC), Wake, Active and Other. Samples that are classified as "Active" or "Wake" are re-inserted into the corresponding session.
	}
		\label{fig:cat}
\end{figure*}

Using predefined features HR MED, TEMP MED, and ACC SD, abnormality classification was performed using a decision tree using marginal quantile thresholding. For each non-normal sample, we applied a standard 1.5 interquartile range (IQR) test \cite{tukey1977exploratory,hoaglin1986performance} to each feature independently. If a feature falls outside of this range it is declared an outlier and further classified. The lower and upper endpoints of the 1.5 IQR interval are defined as: 
\begin{align}
	\mbox{LOWER} &= Q_{0.25} - 1.5 * (Q_{0.75}-Q_{0.25}), \nonumber\\
	\mbox{UPPER} &= Q_{0.75} + 1.5 * (Q_{0.75}-Q_{0.25}).
\end{align}
Here $Q_{0.25}$ and $Q_{0.75}$ are the 25\% and 75\% sample quantiles of the feature empirical distributions computed over the detected sessions. All samples that were categorized as "Wake" or "Active" were incorporated back into the corresponding session.
Abnormal samples assigned to other categories were discarded. The decision tree for this procedure is shown in Figure~\ref{fig:cat}.


\vskip0.1in
\noindent{\em Stage 2: Median filtering}
\vskip0.1in

After processing by HMM-FLDA and reinsertion of physiologically meaningful abnormalities, there commonly exists short bursts of sleep sessions. We expect that some of these are actually sleep while others correspond to resting without sleep. While such bursty behavior could possibly be directly incorporated into an HMM model, e.g., using a semi-Markov switching process \cite{anisimov2008switching}, we took a simpler approach that applies a modified median filter with 90  min smoothing window (median filter of order 9) that has the effect of merging sleep sessions shorter than 60 mins into a wake session. 
We choose to eliminate such short sessions in order to eliminate disambiguate resting and light naps from sleep sessions, defined as a session having deep sleep stages, i.e., Stage 3 \& 4 of non-rapid eye movement (NREM), usually starting 30 minutes after sleep onset and lasting approximately 20 to 40 minutes in the first sleep cycle \cite{carskadon2005normal}. Hence, the 60-minute threshold is designed to eliminate putative sleep sessions that had no deep sleep stages. 

\subsection{Offline feature extraction pipeline: applied to 0-270 hours of data}
\label{sec:pipeline0_270}

\subsubsection{Stage 1: pre-processing}
\label{sec:epochlengthHVC}

\begin{figure*}[!ht]
	\centering
	\includegraphics[width=.6\linewidth]{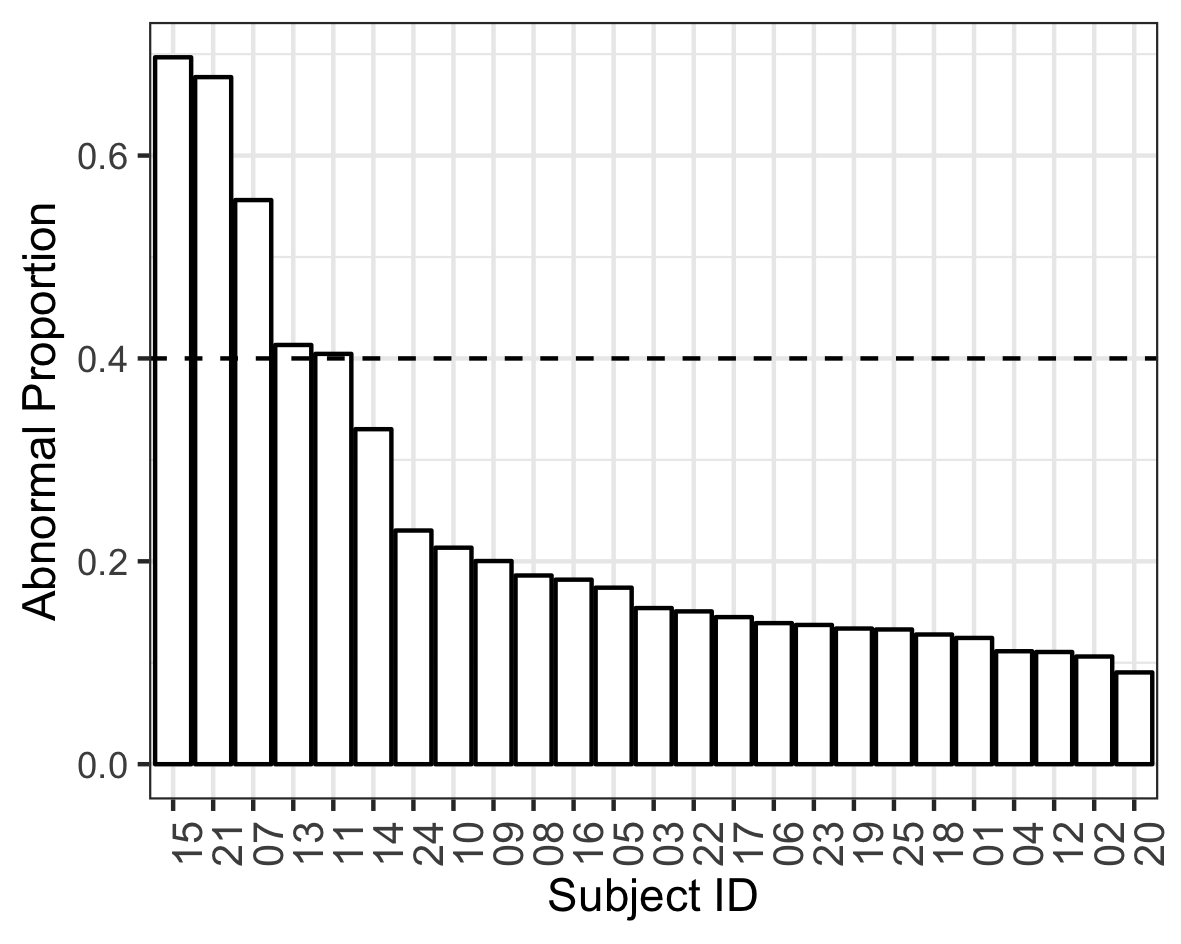}
	\caption{The bar-plot of abnormal data proportions over all time points (0-270) in HVC study data. Bar-plot of availability (analog to Fig. 	\ref{fig:subjectavailability_pre_inf}) is not shown since no subjects had more than 40\% missing data over this 270 hour time period. }
	\label{fig:abn_hist}
\end{figure*}

When evaluated over the time period 0-270 hours, no subjects had missingness greater than 40\% and thus none were filtered out due to inadequate data availability. However, five subjects were found to have greater than 40\% abnormal samples and were filtered out. 
Figure 	\ref{fig:abn_hist} shows the bar-plot of the abnormal proportions for all 25 subjects, showing the 5 subjects exceeding the 40\% threshold for inclusion.

\begin{figure*}[ht!]
	\centering
	\includegraphics[width=.6\linewidth]{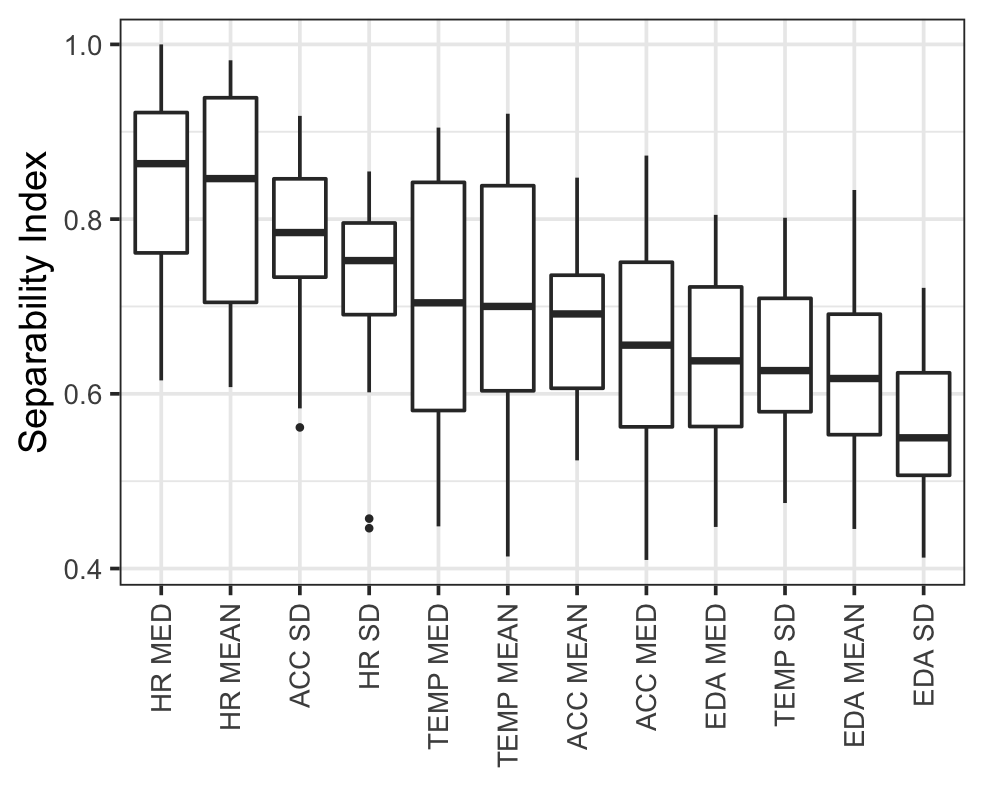}
	\caption{ Box-plot of sleep/wake separability index (\ref{eqn:swsi}) for full (0-270 hour) data, showing the spread of the 12 temporally localized features extracted from wearable data in the first stage (pre-processing) of the processing pipeline. 
	Excluding HR MEAN, which is highly correlated with HR MED and has smaller 25\% quantile, three features (HR MED, ACC SD, HR SD) are selected for training HMM-FLDA in the second stage  of the pipeline.   
	}
		\label{fig:marginal_gsi}
\end{figure*}

To select features for the HMM-FLDA sleep/wake segmentation algorithm over the full time course of 0-270 hours, we designated two one hour periods of the day, 3:00-4:00 and 21:00-22:00, respectively, as sleep and wake (resting). As contrasted to the pre-infection segmentation in which we used 3 hour periods in the 60 hours (2 nights) of available data, here we could take advantage of the higher specificity of one hour periods since 11 nights are available over the 270 hours.    The sleep/wake separability indices (SI), defined in (\ref{eqn:swsi}), of each of the 12 temporally localized features were computed over all time points  in order to evaluate their discrimination power. Figure~\ref{fig:marginal_gsi} shows box-plots of the SWSI values for the 12 local features. 

Four local features had SWSI greater than 0.7 for over over 75\% of the subjects: HR MEAN, HR MED, HR SD, ACC SD.  Since HR MEAN and HR MED are highly correlated, HR MEAN was excluded from the top 4 SI features as its 25\% quantile is lower than that of HR MED. The remaining features were then used as feature variables in the second stage of the pipeline to identify sleep/wake sessions for each subject.  Like in the 0-60  hour segmentation, we again note that the EDA features have lower sleep/wake separability indices than the other E4 variables. 
 

\subsubsection{Stage 2: adaptive segmentation}
 The following table summarizes the results of applying abnormality classification (Fig. \ref{fig:cat}) in Stage 2  to the abnormal 10 min epochs identified by the abnormality filter in Stage 1. For the purposes of illustration, the table is restricted to epochs falling within the typical resting period (21:00 - 22:00) and the typical sleeping period (03:00 - 04:00). The counts in the table represent abnormality classification over the full 270 hours of the HVC experiment summed over the 20 subjects falling under the 0.4 abnormal proportion threshold in Fig. \ref{fig:marginal_gsi}. The Normal category are the Wake and Active classes in Fig. \ref{fig:cat} and represent epochs that were re-inserted into the segmented event stream after termination of the HMM-FLDA algorithm. No epochs in these one hour intervals were labeled as "Other." 
 
 \vspace{0.1in}
 \begin{center}
\begin{tabular}{c c c c c} 
 \hline
 \hline
          & Total & NW & Loss-of-contact & Normal \\
 \hline
  Resting (21:00-22:00) & 1194 & 0   & 122      & 1072 \\ 
  Sleep  (03:00-04:00) & 1194 & 0  & 46        & 1148 \\ 
 \hline
 \hline
\end{tabular}
\end{center}


\clearpage

\subsubsection{Sensitivity of pipeline to tuning parameters}
Here we illustrate the lack of sensitivity of the segmentation to our choice of two tuning parameters: epoch length $\delta_t$ and the number $k$ of nearest neighbors in the $k$-means classifier used for abnormality filtering. 

\begin{figure*}[!ht]
	\centering
	\subfloat[]{\includegraphics[width=.24\linewidth]{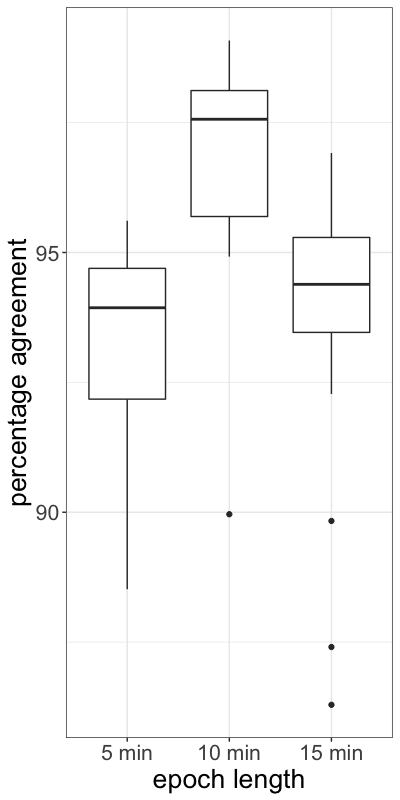}
	\label{fig:epoch_length}}
	\hspace{0.3in}
	\subfloat[]{\includegraphics[width=.225\linewidth]{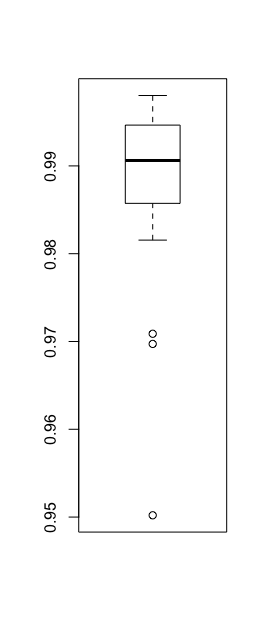}
    \label{fig:kmeans}}
	\caption{\textbf{(a)}  Box-plot of HMM-FLDA event label agreement rates between three different epoch lengths showing that the choice of $\delta_t=10$  min comes closest to consensus.  \textbf{(b)} Box-plotshowing high agreement rates between abnormality detection based on marginal and multivariate k-means clustering among all subjects.}
\end{figure*}

\noindent{\bf Local optimality of selected epoch length for 0-270 hour data}:
 To check the sensitivity of our choice of 10 min epoch length $\delta_t$ we compared three different epoch lengths  5, 10 and 15mins, on the full data (0-270 hours). 
Figure \ref{fig:epoch_length}  shows a consensus criterion that is defined as the {\em agreement rate} between the sleep/wake classifications of the HMM-FLDA pipeline using the 5, 10 and 15 minute epoch lengths, respectively.   Figure \ref{fig:epoch_length} shows that the  10   min epoch length results in sleep/wake labels that achieve the highest agreement rate.  In this sense, $\delta_t=10$ mins. is locally optimal for segmenting the 0-270 hour data.   If other types of events are of primary interest, e.g., higher temporal resolution sleep stage analysis or sleep quality analysis, for which the presence of outliers may be useful information, shorter epoch lengths might work better. 

\noindent{\bf Choice of dimension in the $k$-means clustering used for abnormality filtering (0-270 hours)}: In the context of the HVC data with the full 270 hours of data available, Fig. \ref{fig:kmeans} compares k-means clustering for different feature dimensions in the context of abnormality filtering described in Sec. \ref{subsec:abnormalityfiltering}. Specifically, the case of multivariate k-means, where k-means is applied to local feature vectors in their full 12 dimensions, is compared to the case of marginal k-means, where it is applied to each feature dimension  independently.  For each subject we  calculated the agreement rate (percentage agreement) between the multivariate and marginal k-means implementation, where agreement  is defined as the number of times they agree that a time sample is abnormal over the subject's 270 hours of data. The box plot in Fig. \ref{fig:kmeans}  shows the  distribution over the 25 subjects and indicates that there is greater than 0.98 agreement rate for all but 3 subjects whose agreement rates are all still greater than 0.95. Thus the simpler marginal k-means is virtually equivalent to the multivariate k-means.

\subsubsection{Example: abnormality filtering for two of the  subjects  (0-270 hours)}



\begin{figure}[ht]
  \centering
  \subfloat[]{\includegraphics[width = 0.476\linewidth, valign=t]{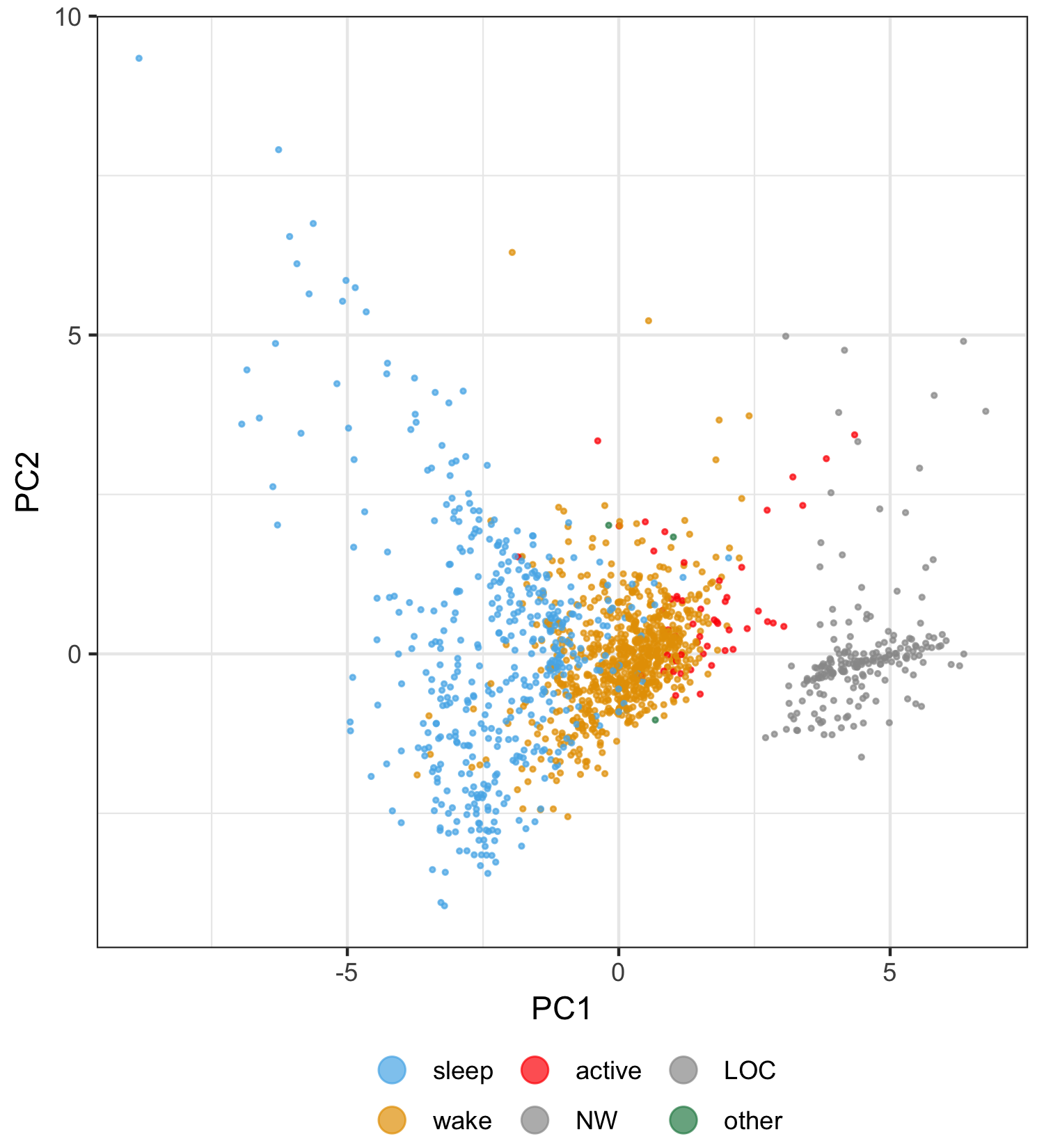}
  \label{fig:norm}}
  \subfloat[]{\includegraphics[width = 0.48\linewidth, valign=t]{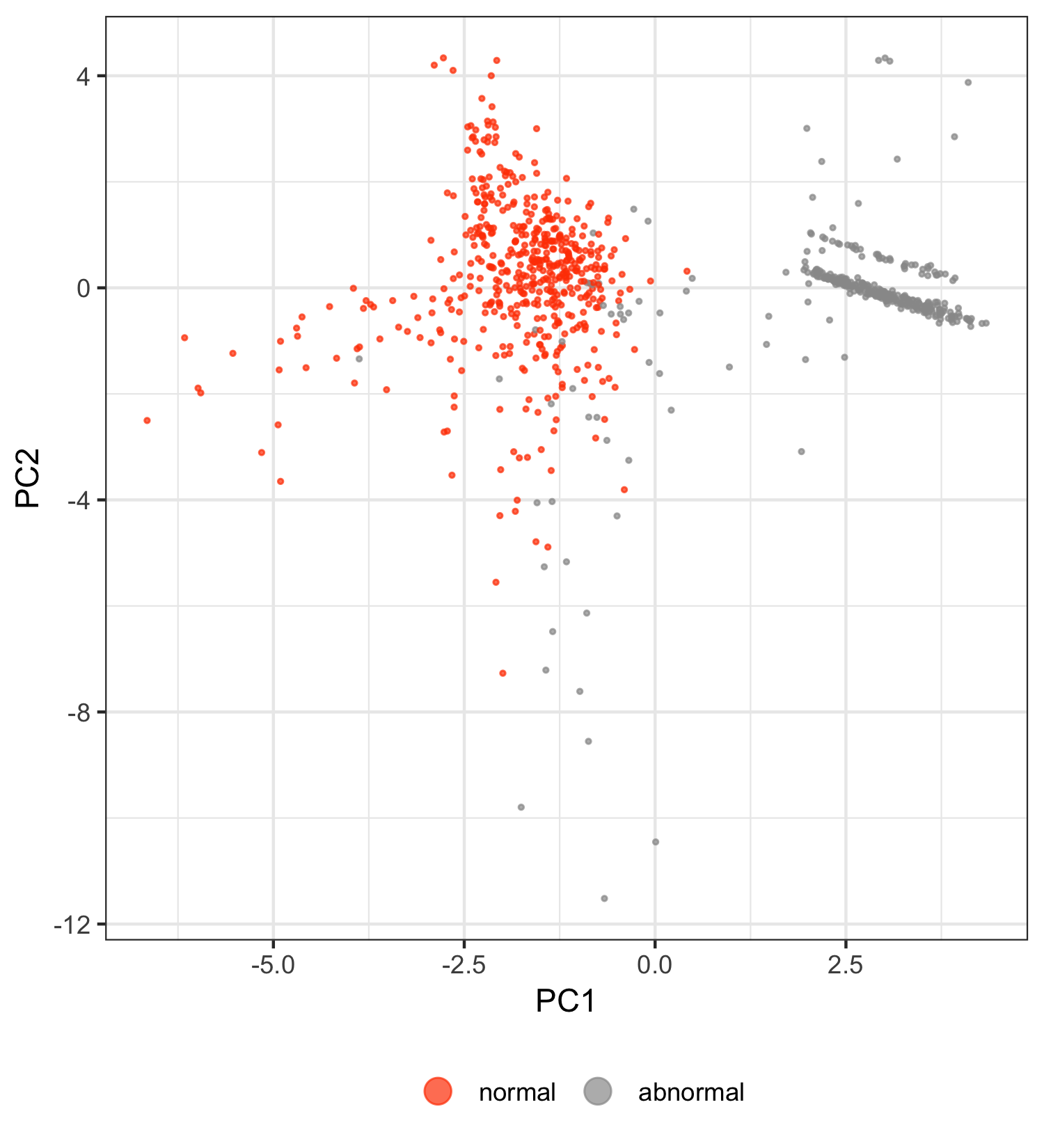}
  \label{fig:abnorm}}
  \caption{(\textbf{a}) scatter plot of the first two PCs for a subject with usable data (abnormal proportion < 40\%); (\textbf{b}) scatter plot of the first two PCs for a subject with unusable data (abnormal proportion > 40\%)}. 
  \label{fig:pca_subj}
\end{figure}

For the purposes of visualization of the differences between the features of subjects with usable and  unusable data, we show the results of principal components analysis (PCA) of two subjects  in Figures~\ref{fig:norm} and \ref{fig:abnorm}. These figures show scatter plots of the first two principal components of all 12 temporally localized features over the full 0-270 hour data.
Both participants have clusters of outliers (grey colored points falling on the right side of the figures) that were identified as non-normal by the abnormality filter in our processing pipeline of Fig. \ref{fig:pipe}. For the subject in  Figures~\ref{fig:abnorm} the non-normal data corresponded to a HR and TEMP features that are highly abnormal (mean TEMP $< 20^o$C and mean HR $>160bpm$) over long time periods, suggestive of a device that is either malfunctioning, not being worn properly, or not being worn. This latter subject was one of the 5 subjects that was eliminated from the analysis since more than $40\%$ of all his time points were detected as abnormal.   For the subject in Figure~\ref{fig:norm} the detected non-normal data corresponded to a short time period where the HR and ACC features exhibited high median and high standard deviation, suggestive of a short session of intense physical exercise. This subject was included in the analysis since less than $40\%$ of the time points were detected as abnormal. For this subject, the abnormality classifier described in Appendix \ref{sec:abnormalityclassification} classified these non-normal data points as "Active," and,  based on their time stamps, inserted them into back into the corresponding wake session determined by the HMM-FLDA procedure. 


\noindent{\bf Example: segmentation results for two subjects (0-270 hours)}

\begin{figure*}[!ht]
	\centering
	\subfloat[]{\includegraphics[width=.6\linewidth]{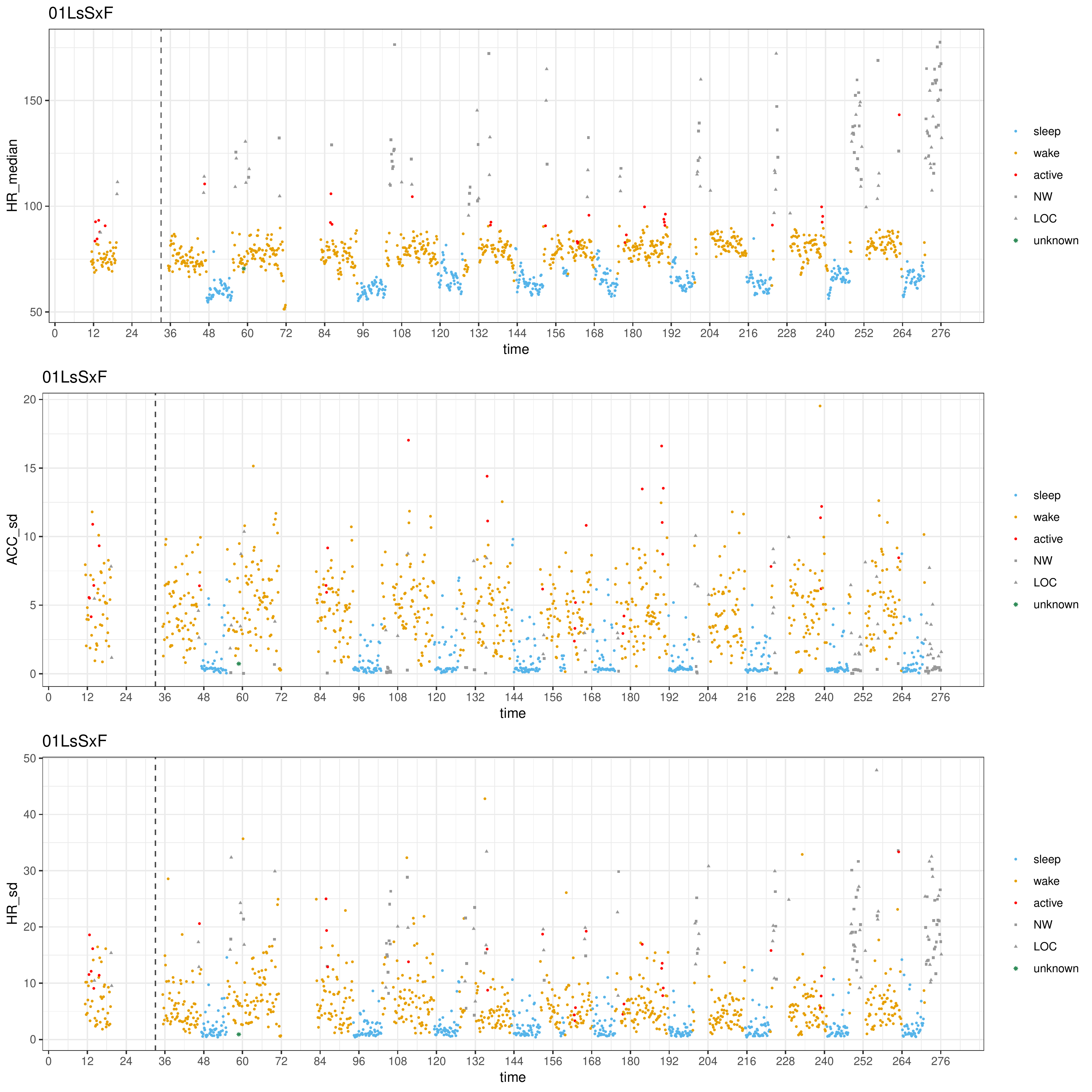}
	\label{fig:1_sleep}}
	\hfil
	\\
	\subfloat[]{\includegraphics[width=.6\linewidth]{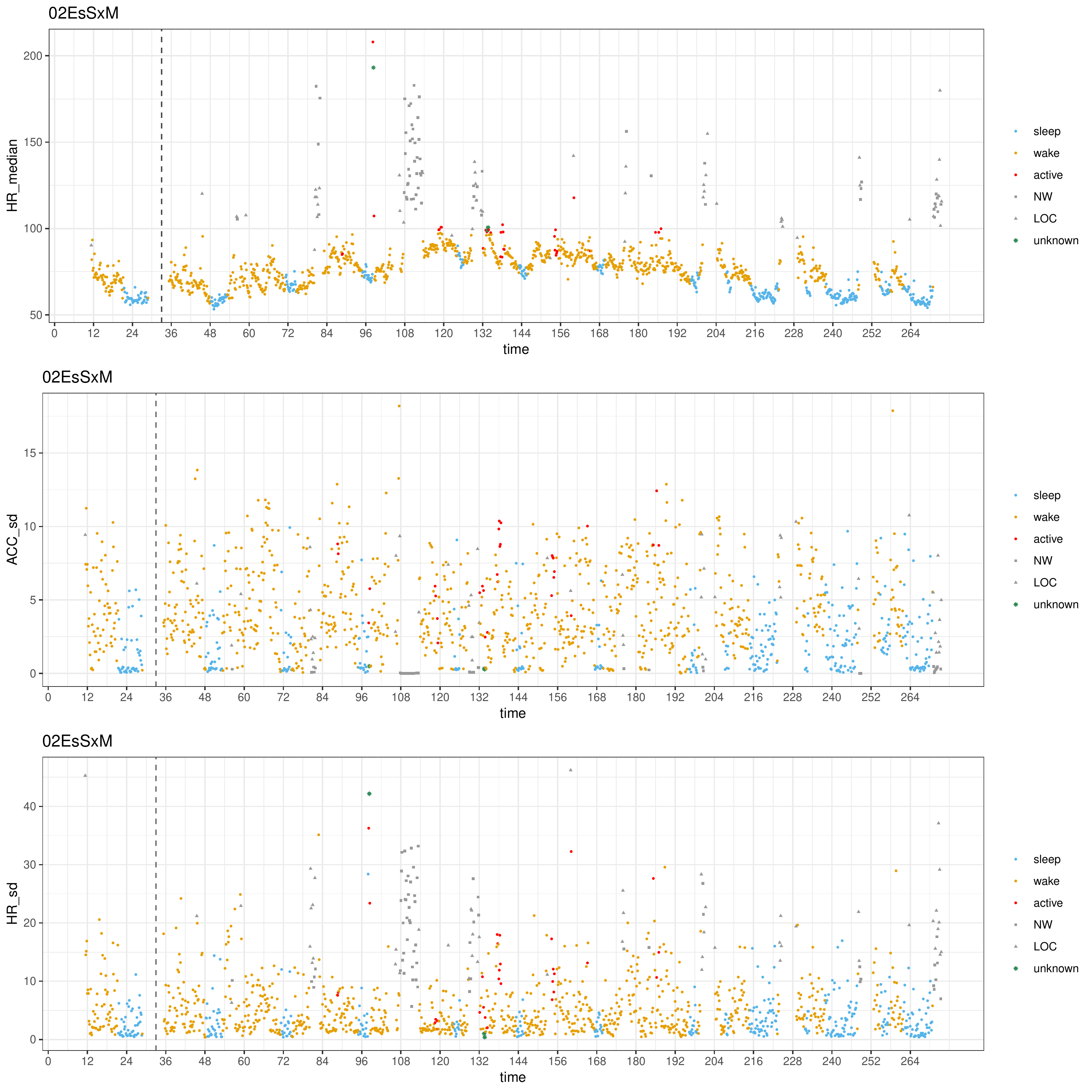}
	\label{fig:2_sleep}}
	\caption{Processed segmented time courses for Subjects 1 and 2 shown in Fig. \ref{fig:timecourseHVC} for the 3 event classification features, HR median, ACC sd and HR.sd, color coded according to the abnormality classification categoires: sleep, wake, active, not-worn (NW), loss-of-contact (LOC), and other (Unknown).   \textbf{(a)} segmented data from non-infected Subject 1; 
	\textbf{(b)} segmented data from  infected Subject 2. 
	} 
	\label{fig:timecoursesegmented}
\end{figure*}

Figure \ref{fig:timecoursesegmented} shows the sleep segmented data for two subjects with usable data (Subject 1 and Subject 2 shown in Fig. \ref{fig:timecourseHVC}) at the output of the pipeline  of Fig. \ref{fig:pipe} when all time points (0-270 hours) of all subjects are available for processing.    

\clearpage
\section{Clinical outcome prediction}
\label{sec:clinicaloutcome_APP}


Here we give additional information relevant to  prediction of clinical outcomes, infection/non-infection status and shedding onset time,  as discussed in Section \ref{sec:analysis}. 

Recall that the pipeline produces high dimensional sleep/wake labeled features obtained as statistical summarizations of the local E4 features. Two implementations are evaluated: offline and online feature extraction. The offline implementation accesses the subject's full data (all sampling times, 0-270 hrs), using it to perform anomaly detection, HMM-FLDA sleep/wake segmentation and feature extraction. The online implementation only has access to the pre-infection data (sample times over 0-60 hours) to perform these operations. The offline and online versions of the pipeline both ended up rejecting 5 of the subjects with excessive missingness and abnormality, and for each non-rejected subject  generates a set of features described in Table \ref{table:featuresHVC}.  

In Table \ref{table:infectionstatus} are shown the distribution of subjects over the clinical outcomes, in the separate categories of infection status and infection onset time. Here infection status is the binary outcome that a subject will or will not shed virus over the course of 270 hours of the confinement phase of the challenge study. Infection onset time is a ternary response variable indicating the categories "Early onset" (shedding begins on day 4), "Late onset" (shedding begins on day 5), and "No onset" (shedding not detected on any day).

The 1st row of
Table  \ref{table:infectionstatus} shows numbers of subjects in each infection category among all 25 subjects having E4 data records. The 2nd row shows the same statistics for the 20 subjects with sufficient quantity and quality data as determined by the missingness and abnormality filtering implemented by the offline feature extraction pipeline having access to the full time course (0-270 hrs). The 3rd row shows these statistics for the 20 subjects with sufficient quantity and quality data determined by the online feature extraction pipeline (0-60hrs). 

Note that the classes are moderately well balanced for the binary infection status class but are unbalanced for the ternary shedding onset time, with the early class having only 2 or 3 subjects. We address this class imbalance using a compensation method in Section \ref{sec:clinicaloutcome_APP}.


\begin{table*}[ht]
\centering
  \caption{First row: infection status shedding onset time of all 25 subjects having E4 data records. Second row: infection status of the 20 subjects with sufficient non-anomalous  data over 0-270 hours as determined by the offline feature extraction pipeline having access to the full time course  data (0-270 hours). Third row: infection status of the 20 subjects with sufficient non-anomalous data as determined by the online feature extraction pipeline having access only to the pre-infection time course data (0-60 hours).}
  \resizebox{0.75\textwidth}{!}{
\begin{tabular}{c c c c c c c}
\hline\hline
 \multirow{2}{*}{Subject group} & \multicolumn{2}{c}{Infection Status} &  & \multicolumn{3}{c}{Shedding onset time} \Tstrut\\ [+3pt]\cline{2-3} \cline{5-7} 
                           & Infected        & Not infected       &  & Early     & Late     & No onset   \Tstrut \\[+3pt] \hline
25 subjects w/ E4 data records               & 11              & 14                 &  & 3         & 8          & 14      \Tstrut \\[+3pt]
20 subjects w/ sufficient data (0-270 hrs)       & 8               & 12                 &  & 2         & 6          & 12      \\[+3pt]
20 subjects w/ sufficient data (0-60 hrs)     & 9               & 11                 &  & 3         & 6          & 11      \\[+3pt]
\hline\hline
\end{tabular}
}
  \label{table:infectionstatus}
\end{table*}


To assess the value of each of the 196 features for predicting clinical outcomes in Table \ref{table:infectionstatus} to types of univariate classifiers were implemented on each feature and their performance was evaluated using the area under the receiver operating characteristic (ROC) curve. 

For classifying the binary infection status the logistic regression (LR) classifier  
was implemented. The LR classifier fits the probability $Pr(Y=y)$, $y\in \{0,1\}$, to the generalized linear model:
\[
  \mbox{logit}\bigl[Pr(Y = 1|X)\bigr] = \beta_0 + \beta_1X \,,
\]
where $\mbox{logit}(x) = \log[x/(1-x)]$, $y\in \{0,1\}$ is the shedding status, and $X$ is one of the 196 features produced by the pipeline. 

For classifying the onset time we fit a \textit{continuation-ratio} (CR) regression model \cite{agresti2010analysis} to the conditional probability $P(Y=y|Y\geq y)$, where $y = 1$ for early shedders, $y=2$ for late shedders and $y=3$ for no onset (non-) shedders). This becomes an ordinal censured response model when "no onset" is identified as onset at infinity.  Mathematically, the continuation ratio is a logistic regression model for this conditional probability: 
\[
\mbox{logit}\bigl[Pr(Y = y \,|\, Y \geq y)\bigr] = \beta_0 + \beta_1X \,,
\]
for $y = 1, 2,3$.  

\subsection{Clinically discriminating features: online pre-infection timecourse feature extraction}
\label{sec:boxpath_APPpre}

We emulate online implementation of the feature extraction pipeline for infection status prediction by applying the pipeline to only the pre-infection timecourse (0-60 hours). In this case, only the first 60 hours affect the sleep segmentation and the sleep features. Some of the 196 resulting sleep/wake features in Table \ref{table:featuresHVC} are useful for predicting the the clinical outcome, i.e., infection status, of exposure to the viral challenge. Table \ref{table:top_featuresHVCpreinfection_app} shows the ten top ranking features for predicting infection vs. non-infection (left column) and for predicting infection onset (right column) where the predictors use features from inoculation day only. To obtain the rankings on the left column of the table, the AUC attained by the univariate logistic regression (LR) was computed using leave-one-out cross validation (LOOCV) on the 20 subjects who were not filtered out by Stage 1 of the pipeline, i.e., these subjects do not have excessive missingness or abnormal samples.  We also include the Accuracy (average classification error), computed using the same LOOCV procedure as used for computing AUC. Observe that the top 3 features ranked by accuracy are sleep features. Feature rankings by Accuracy and AUC are not the same since AUC is more stringent as it measures accuracy over the full range of sensitivities and specificities of the LR model. The right column of the table is obtained by ranking the features in decreasing order of min $\{$AUC(Early), AUC(Late), AUC(No onset)$\}$.



\begin{table*}[ht]
  \centering
  \caption{
 Top 10 E4 sleep/wake features capable of predicting clinical outcome (shedding)
 using only the extracted features from inoculation day (day 2) when these features are obtained from the online implementation of HMM-FLDA, for which only pre-infection data (0-60 hrs) is available to the HMM-FLDA  procedure.}
  \resizebox{.9\textwidth}{!}{%
  \begin{tabular}{l c c c c l c c c c}
    \hline\hline
    \multicolumn{4}{c}{\textit{logistic regression model}} &   & \multicolumn{5}{c}{\textit{continuation-ratio regression model}}\Tstrut\\[+3pt]
    \cline{1-4}\cline{6-10}
    \multirow{2}{*}{Feature}   & \multirow{2}{*}{Coef.} & \multirow{2}{*}{Accuracy}  & \multirow{2}{*}{AUC}   & \multirow{2}{*}{ }   & \multirow{2}{*}{Feature}   & \multirow{2}{*}{Coef.}   & \multicolumn{3}{c}{AUC}\Tstrut\\[+3pt]
    \cline{8-10}
    & & & & & & & Early & Late & No onset \Tstrut\\[+3pt]
    \cline{1-4}\cline{6-10}
    HR MED.sd (sleep) & -3.921 & 0.842 & 0.758 & & HR MED.sd (sleep) & -5.073 & 0.882 & 0.718 & 0.864 \Tstrut\\[+3pt]
    ACC SD.linear.coef1 (wake) & -14.468 & 0.750 & 0.737 & & HR MEAN.sd (sleep) & -4.466 & 0.706 & 0.628 & 0.773 \\[+3pt]
    HR MED.quad.coef2 (sleep) & 4.318 & 0.789 & 0.707 & & Total duration & -0.599 & 0.676 & 0.551 & 0.750 \\[+3pt]
    Offset & -1.452 & 0.789 & 0.697 & & Night duration & -0.616 & 0.647 & 0.500 & 0.705 \\[+3pt]
    EDA MED.linear.coef0 (wake) & 0.802 & 0.600 & 0.697 & & ACC MEAN.sd (sleep) & -2.303 & 0.500 & 0.449 & 0.602 \\[+3pt]
    EDA MEAN.linear.coef0 (wake) & 0.815 & 0.650 & 0.687 & & Offset & -0.774 & 0.618 & 0.436 & 0.750 \\[+3pt]
    HR MEAN.sd (sleep) & -4.077 & 0.632 & 0.677 & & ACC SD.linear.coef1 (wake) & -10.394 & 0.412 & 0.679 & 0.747 \\[+3pt]
    HR MED.linear.coef0 (wake) & 0.117 & 0.650 & 0.677 & & HR MEAN.linear.coef0 (wake) & 0.091 & 0.412 & 0.488 & 0.646 \\[+3pt]
    Total duration & -0.655 & 0.632 & 0.667 & & ACC MEAN.linear.coef1 (wake) & -13.907 & 0.412 & 0.512 & 0.636 \\[+3pt]
    HR SD.quad.coef2 (sleep) & 3.949 & 0.737 & 0.646 & & HR SD.sd (sleep) & -1.831 & 0.471 & 0.410 & 0.568 \\[+3pt]
    \hline\hline
  \end{tabular}%
  }
    \label{table:top_featuresHVCpreinfection_app}
\end{table*}

\begin{figure}[!ht]
 \centering
\includegraphics[width = 0.9\linewidth,valign=f]{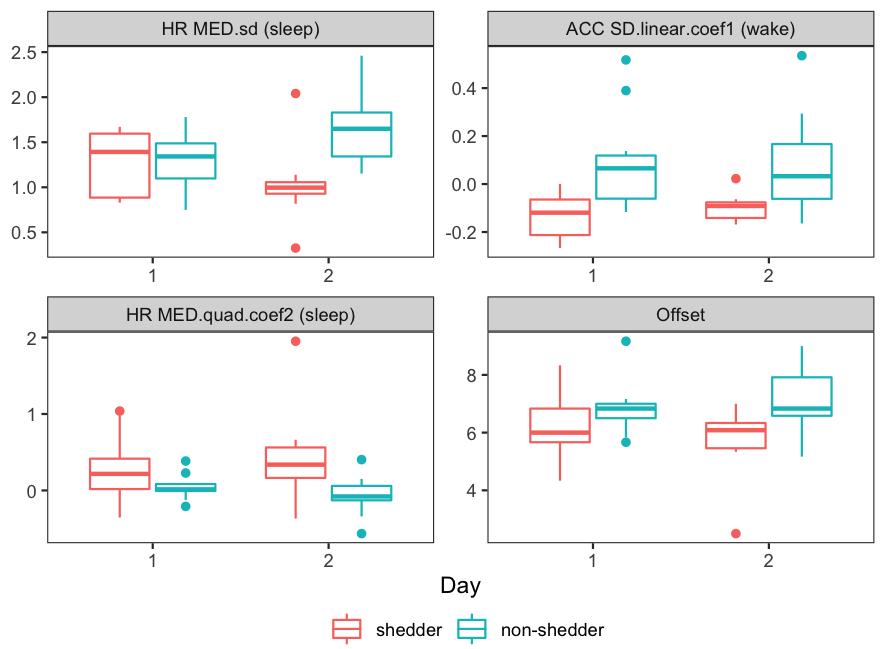}
  \caption{Boxplot of top 4 features for predicting binary infection status using univariate logistic regression  on the 196 sleep/wake features on the day of inoculation (day 2) obtained from online HMM-FLDA, which only has access to pre-infection data (0-60 hrs).   }
 \label{fig:LR60_Prometheus}
\end{figure}

\begin{figure}[!ht]
 \centering
\includegraphics[width = 0.9\linewidth,valign=f]{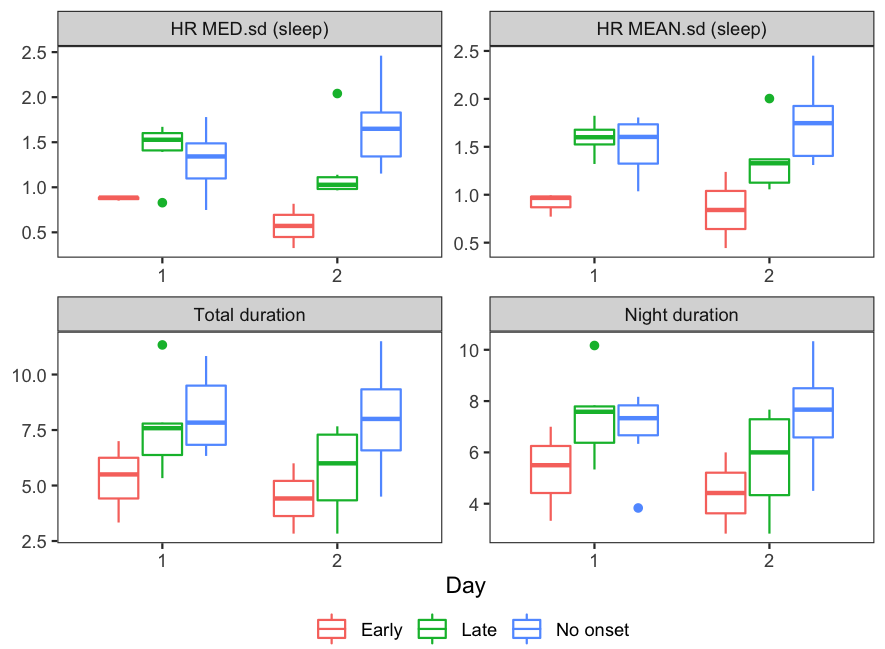}
  \caption{Boxplot of top 4 features for predicting ternary infection onset time using univariate continuation ratio regression on the 196 sleep/wake features on the day of inoculation (day 2) obtained from online HMM-FLDA, which only has access to pre-infection data (0-60 hrs).   }
 \label{fig:CR60_Prometheus}
\end{figure}

Figures \ref{fig:LR60_Prometheus} and \ref{fig:CR60_Prometheus}  show box-plots of the four top features in Table \ref{table:top_featuresHVCpreinfection_app} for logistic regression and continuation ratio regression stratified over shedders and non-shedders.

\clearpage

\subsection{Clinically discriminating features: offline full timecourse feature extraction}
\label{sec:boxpath_APPfull}

In the offline implementation, the full timecourse 
(0-270 hours) is provided to the pipeline for sleep/wake segmentation and the session-level feature set depends on all past and future time points. To these features we trained the same LR and CR infection status classification algorithms status as in the online implementation described above.
Table \ref{table:top_featuresHVCfull_app} shows the top 10 features in terms of maximizing the AUC. Many of the features in this offline table are similar to the features found by the online implementation shown in Table \ref{table:top_featuresHVCpreinfection_app}. The offline AUC tends to be higher than the online AUC in Table \ref{table:top_featuresHVCpreinfection_app}, possibly due to improved wake/sleep segmentation when more data is available for training (270 hours of data instead of 60 hours of data).

\begin{table*}[ht]
  \centering
  \caption{
  Top 10 E4 sleep/wake features capable of predicting clinical outcome (shedding)
  using only the extracted features from inoculation day (day 2) when these features are obtained from the offline implementation of HMM-FLDA, for which the full data (0-270 hrs) is available to the HMM-FLDA  procedure. 
 }
  \resizebox{.9\textwidth}{!}{%
  \begin{tabular}{l c c c c l c c c c}
    \hline\hline
    \multicolumn{4}{c}{\textit{logistic regression model}} &   & \multicolumn{5}{c}{\textit{continuation-ratio regression model}}\Tstrut\\[+3pt]
    \cline{1-4}\cline{6-10}
    \multirow{2}{*}{Feature}   & \multirow{2}{*}{Coef.}  & \multirow{2}{*}{Accuracy}   & \multirow{2}{*}{AUC}   & \multirow{2}{*}{ }   & \multirow{2}{*}{Feature}   & \multirow{2}{*}{Coef.}   & \multicolumn{3}{c}{AUC}\Tstrut\\[+3pt]
    \cline{8-10}
    & & & & & & & Early & Late & No onset \Tstrut\\[+3pt]
    \cline{1-4}\cline{6-10}
    HR MED.sd (sleep) & -6.196 & 0.750 & 0.833 & & HR MED.sd (sleep) & -7.195 & 0.944 & 0.631 & 0.844\Tstrut\\[+3pt]
    Offset & -1.629 & 0.650 & 0.802 & & Total duration & -1.139 & 0.750 & 0.619 & 0.813\\[+3pt]
    Total duration & -1.211 & 0.750 & 0.781 & & HR MEAN.sd (sleep) & -4.597 & 0.667 & 0.595 & 0.750\\[+3pt]
    HR MEAN.sd (sleep) & -4.958 & 0.650 & 0.750 & & Night duration & -0.854 & 0.667 & 0.560 & 0.698\\[+3pt]
    Night duration & -0.896 & 0.600 & 0.688 & & Offset & -0.864 & 0.611 & 0.488 & 0.802\\[+3pt]
    HR MED.mean (wake) & 0.147 & 0.750 & 0.625 & & HR MED.mean (wake) & 0.154 & 0.472 & 0.619 & 0.646\\[+3pt]
    HR MED.median (wake) & 0.149 & 0.750 & 0.625 & & HR SD.sd (sleep) & -1.877 & 0.500 & 0.464 & 0.573\\[+3pt]
    HR MEAN.median (wake) & 0.142 & 0.750 & 0.615 & & HR MED.median (wake) & 0.150 & 0.444 & 0.595 & 0.625\\[+3pt]
    ACC SD.linear.coef1 (wake) & -7.160 & 0.550 & 0.615 & & HR MEAN.mean (wake) & 0.151 & 0.417 & 0.607 & 0.635\\[+3pt]
    HR MEAN.mean (wake) & 0.141 & 0.750 & 0.604 & & HR MEAN.median (wake) & 0.142 & 0.417 & 0.607 & 0.625\\[+3pt]
    \hline\hline
  \end{tabular}%
  }
    \label{table:top_featuresHVCfull_app}
\end{table*}

\begin{figure}[!ht]
 \centering
 \includegraphics[width = 0.9\linewidth, valign=t]{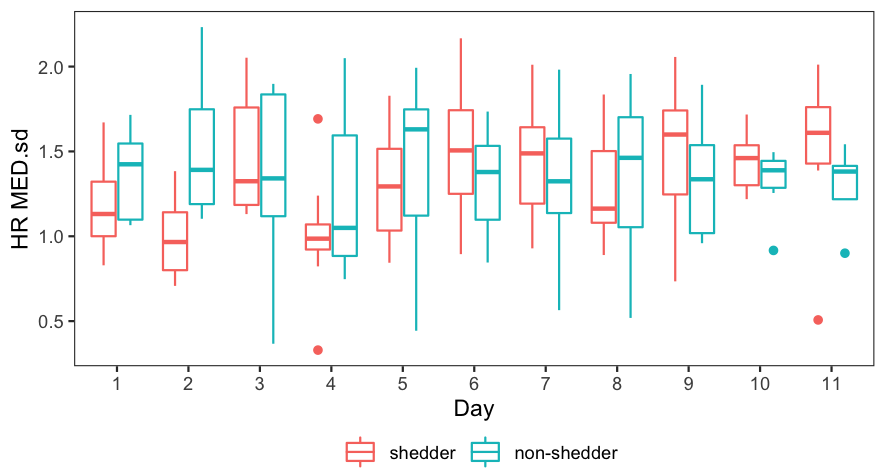}
   \caption{ HR MED.sd (Standard deviation of median heart rate)  sleep feature generated by the proposed HMM-FLDA pipeline when implemented offline, i.e., the HMM-FLDA sleep/wake segmentation is computed assuming availability of the full time course (0-270 hours) human viral challenge study (HVC) data.
   Viral inoculation  took place on day 2 and all viral shedders started shedding on day 4 or day 5.
   Note that, as compared to the non-shedders, the shedders tend to have lower sleeping heart rate variation in the first two days. 
   }
 \label{fig:box_duration1}
\end{figure}
\begin{figure}[!ht]
 \centering
 \includegraphics[width = 0.9\linewidth, valign=t]{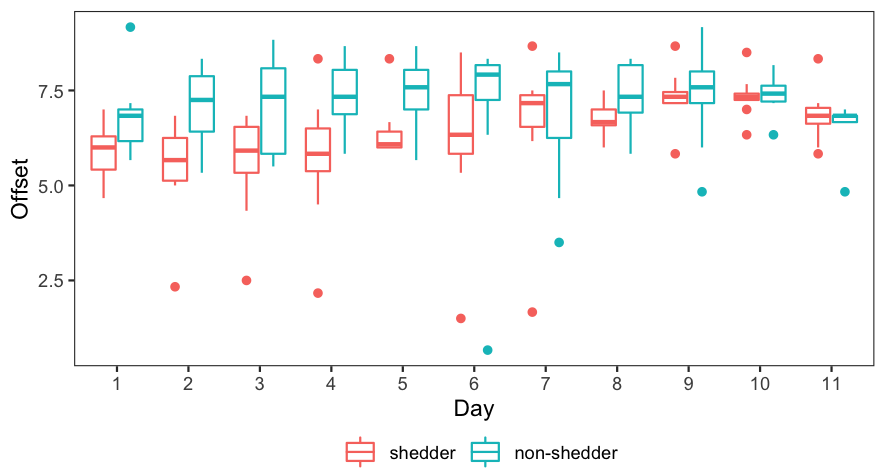}
   \caption{Sleep offset (wakeup time from night sleep) feature generated by the proposed HMM-FLDA pipeline when implemented offline, i.e., the HMM-FLDA sleep/wake segmentation is computed assuming availability of the full time course (0-270 hours) human viral challenge study (HVC) data.
   Viral inoculation  took place on day 2 and all viral shedders started shedding on day 4 or day 5.
   Note that, as compared to the non-shedders, during the first 8 days of the study the shedders tend to wake up earlier than  the non-shedders.  
   }
 \label{fig:box_duration2}
\end{figure}
\begin{figure}[!ht]
 \centering
 \includegraphics[width = 0.9\linewidth, valign=t]{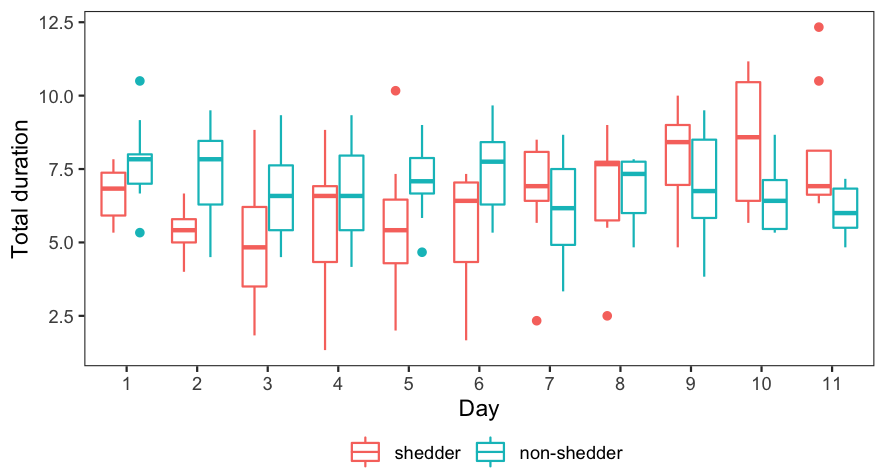}
   \caption{Total duration (the sum of all night and day sleep hours) sleep feature generated by the proposed HMM-FLDA pipeline when implemented offline, i.e., the HMM-FLDA sleep/wake segmentation is computed assuming availability of the full time course (0-270 hours) human viral challenge study (HVC) data.
   Viral inoculation  took place on day 2 and all viral shedders started shedding on day 4 or day 5.
As compared to the non-shedders, the shedders tend to have a sleep deficit until the 7th day. 
   }
 \label{fig:box_duration3}
\end{figure}
\begin{figure}[!ht]
 \centering
 \includegraphics[width = 0.9\linewidth, valign=t]{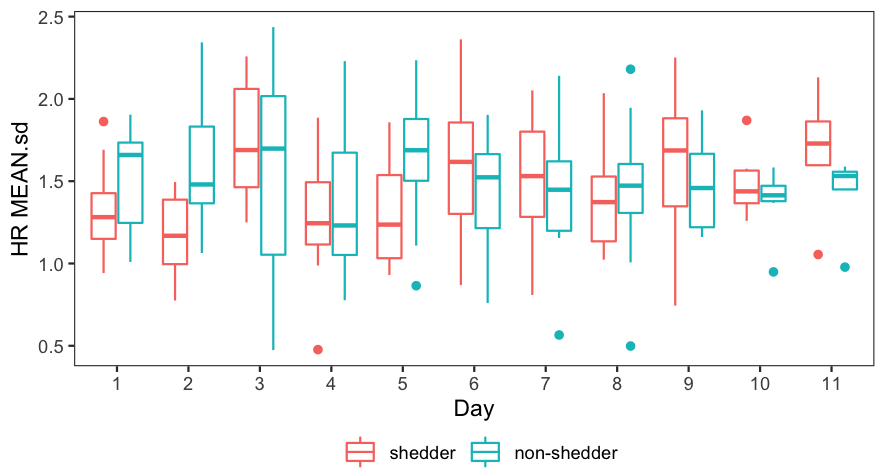}
 \caption{HR MEAN.sd (standard deviation of the heart rate mean) sleep  feature generated by the proposed HMM-FLDA pipeline when implemented offline, i.e., the HMM-FLDA sleep/wake segmentation is computed assuming availability of the full time course (0-270 hours) human viral challenge study (HVC) data.    Viral inoculation  took place on day 2 and all viral shedders started shedding on day 4 or day 5.     }
\label{fig:box_duration4}
\end{figure}
\begin{figure}[!ht]
 \centering
 \includegraphics[width = 0.9\linewidth, valign=t]{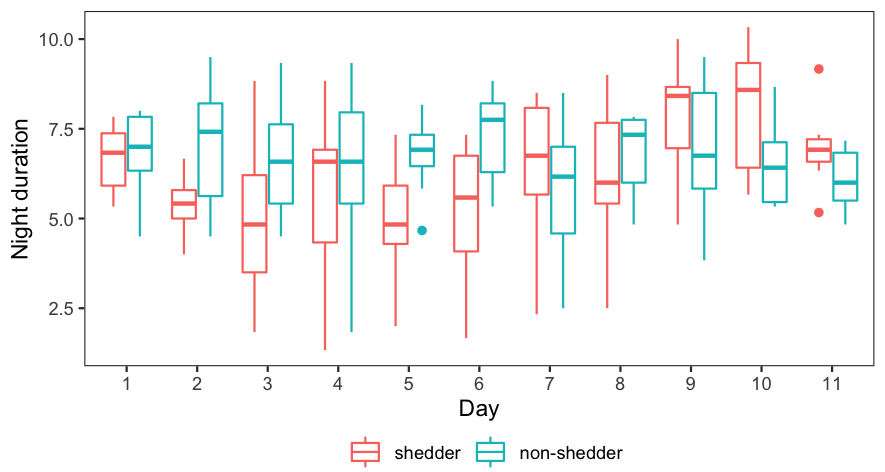}
   \caption{Night sleep duration feature generated by the proposed HMM-FLDA pipeline when implemented offline, i.e., the HMM-FLDA sleep/wake segmentation is computed assuming availability of the full time course (0-270 hours) human viral challenge study (HVC) data.
   Viral inoculation  took place on day 2 and all viral shedders started shedding on day 4 or day 5.
   As compared to the non-shedders, the shedders tend to have a night sleep deficit until the 7th day. Compare to the total duration of sleep feature shown in \ref{fig:box_duration3} that includes duration of both night-time and daytime sleep. 
   }
 \label{fig:box_duration5}
\end{figure}

Figures \ref{fig:box_duration1}-\ref{fig:box_duration5}  show box-plots of the top 5 top logistic regression features (first column) in Table \ref{table:top_featuresHVCfull_app} stratified over shedders and non-shedders.

\clearpage
\subsection{Extensions of marginal clinical outcome predictor}
\label{sec:featurecorr}

Two extensions to the univariate regression framework for clinical outcome prediction are presented here. First a multivariate logistic regression and continuation ratio classifier are implemented on the top three univariate features for the full timecourse segmentation data. Then we apply a method that improves (univariate or multivariate) classifier performance for data with imbalanced class sizes.   


\subsubsection{ Multivariate classifiers}
When the top three univariate logistic regression features in Table \ref{table:top_featuresHVCfull_app} (HR MED.sd, offset and total duration) are combined into  a single model, the multivariate logistic model achieves  $\mbox{AUC} = 0.844$, and the continuation-ratio model achieves $\mbox{AUC} = 0.885$ for classifying ``Late shedder versus others.''  This represents only a moderate improvement relative to the corresponding marginal AUC's listed in the table. 

Table  \ref{tab:cor_top} shows the Pearson correlation between the 3 features having maximum marginal area-under-the-curve (AUC) of the receiver operating characteristic curve (ROC) for classifying clinical outcome, presented in Section \ref{sec:analysis}.  The presence of significant correlation between feature over the population explains why multivariate LR and CR classifiers of clinical outcome can only attain marginally better AUC than do the univariate LR and CR classifiers.

\begin{table}[ht!]
  \centering
  \caption{Pearson correlation coefficients (p-values) between top 3 features selected using AUC criterion for univariate logistic regression and continuation ratio classifiers of clinical outcome.}
  \begin{tabular}{ l  c c c}
    \hline\hline
                   & Total duration & Offset & HR MED.sd \Tstrut\Bstrut\\[+3pt]
    \hline
    Total duration & 1               & 0.778 (< 0.001)  & 0.434 (0.056) \Tstrut\\[+3pt]
    Offset         &                 & 1      & 0.485 (0.030) \\[+3pt]
    HR MED.sd      &                 &        & 1 \\[+3pt]
    \hline\hline
      \end{tabular}
    \label{tab:cor_top}
\end{table}

\subsubsection{Imbalanced class sizes}: It is well known that imbalanced class size in the training data can negatively affect performance, especially for the under-represented classes. 
To mitigate the impact of these imbalanced shedding class sizes, we applied the synthetic minority over-sampling technique (SMOTE) \cite{chawla2002smote} to rebalance the class sizes. 
SMOTE is a method for equalizing minority class sizes by introducing synthetic data samples that rebalances to match the majority class size. It does this rebalancing as follows. First a feature instance $x$ is randomly selected from the minority class. Then for this point $x$ the $k$ minority class nearest neighbors $\{x_i\}_{i=1}^k$ are identified. One of these nearest neighbors $x_R$ is selected as random and a random variable $U$ is drawn from the uniform distribution over $[0,1]$. Finally, the new SMOTE point in the minority class is defined as $s=x+U*(x_R-x)$. This process is repeated until the minority class has as many samples as the majority class. 

SMOTE can be applied to any classifier but can be expected to work best when the classification regions are convex in feature space.   SMOTE is commonly applied to cases where the class imbalance is moderately small, e.g., the ratio of the smallest to largest class sizes is greater than 0.5. For the HVC data, Table \ref{table:infectionstatus} indicates that for the binary infection status outcome the imbalance is moderate (class size ratio $>$ 2/3) but for the ternery infection onset outcome the imbalance is severe (class size ratio can be as low as $1/3$).  

First we considered compensation of class imbalance in the univariate logistic and continuation ratio regression predictors of shedding and shedding onset 24 hours before any subject starts to shed virus. SMOTE was applied using $k=3$ nearest neighbors for each minority class except for the early onset shedder group where only 1 nearest neighbor was used since this group only contains 2 or 3 samples. Since SMOTE is a randomized algorithm,  100 SMOTE runs were performed. Recall that HR MED.sd was the top ranked feature for both LR and CR and for both the full data and the pre-infection data. 

We implemented SMOTE for LR and CR on the univariate predictor using HR.MED.sd and on the multivariate predictor using HR.MED.sd, Total Duration and Offset. We considered both the offline case (sleep/wake segmentation was performed by FLDA using the full data, 0-270hrs) and the online case (sleep/wake segmentation was performed by FLDA using only the pre-infected data, 0-60hrs).  

The first 3 rows  and the last 3 rows of Table \ref{table:SMOTE} shows the Accuracy (average classification error rate) and the AUC for logistic regression (LR), and the AUC for continuation ratio regression (CR) in the univariate case. The right columns of the table indicate significant AUC performance improvement for CR, for which class size is severely imbalanced (P-VALUE is the result of double sided paired t-test). In the case of the LR, on left side of table, the class imbalance is not as severe and SMOTE leads to no improvement in AUC or Accuracy for the univariate case. Indeed, SMOTE compensated univariate LR has worse Accuracy (p-value $<$ 0.01 according to double sided t-test) when either offline or online sleep/wake segmentation is used.

The middle 3 rows of the table show the results of applying SMOTE to multivariate logistic and continuation ratio regression for the offline case. 
For multivariate CR, SMOTE improves the AUC (p-value $< $ 0.01 according to double sided t-test). For multivariate LR, SMOTE improves the AUC and the Accuracy only in the offline case.


\begin{table}[ht!]
  \centering
  \caption{SMOTE compensation for imbalanced sample size for predictors of shedding and shedding onset time 24 hours before first shedding occurs.}
\begin{tabular}{l l c l c c l c c c}
\hline\hline
& \multirow{2}{*}{Feature(s)} \Tstrut                                    &     &   & \multicolumn{2}{c}{LR} \Tstrut &       & \multicolumn{3}{c}{CR AUC} \Tstrut     \\ [+3pt] \cline{5-6} \cline{8-10} 
 &     &   &    & Accuracy   & AUC       & \multicolumn{1}{c}{} & Early    & Late     & No onset \Tstrut \\[+3pt] \hline
\multirow{6}{*}{\begin{tabular}[c]{@{}c@{}}Offline training \\(FLDA trained on 0-270hrs)\end{tabular}}   & \multirow{3}{*}{HR MED.sd (sleep)}                             & w/o SMOTE              &         & 0.750      & 0.833     &                      & 0.944    & 0.631    & 0.844    \Tstrut \\[+3pt]
 &     & \multirow{2}{*}{SMOTE} & MEAN    & 0.703      & 0.846     &                      & 0.951    & 0.656    & 0.907    \\[+3pt]
  & &  & P-VALUE & 4.68E-11   & 0.015     &                      & 1.38E-06 & 6.68E-07 & 2.80E-56 \\[+3pt]
                                          & \multirow{3}{*}{\begin{tabular}[c]{@{}c@{}}Total    duration + Offset\\  + HR MED.sd (sleep)\end{tabular}} & w/o SMOTE              &         & 0.800      & 0.844     &                      & 0.889    & 0.655    & 0.885    \\[+3pt]
                                          &                                                                & \multirow{2}{*}{SMOTE} & MEAN    & 0.838      & 0.865     &                      & 0.974    & 0.811    & 0.937    \\[+3pt] 
                                          &                                                                &                        & P-VALUE & 2.21E-21   & 6.78E-35  &                      & 1.66E-71 & 3.28E-83 & 1.38E-75 \\ [+3pt] \hline
\multirow{3}{*}{\begin{tabular}[c]{@{}c@{}}Online training \\(FLDA trained on 0-60hrs)\end{tabular}} & 
                        \multirow{3}{*}{HR MED.sd (sleep)}                             & w/o SMOTE              &         & 0.842      & 0.758     &                      & 0.882    & 0.718    & 0.864  \Tstrut  \\ [+3pt] 
                                          &                                                                & \multirow{2}{*}{SMOTE} & MEAN    & 0.781      & 0.727     &                      & 1.000    & 0.751    & 0.899    \\[+3pt] 
                                          &                                                                &                        & P-VALUE & 8.23E-13   & 3.00E-03  &                      & -        & 5.98E-06 & 1.69E-25 \\ [+3pt] 
                                          \hline\hline
\end{tabular}
\label{table:SMOTE}
\end{table}




\end{document}